\documentclass[a4paper,11pt]{article}
\pdfoutput=1 
\usepackage{paper_jheppub}
\usepackage{subfigure}
\usepackage{tabularx}
\usepackage[T1]{fontenc}

\newcommand{\softQCD}{\textit{softQCD}\,}
\newcommand{\hardQCD}{\textit{hardQCD}\,}

\newcommand{\smin}{\ensuremath{s'_{\mbox{min}}}}
\newcommand{\Ntrk}{\ensuremath{N_{\mbox{Trk}}}}
\newcommand{\Etatrk}{\ensuremath{\eta_{Trk}}}

\newcommand{\mInst}{\ensuremath{m_{I}}}

\newcommand{\NJets}{\ensuremath{N_{\mbox{Jets}}}}
\newcommand{\NDisplaced}{\ensuremath{N_{\mbox{Displaced}}}}
\newcommand{\BTracks}{\ensuremath{{\cal B}_{\mbox{Tracks}}}}
\newcommand{\TTracks}{\ensuremath{{\cal T}_{\mbox{Tracks}}}}

\newcommand{\inpb}{\ensuremath{\mbox{pb}^{-1}}}

\newcommand{\pT}{\ensuremath{p_{\mathrm{T}}}}

\def\ifb{\mbox{fb$^{-1}$}}
\def\ipb{\mbox{pb$^{-1}$}}
\def\GeV{\ifmmode {\mathrm{\ Ge\kern -0.1em V}}\else \textrm{Ge\kern -0.1em V}\fi}%
\def\MeV{\ifmmode {\mathrm{\ Me\kern -0.1em V}}\else \textrm{Me\kern -0.1em V}\fi}%
\def\TeV{\ifmmode {\mathrm{\ Te\kern -0.1em V}}\else \textrm{Te\kern -0.1em V}\fi}%

\title{\boldmath How to discover QCD Instantons at the LHC\footnote{Work was prepared in the context of the workshop \href{https://indico.cern.ch/event/965112}{"Topological Effects in the Standard Model: Instantons, Sphalerons and Beyond at LHC"} in December 2020. A higher quality version of this work will become available shortly}}

\author[a]{Simone Amoroso}
\author[b]{Deepak Kar}
\author[c]{Matthias Schott\footnote{corresponding author}}

\affiliation[a]{DESY, Hamburg, Germany}
\affiliation[b]{University of Witwatersrand, South Africa}
\affiliation[c]{Johannes Gutenberg-University, Mainz, Germany}

\emailAdd{matthias.schott@cern.ch}

\abstract{
The Standard Model of particle physics predicts the existence of quantum tunnelling processes across topological inequivalent vacua, known as Instantons.
In the electroweak sector, instantons provide a source of baryon asymmetry within the Standard Model. In Quantum Chromodynamics they are linked  to chiral symmetry breaking and confinement. The direct experimental observation of Instanton-induced processes would therefore be a breakthrough in modern particle physics. Recently, new calculations for QCD Instanton processes in proton-proton collisions became public, suggesting sizable cross sections as well as promising experimental signatures at the LHC. In this work, we study possible analysis strategies to discover QCD Instanton induced processes at the LHC and derive a first limit based on existing Minimum Bias data. 
}

\begin{document}

\maketitle

\newpage
\section{\label{Sec:Intro}Introduction}
Yang-Mills theories~\cite{Yang:1954ek}, embedded in the Standard Model (SM) of particle physics, form the basis of our understanding of the strong and electroweak interactions.
The beauty and success of the SM lies in its predictive power, which is however only achieved in the weakly coupled regime.
Perturbation theory, developed in order to describe hadron collisions at high energies,
relies on the smallness of the strong coupling at high momentum transfers and short distances.
The study of Quantum Chromodynamics (QCD) in this perturbative regime has seen tremendous advancements in the last decades.  
Hard processes have been calculated up to the third order in the strong coupling~\cite{Mistlberger:2018etf,Duhr:2020seh},  and perturbative QCD predictions have been verified to incredible precision over many order of magnitudes of momentum transfer at high energy colliders~\cite{denterria20}. 
At the same time a fundamental understanding of Yang-Mills theories in the strongly coupled limit is still lacking, and remains one of the biggest challenges for particle physics to date.

\begin{figure}[th]
\centering
\begin{minipage}{9.8cm}
\centering
\includegraphics[width=1.0\linewidth]{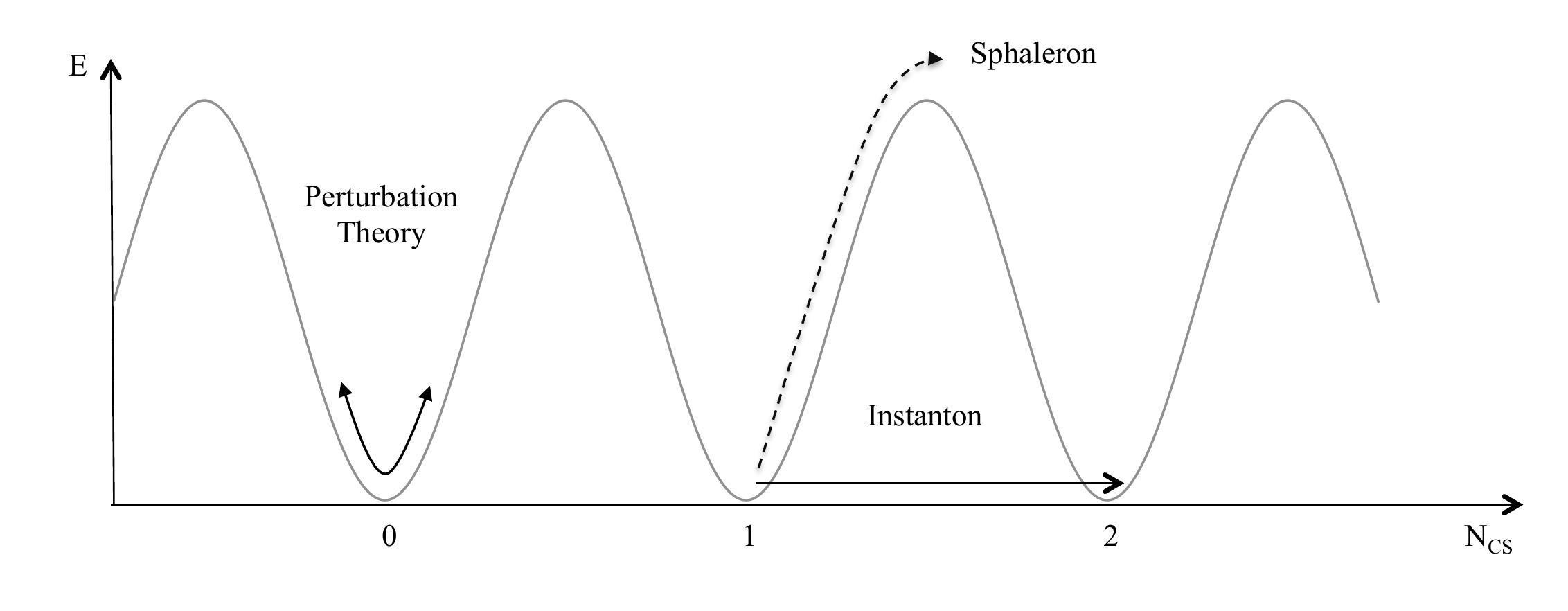}
	\caption{Instanton and Sphaleron processes in the topology of a Yang-Mills vacuum; energy density of the gauge field (y-axis) vs.  winding number $N_{CS}$ (x-axis). }
	\label{fig:YMVacua}
\end{minipage}
\end{figure}

Unlike the abelian case, Yang-Mills theories exhibit a rich and non-trivial vacuum structure. In particular, they admit semi-classical solution corresponding to fluctuations of the gauge fields across topologically non-equivalent vacua, the Instantons~\cite{Belavin:1975fg}.
These inherently non-perturbative phenomena are of great theoretical interest (introductory reviews on the physics of instantons can be found in \cite{Shuryak:1981ff, Arnold:1987mh, Forkel:2000sq, Shuryak:2018fjr}). 
The vacuum structure of a Yang-Mills theory is depicted in Figure \ref{fig:YMVacua}, showing the energy density of the gauge field as a function of the Chern-Simons (or winding number), $N_{CS}$, characterising the topological charge of a system. Instantons describe tunneling transitions in Minkowski spacetime between classically degenerate vacua, which only differ by their winding number by one unit, i.e. $\Delta N_{CS} = 1$. Instanton solutions are not only localised in time, but also in space, i.e. they have a certain spatial extension. 
There exists also a second class of classical solutions, known as Sphalerons, corresponding to transitions from one vacuum by a half-integer winding number on top of the energy barrier (also shown in Figure \ref{fig:YMVacua}), where its static energy corresponds to the barrier height.

These tunnelling solutions differ significantly from the solutions known from ordinary perturbation theory, where only those field configurations are accessible which correspond to small changes of the vacuum field at $N_{CS} = 0$,  while other minima, which are not accessible by continuous transformation of the gauge field, are ignored. Clearly, this approximation holds only as long as the energy barrier between the vacua is sufficiently large. Instanton and Sphaleron solutions provide crucial ingredients for an understanding of a number of non-perturbative issues in the SM.  In the electroweak theory, Instanton and Sphaleron transitions  are associated to $B+L$ violation. These become highly relevant at high temperatures \cite{Kuzmin:1985mm, Arnold:1987mh,Farrar:1993sp} and  have a crucial impact on the evolution of the baryon and lepton asymmetries of the universe (see also Ref. \cite{Rubakov:1996vz} for a review). In QCD these topological solutions have been argued to play an important role in various long-distance aspects of the theory. They provide a possible solution to the axial $U(1)$ problem \cite{tHooft:1986ooh} and are associated to  chiral symmetry breaking 
\cite{tHooft:1976snw, tHooft:1976rip,Diakonov:1985eg}.

The  height of the energy barrier between two vacua, called Sphaleron mass $M_{Sp}$,
in the electroweak theory is of the order of $ M_{Sp} \sim \frac{\pi}{\alpha \rho_{\mathrm{eff}}} \sim \pi \frac{M_W}{\alpha_W} \sim 10\mbox{\,TeV}$~\cite{Klinkhamer:1984di}, where $\alpha_{W}$  is the weak  coupling constant and $\rho_{\rm eff}$ the effective Instanton size.
As the energy barrier is lower than the LHC center-of-mass energy,  one might think that  electroweak Sphalerons should be produced, and could be observed at the LHC. However it was shown that the difficulty of obtaining a coherent state makes these processes  likely to remain unobservably small at current and future colliders \cite{Ringwald:2002sw, Khoze:2020paj}. 
The situation is different for QCD Instanton processes, for which the energy barrier, $M_{Sp}\sim \frac{3 \pi}{4\alpha_s\rho_{\rm eff}} \sim Q $~ \cite{Moch:1996bs}, 
with $\alpha_s$ the strong coupling and the parameter $Q$ related to the energy scale of the underlying process, can be as low as a few \GeV. 
Searches for Instanton processes have been performed in Deep Inelastic Scattering at the HERA collider~\cite{Adloff:2002ph,Chekanov:2003ww,H1:2016jnv} already excluding the lower range of the predicted cross-sections.
It is then interesting to understand if these processes could be measured also at the LHC.
Recent works have provided first calculations for LHC cross-sections~\cite{Khoze:2019jta}, and some discussions on the expected phenomenology~\cite{Khoze:2020tpp}.
In this work we explore in further details suitable analysis strategies at the LHC, in particular exploring the (relatively) small-size  regime with Instanton masses of few tenths of GeVs, where the cross-section is the highest.
In this regime the challenge lies in finding suitable observables that, while retaining sensitivity to the soft decay products of the Instanton, can also be described to an acceptable level of accuracy by the non-perturbative models of soft QCD activity.

The paper is structured as follows: In section \ref{Sec:Processes} we briefly provide a review on Instanton processes at the LHC, covering their expected production cross sections and experimental signature. This is followed by an overview of the Monte Carlo samples used  in section \ref{Sec:MCSamples}. Possible search strategies and the optimisation of the event selection are described in section \ref{Sec:Search}. The expected sensitivity of the proposed analysis, as well as first limit on QCD Instanton processes are presented in section \ref{Sec:Limit}. The paper concludes in section \ref{Sec:Conclusion}.

\section{\label{Sec:Processes}QCD Instanton Processes at the LHC}

\subsection{Production of the Instanton Pseudo-Particle in Proton-Proton Collisions}

The expected inclusive cross sections of QCD Instanton-induced processes in Instanton perturbation theory \cite{Moch:1996bs, Ringwald:1998ek} can schematically be written as

\begin{equation} 
\sigma^{(I)}_{parton, parton} \sim \int_0^\infty \int_0^\infty D(\rho) D(\bar \rho) d\rho d\bar \rho \cdot e^{-(\rho+\bar \rho) Q'} \cdot \int \cdot e^{-\frac{4\pi}{\alpha} \Omega(E/M_{I})} \cdot \mbox{(further terms),}
\end{equation}

\noindent where $D(\rho)$ and $D(\bar \rho)$ denote the Instanton and anti-Instanton size distributions, $E$ the available energy of the process and $\Omega$ describes the Instanton anti-Instanton interaction, with $\Omega(x)=1$ for $x\rightarrow 0$ and $\Omega(x)=0$ for $x\rightarrow \infty$. The Instanton size distribution is proportional to $D(\rho)\sim \rho^{11-2/3 n_f - 5}$ \cite{tHooft:1976snw, Bernard:1979qt, Morris:1984zi, Moch:1996bs, Ringwald:1999ze}, thus an integral over $\rho$ would diverge. However, it was shown that the additional term $e^{-(\rho+\bar \rho) Q'}$ has to be taken into account \cite{Moch:1996bs}, where $Q'$ describes a generic hard scale of the Instanton process. This form factor effect renders the $\rho$ integration convergent. In order to make reliable calculations of cross sections in QCD, Instanton perturbation theory \cite{Moch:1996bs} has to be applied. This requires the validity of the diluted gas approximation \cite{Moch:1996bs}, i.e. requires that the extensions of Instantons and anti-Instantons are not overlapping. Therefore, the validity of Instanton perturbation theory requires Instantons to be sufficiently localized in space-time. In QCD, a generic hard scale $Q$ of the underlying process can be defined which reduces the Instanton size, justifying the diluted gas approximation and enabling the Instanton perturbation theory approach. Here, the cross section can become sizeable at high energies. The reason for the increasing cross section can be intuitively understood \cite{Schrempp:2002kd} by changing the picture from a tunnelling between vacua at E=0 to that of the actual creation of a Sphaleron-like configuration \cite{Klinkhamer:1984di} on top of the potential barrier of height. Therefore, in a naive (but not fully correct) picture, the Instanton process can be interpreted as the creation and the decay of a Sphaleron pseudo-particle, where the pass of the pseudo-particle depends directly on the height of potential barrier.

\begin{figure}[th]
\centering
\begin{minipage}{9.8cm}
\centering
\includegraphics[width=1.0\linewidth]{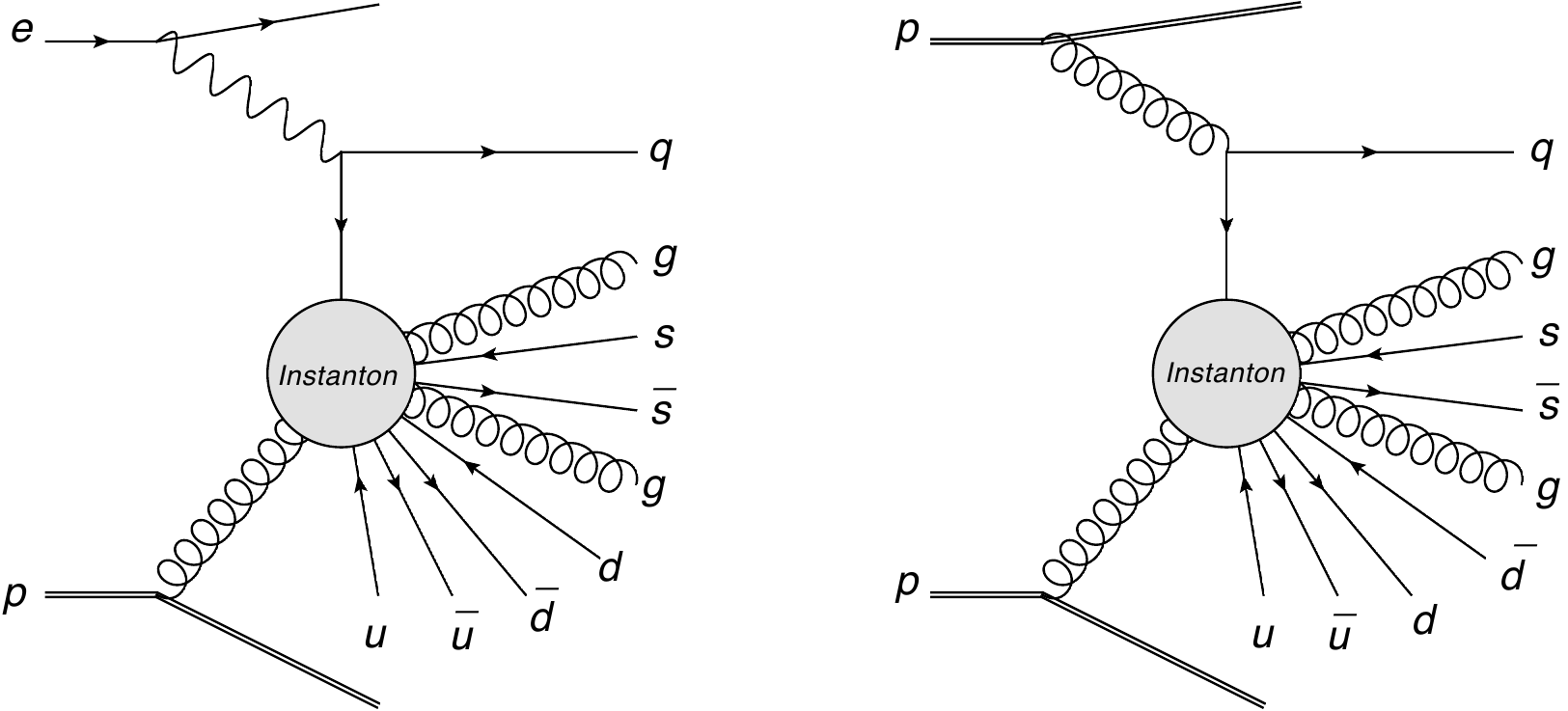}
\caption{Depiction of a QCD Instanton processes in electron-proton (left) and proton-proton (right) collisions, where an external scale parameter $Q'$ is required. \vspace{1.4cm} \label{fig:InstProcess1}}
\end{minipage}
\hspace{0.3cm}
\begin{minipage}{4.5cm}
  \centering
  \includegraphics[width=1.0\linewidth]{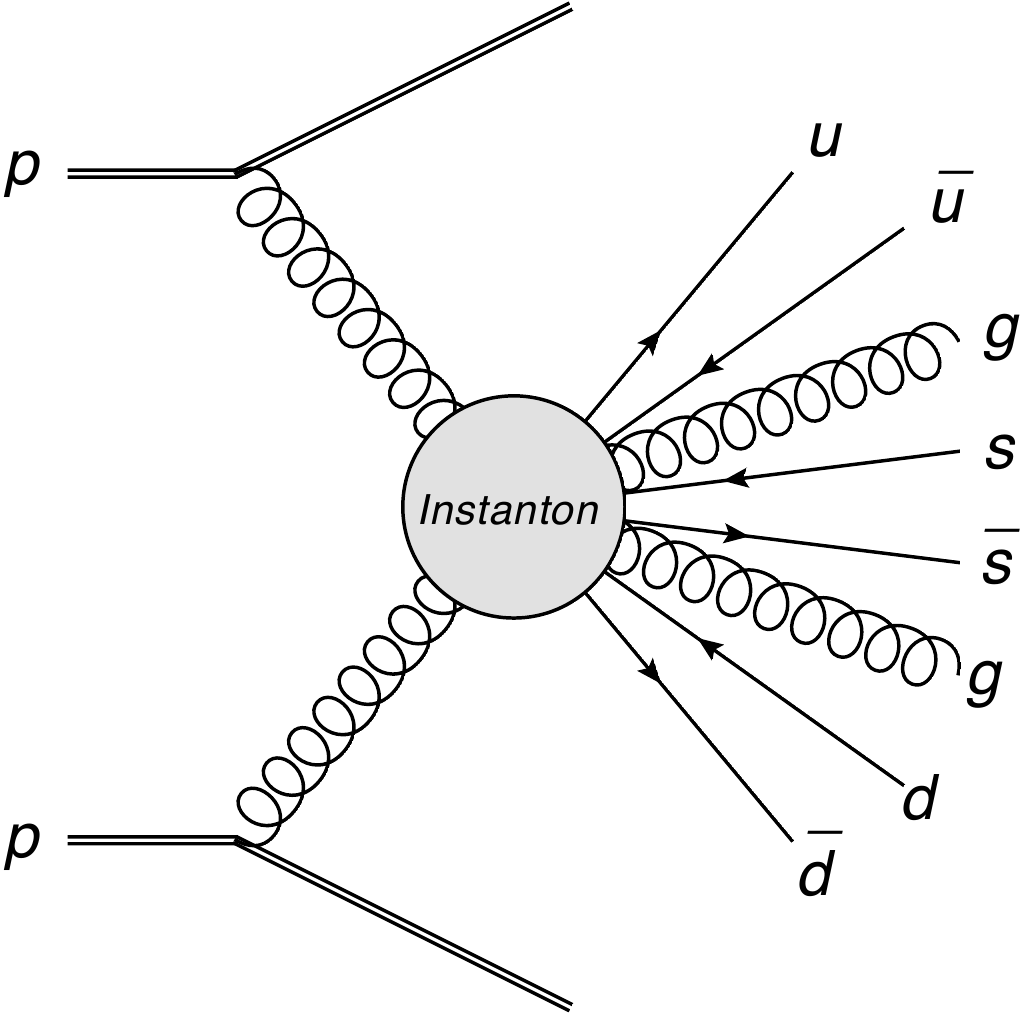}
	\caption{Depiction of a QCD Instanton processes in proton-proton (right) collisions without the requirement of an external scale parameter $Q'$ \label{fig:InstProcess2}}
\end{minipage}%
\end{figure}

For deep inelastic scattering processes, the necessary hard scale $Q'$ was defined by a highly virtual momentum transfer by a photon, emitted by the incoming electron, yielding a high energetic final state quark in addition to the Instanton process, as shown in Figure \ref{fig:InstProcess1}. This concept can be easily transferred to proton-proton collisions where the photon exchange is simply replaced by a gluon as shown in Figure \ref{fig:InstProcess1}. A first calculation of the the latter processes became recently public \cite{Khoze:2020tpp}, suggesting negligible cross sections at the LHC for typical energies at LHC. 

However, an alternative strategy to calculate finite cross sections was also recently published \cite{Khoze:2019jta}. Here, no a second independent kinematic scale, as the DIS highly virtual momentum scale $Q$, is required (Fig. \ref{fig:InstProcess2}). In this approach, only small Instantons contribute to the scattering processes in QCD and potentially problematic contributions of Instantons with large size are automatically cut-off by the inclusion of quantum effects due to interactions of the hard initial states that generate the factor
 $e^{- \alpha_s \, \rho^2 s' \log s'}$. The latter argumentation was already brought in 1991 \cite{Mueller:1990ed, Khoze:1991mx} and provides a dynamical solution to the problem of IR divergences arising from Instantons of large scale-sizes in QCD. Basically, the scale invariance of the classical Yang-Mills theories is broken by those quantum effects which lead to a suppression of all 
but small Instantons with sizes $\rho \lesssim (10-30) / \sqrt{s'}$. The characteristic QCD Instanton size is therefore inversely proportional to the centre-of-mass (CoM) energy of two colliding partons $\sqrt{s'}$. Table \ref{tab:crosssections} shows the proton-proton cross sections for Instanton processes at a center of mass energy of 13~\TeV{} for various choices of minimal values of $\sqrt{\smin}$ as calculated in \cite{Khoze:2019jta}. The cross section at $\sqrt{\smin}=20 \GeV$ contributes already several percent to the total proton-proton cross section and at even lower values would saturate. This implies that the calculation breaks down at some small value of $\sqrt{\smin}$. 

In the following it is assumed that the cross section estimate is reliable for $\sqrt{\smin}>20 \GeV$, keeping in mind that the uncertainty on this prediction could be of several order of magnitudes. The cross section dependence can be interpolated by a phenomenological formula $\sigma \sim e^{a\cdot e^{b\cdot x}+c\cdot x^2 + d\cdot x + e}$, implying an exponential decrease vs. $\sqrt{\smin}$. In the pseudo-particle picture of the Instanton process, the available energy $\sqrt{\smin}$ can be interpreted as the Instanton mass $\mInst$. The production of Instantons is therefore not a resonant but a continues processes, with large production rates of Instantons of low mass and small production rates of  Instantons of high masses  expected.

\begin{table}[b]
\centering
\small
\begin{tabular}{c | l| l| l| l| l| l}
\hline
$\sqrt{\smin}$ [GeV]							& 10					&	20			& 50				& 100			& 200			& 500		\\
\hline
$\sigma(pp\rightarrow I \rightarrow X)$ [pb]		& $1.7\cdot 10^{11}$		& $6.3\cdot 10^{9}$	& $4.1\cdot 10^{7}$	& $8.0\cdot 10^{4}$	& $1.1\cdot 10^{2}$	& $3.5\cdot 10^{-3}$	\\		
\hline
\end{tabular}
\caption{Overview of expected cross sections of Instanton processes with different values of $\sqrt{\smin}$ \cite{Khoze:2019jta}.\label{tab:crosssections}}
\end{table}

\subsection{Decay of the Instanton Pseudo-Particle and Experimental Observables \label{sec:InstObs}}

The question whether manifestations of such topological fluctuations, i.e. Instanton processes, can be directly observed in high-energy experiments was already raised in the 1980s in the context of the electroweak sector \cite{Ringwald:1989ee, Aoyama:1986ej, Espinosa:1989qn, Ringwald:2002iy}. A QCD Instanton tunneling process between  $\Delta N_{CS}=1$ vacua leads to the creation of  a quark-antiquark pair of different chirality for each flavour, $N_f$ in association with  a number  $n_g$ of additional gluons.
Following the approach of~\cite{Khoze:2019jta} we only consider the dominant contribution from gluon-gluon transition, which proceeds through the process:
\begin{equation}
    g+g\to \sum_{f=1}^{N_f}(q_R^f+\bar{q}_L^f)+n_g g
\end{equation}
In the pseudo particle picture, this can be interpreted as a decay process of a Instanton pseudo particle with a mass $m_I$. For  low Instanton masses, e.g. in the 50~\GeV{} range, we expect therefore an isotropic decay into up to 5 quarks, 5 anti-quarks as well as 5-10 gluons. The number of gluons is assumed to be Poisson distributed around $\langle n_g\rangle$, which has been calculated in~\cite{Khoze:2019jta} in turn depends on $m_I$ and varies between 5 and 13 over the broad 10~\GeV < $m_I$ < 4~\TeV. As a consequence, QCD Instanton-induced scattering processes produce {\it soft bombs} -- very high-multiplicity spherically symmetric distributions of relatively soft particles \cite{Harnik:2008ax}. 

These generic properties of QCD Instanton decays reflect into the experimental observables that can be used to search for these processes. Most importantly the Instanton cross section falls rapidity with increasing mass of the Instanton pseudo-particle, \mInst, or equivalently with the center of mass energy at parton level, $\sqrt{s'}$ of the Instanton process. The cross section dependence on $\sqrt{s'}$ is expected to very different from other SM processes, hence it is expected that different regimes of \mInst will have on the one hand different signal to background ratios and on the other hand, different processes which contribute to the background. Experimentally, \mInst can be approximated by the 4-vector sum of all charged particles with a certain minimal transverse momentum.

\section{\label{Sec:MCSamples}MC Samples and Detector Simulation}

While the decay of low mass Instantons, e.g. $\mInst=30$ \GeV, results  in events with a high multiplicity of low energetic charged particle tracks, the decay of Instantons with masses larger than 200~\GeV\ results in numerous reconstructed particle jets. Depending on the range of Instanton processes masses considered, different SM processes can thus act as background. In the low mass regime, \softQCD events originating from inelastic, non-diffractive processes play the dominant role due to their large production cross section.  For higher Instanton masses,  high-\pT\ jet production processes (\hardQCD), as well as vector-boson and top anti-top quark pair production in the hadronic decay channels contribute. 

The different signal and background samples used in this study are summarized in Table \ref{tab:samples} and discussed in the following sections. A typical detector response for the different generated samples has been simulated through the \textsc{Delphes} framework \cite{deFavereau:2013fsa} with settings corresponding to the ATLAS experiment, and without considering  additional pile-up interactions. Pile-up activity could become an important source of background when selecting events with high multiplicity final states, and will require dedicated studies in LHC analyses that cannot be performed with parametrised simulations.

Several observables sensitive to Instanton processes have been defined at detector level. Basic reconstructed quantities are the 4-vectors of charged particles reconstructed as particle tracks as well as particle jets, reconstructed using an anti-k$_T$ algorithm with a radius parameter of 0.4. Tracks are assumed to be massless and hence their 4-vector is defined by their transverse momentum, $p_T$, in the $x-y$ plane\footnote{transverse to the beam axis}, the polar angle $\theta$  measured from the positive $z$ axis, as well as the azimuthal angle $\phi$ in the $x-y$ plane.  The polar angle is mostly expressed in terms of the pseudorapidity $\eta$, defined by $\eta = - \ln(\tan\theta/2)$. We require  transverse momenta greater than 500~\MeV{} for the reconstructed tracks, as well as a maximal absolute pseudo-rapidity of 2.5. Particle jets are required to have at least a transverse momentum of 20 GeV and $\eta$<2.5.

A first experimental observable for the selection of Instanton processes is the number of reconstructed tracks, $\Ntrk$, as well as the number of reconstructed jets $N_{jet}$. Since many charged decay particles are expected from the decay of an Instanton pseudo-particle with a given mass, also the ratio of the mass and the number of tracks, $\mInst/\Ntrk$ is of interest. Similarly, the scalar sum of the transverse momenta $\pT$ of all charged tracks (or particle-jets), $S_T=\sum |\pT^i|$ in dependence of $\mInst$ is studied. Isotropic decays of resonances are expected to have more central than forward activity, i.e. the pseudo-rapidity distribution of all charged tracks, $\Etatrk$, as well as the average pseudo-rapidity of charged tracks per event $\langle\Etatrk\rangle$ are expected to be sensitive. Due to the presence of   $c$- and $b$- quarks as decay products of the Instanton and the relatively long life-times of the corresponding hadronized mesons, one might expect a higher number of charged particles displaced  vertices  compared to other Standard Model processes. The number of reconstructed charged particles tracks with a production vertex that has a distance in the transverse plane of more than 0.02~mm to the primary vertex of the collision, $\NDisplaced$, is therefore also studied.

The observables discussed previously   are not  directly related to the expected isotropy of Instanton decays. One variable that targets the isotropy is event-sphericity and defined via the tensor $S$,

\begin{equation}
	S^{\rm \alpha \beta} \,=\,  \frac{\displaystyle\sum_{i} p^{\rm \alpha}_{i} p^{\rm \beta}_{i}}{\displaystyle\sum_{i} |{\vec{p}_{i}}|^{2}}~\mbox{,}\nonumber
\end{equation}

where the indices denote the $x$, $y$, and $z$ components of the momentum of the particle $i$ in its rest-frame. The sphericity of the event is then constructed using the two smallest eigenvalues of this tensor, $\lambda_2$ and $\lambda_3$, i.e.\ $S=\frac{3}{2}({\lambda_2 + \lambda_3})$ and takes values between 0 and 1. A fully balanced dijet events leads to a sphericity of $S=0$, while a fully isotropic event has a sphericity of $S=1$. A similar event shape variable is thrust, defined as

\begin{equation} 
\mathcal{T}=1-\max_{\vec{n}}\frac{\sum_{i} \left\lvert\vec{p}_{i} \cdot \vec{n}\right\rvert}{ \sum_{i}\left\lvert\vec{p}_{i}\right\rvert},
\end{equation} 

where $\vec{n}$ is a unit vector. Fully spherical symmetric events yield $\tau=0.5$, while fully balanced dijet events have $\mathcal{T}=0$. The definition of thrust also defines the thrust axis $\vec{n}$, which maximizes the value of $\mathcal{T}$. The thrust axis defines a left $\mathcal{L}$ and right $\mathcal{R}$ hemisphere for each event, which can be used to define the jet broadening of an event. The left and right broadening is defined as

\begin{equation}
\mathcal{B}_{\mathcal{L}}=\sum_{i\in\mathcal{L}}\frac{\lvert\vec{p}_{i}\times\vec{n}|}{\sum_{i}\lvert\vec{p}_{i}\rvert} \quad \mbox{and} \quad \mathcal{B}_{\mathcal{R}}=\sum_{i\in\mathcal{R}}\frac{\lvert\vec{p}_{i}\times\vec{n}|}{\sum_{i}\lvert\vec{p}_{i}\rvert}.
\end{equation}

The total jet broadening $\mathcal{B}$ is then defined as $\mathcal{B} = \mathcal{B}_{\mathcal{L}} + \mathcal{B}_{\mathcal{R}}$, and behaves similar as $\tau$, i.e. is 0 and 0.5 for dijet and spherically symmetric events, respectively. 

The sphericity $\mathcal{S}$, the thrust $\mathcal{T}$ as well as the total jet broadening $\mathcal{B}$ are calculated using all reconstructed tracks as well as for all reconstructed jets per event. The calculation is based on the code provided in \cite{Bierlich:2019rhm}. It should be noted that these three observables are significantly correlated. An additional correlation is observed between $\mathcal{S}$ and the $\Etatrk$ distribution as events with large values of $\mathcal{S}$ tend to enhance the number of tracks in the central region, i.e. with $|\eta|<1.0$.

\begin{table}[thb]
\medskip
\footnotesize
\centering
\begin{tabular}{l| c|c|c}
\hline
{\bf Process}								& {\bf Generator}	& {\bf Main Generator Setting}					& {\bf \# Events}	\\
\hline
QCD-Instanton (low-mass regime)				& \textsc{Sherpa}	& Instanton\_MIN\_MASS: 25.				& 10,000			\\
QCD-Instanton (low-mass regime)				& \textsc{Sherpa}	& Instanton\_MIN\_MASS: 50.				& 10,000			\\
QCD-Instanton (medium-mass regime)			& \textsc{Sherpa}	& Instanton\_MIN\_MASS: 100.				& 10,000			\\
QCD-Instanton (medium-mass regime)			& \textsc{Sherpa}	& Instanton\_MIN\_MASS: 200.				& 10,000			\\
QCD-Instanton (high-mass regime)				& \textsc{Sherpa}	& Instanton\_MIN\_MASS: 300.				& 10,000			\\
QCD-Instanton (high-mass regime)				& \textsc{Sherpa}	& Instanton\_MIN\_MASS: 500.				& 10,000			\\
QCD-Instanton (high-mass regime)				& \textsc{Sherpa}	& Instanton\_MIN\_MASS: 1000.				& 10,000			\\
\hline
\softQCD									& \textsc{Pythia8}	& \textsc{SoftQCD:all = on} - Monash Tune		& 1,000,000			\\
\softQCD									& \textsc{Pythia8}	& \textsc{SoftQCD:all = on} - A14 Tune 			& 1,000,000			\\
\softQCD									& \textsc{Sherpa}	& 							 			& 1,000,000			\\
\softQCD									& \textsc{Herwig}	& 							 			& 1,000,000			\\
\hline
$qq\rightarrow X, qg\rightarrow X, gg\rightarrow X$ 	& \textsc{Pythia8}	& \textsc{HardQCD:all = on}					& 1,000,000	\\
(\hardQCD)								&				& \textsc{PhaseSpace:pTHatMin = 5.}			&		\\
$qq\rightarrow X, qg\rightarrow X, gg\rightarrow X$	& \textsc{Pythia8}	& \textsc{HardQCDAll=on}					& 1,000,000	\\
(\hardQCD)								&				& \textsc{PhaseSpace:pTHatMin = 100.}			&		\\
$qq\rightarrow X, qg\rightarrow X, gg\rightarrow X$	& \textsc{Pythia8}	& \textsc{HardQCDAll=on}					& 1,000,000	\\
(\hardQCD)								&				& \textsc{PhaseSpace:pTHatMin = 300.}			&		\\
\hline
$W\rightarrow q\bar q+X$						& \textsc{Pythia8}	& \textsc{WeakSingleBoson:ffbar2W = on}			& 1,000,000					\\
$Z\rightarrow q\bar q+X$						& \textsc{Pythia8}	& \textsc{WeakSingleBoson:ffbar2W = on}			& 1,000,000					\\
$WW\rightarrow q\bar q q\bar q + X$					& \textsc{Pythia8}	& 										& 1,000,000					\\
\hline
$t\bar t\rightarrow b q\bar q+\bar b q\bar q+X$		& \textsc{Pythia8}	& \textsc{Top:all = on}						& 1,000,000			\\
\hline
\end{tabular}
\caption{\label{tab:samples} Overview of the MC samples used to model the Instanton signal process and the SM background processes.}
\end{table}

\subsection{Soft QCD Processes}

Due to their non-perturbative nature, \softQCD processes are described by phenomenological models, which have been tuned to data using a wide variety of reference measurements. Within this study, the \softQCD processes in the \textsc{Pythia8} \cite{Sjostrand:2007gs} generator is used as baseline. In total one million events have been generated for proton-proton collisions at a center of mass energy of 13~\TeV{} using the \textsc{NNPDF23lo} PDF set \cite{Ball:2012cx}. 

\begin{figure}[tbh]
\begin{center}
\includegraphics[width=3.5cm]{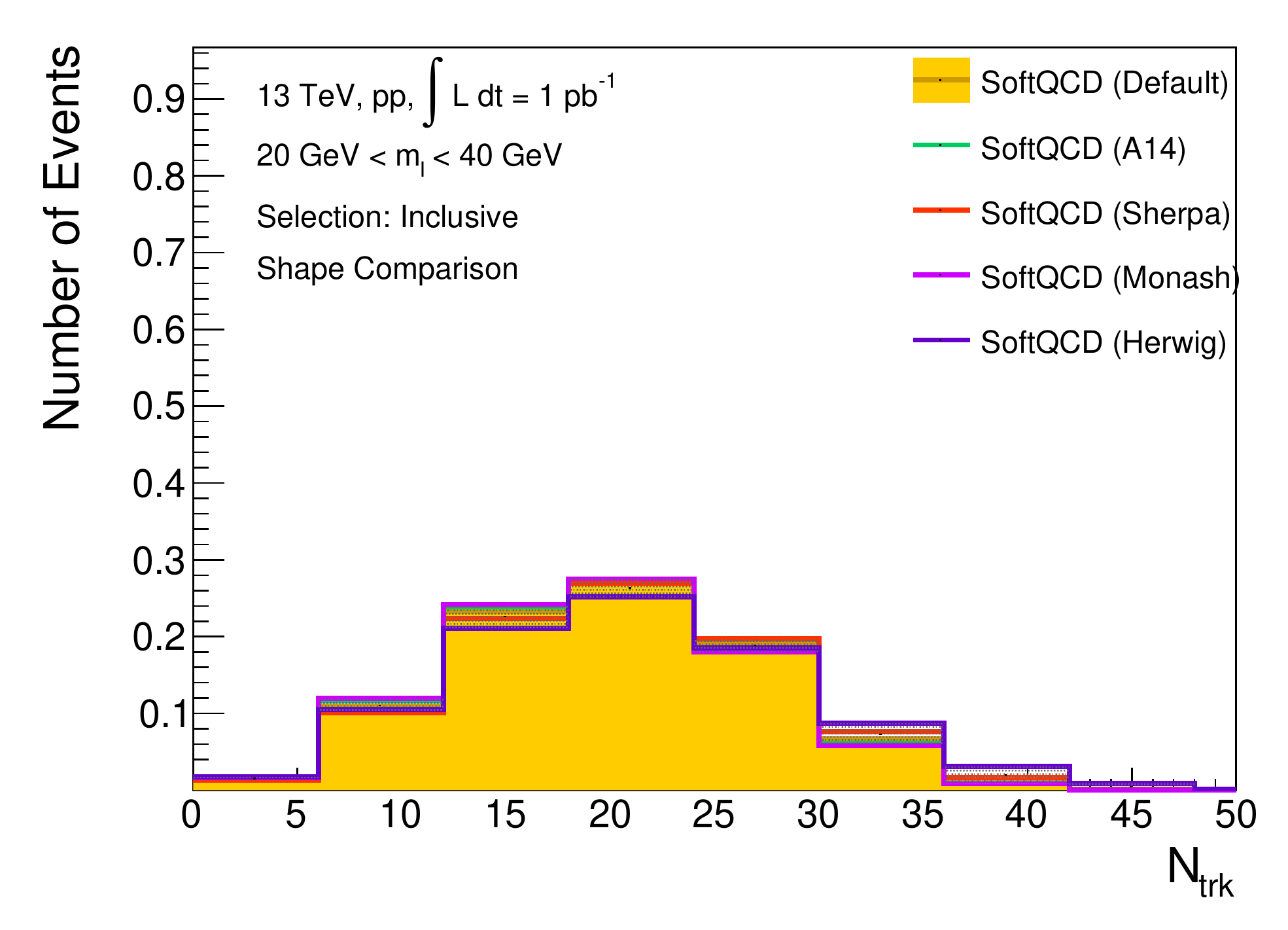} 
\hspace{0.02cm}
\includegraphics[width=3.5cm]{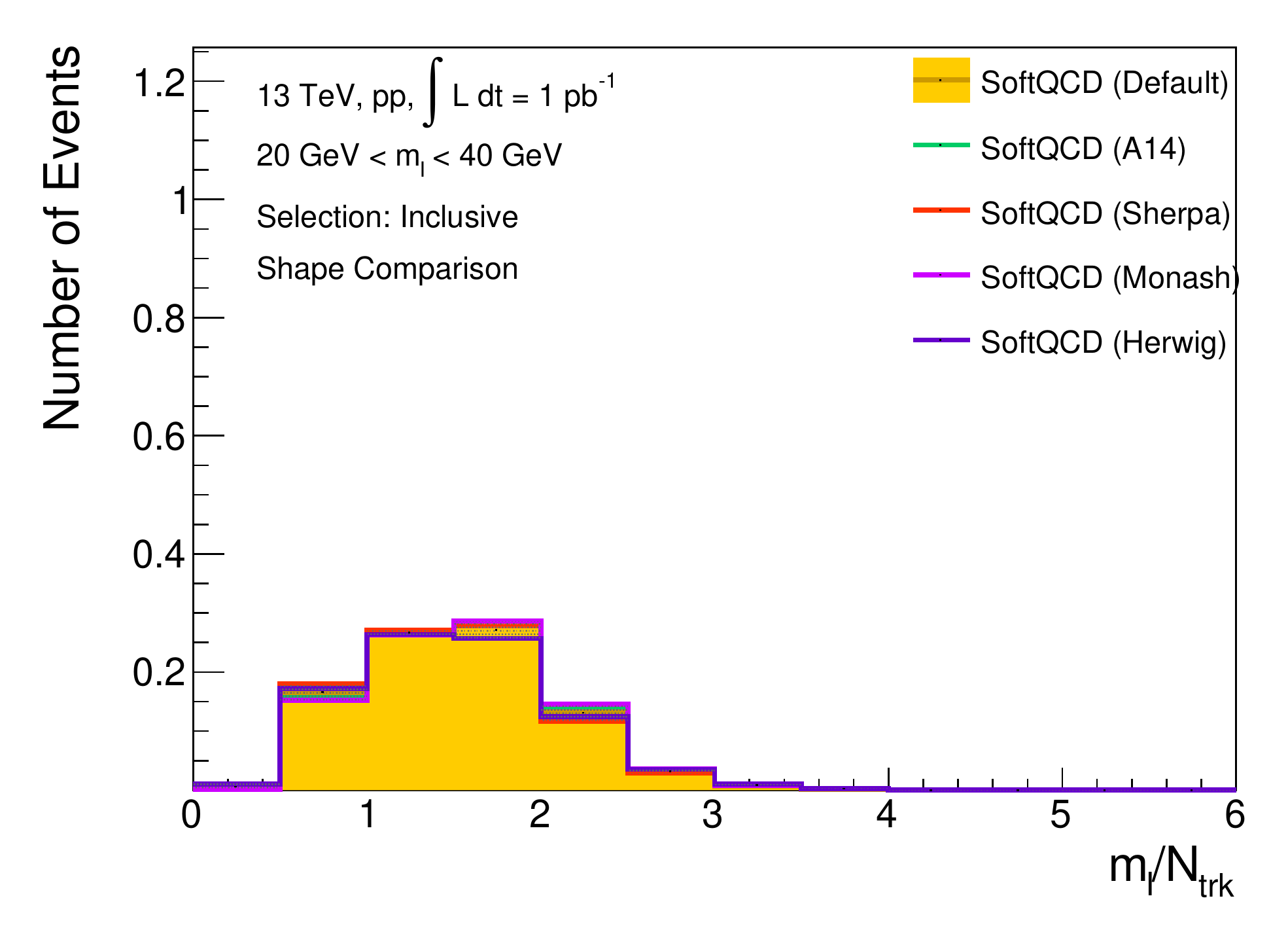}
\hspace{0.02cm}
\includegraphics[width=3.5cm]{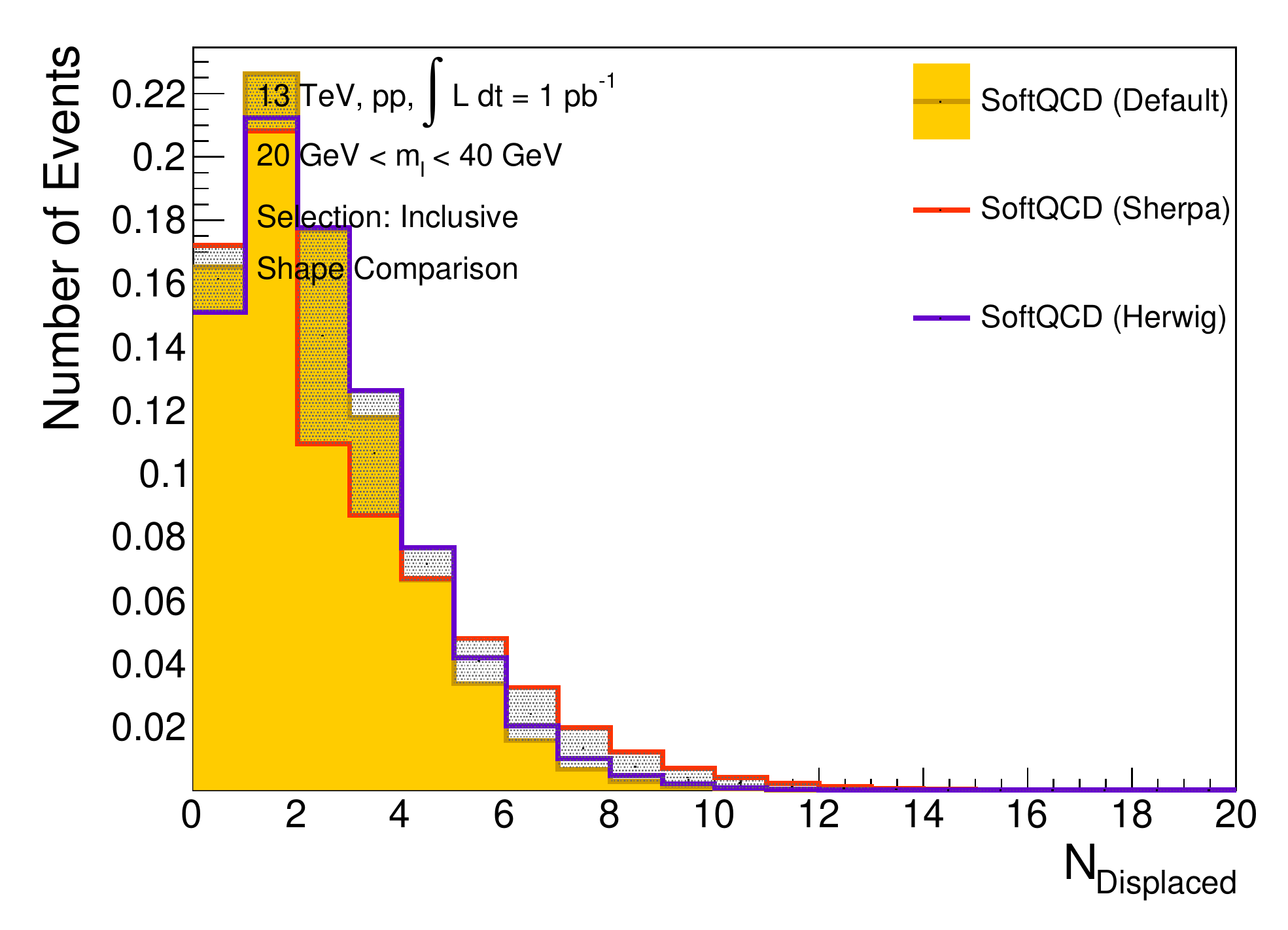} 
\hspace{0.02cm}
\includegraphics[width=3.5cm]{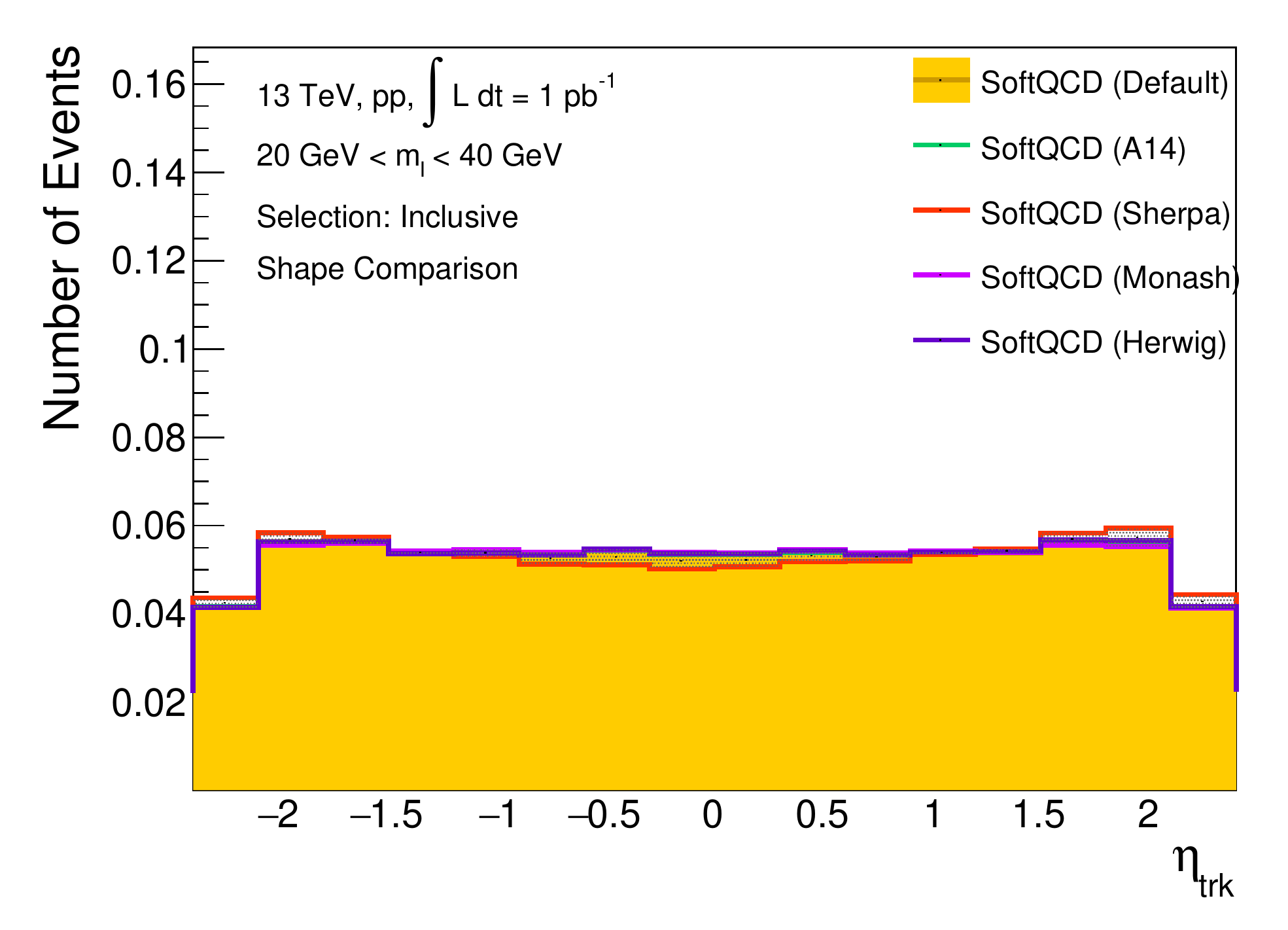}
\\
\includegraphics[width=3.5cm]{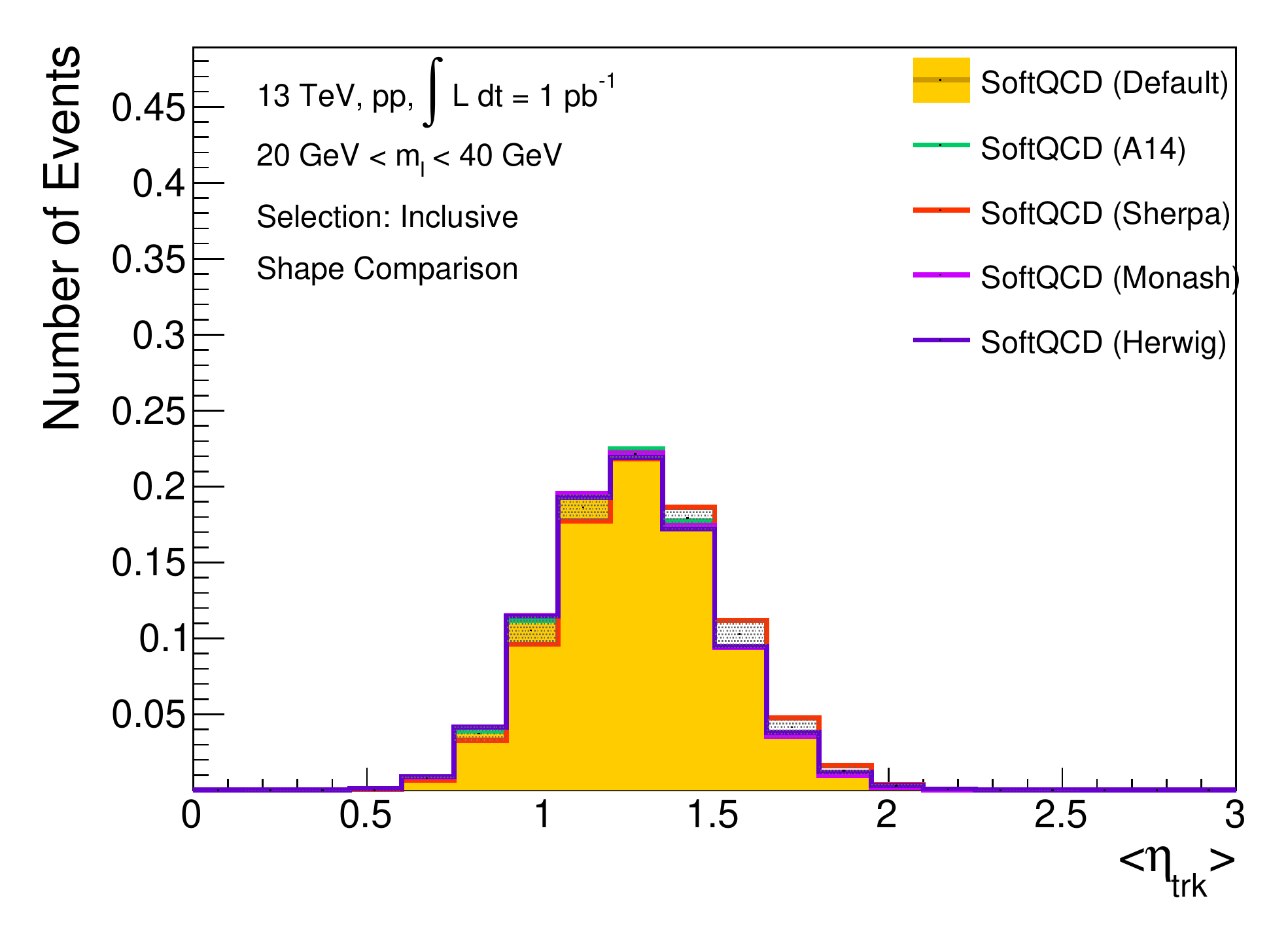} 
\hspace{0.02cm}
\includegraphics[width=3.5cm]{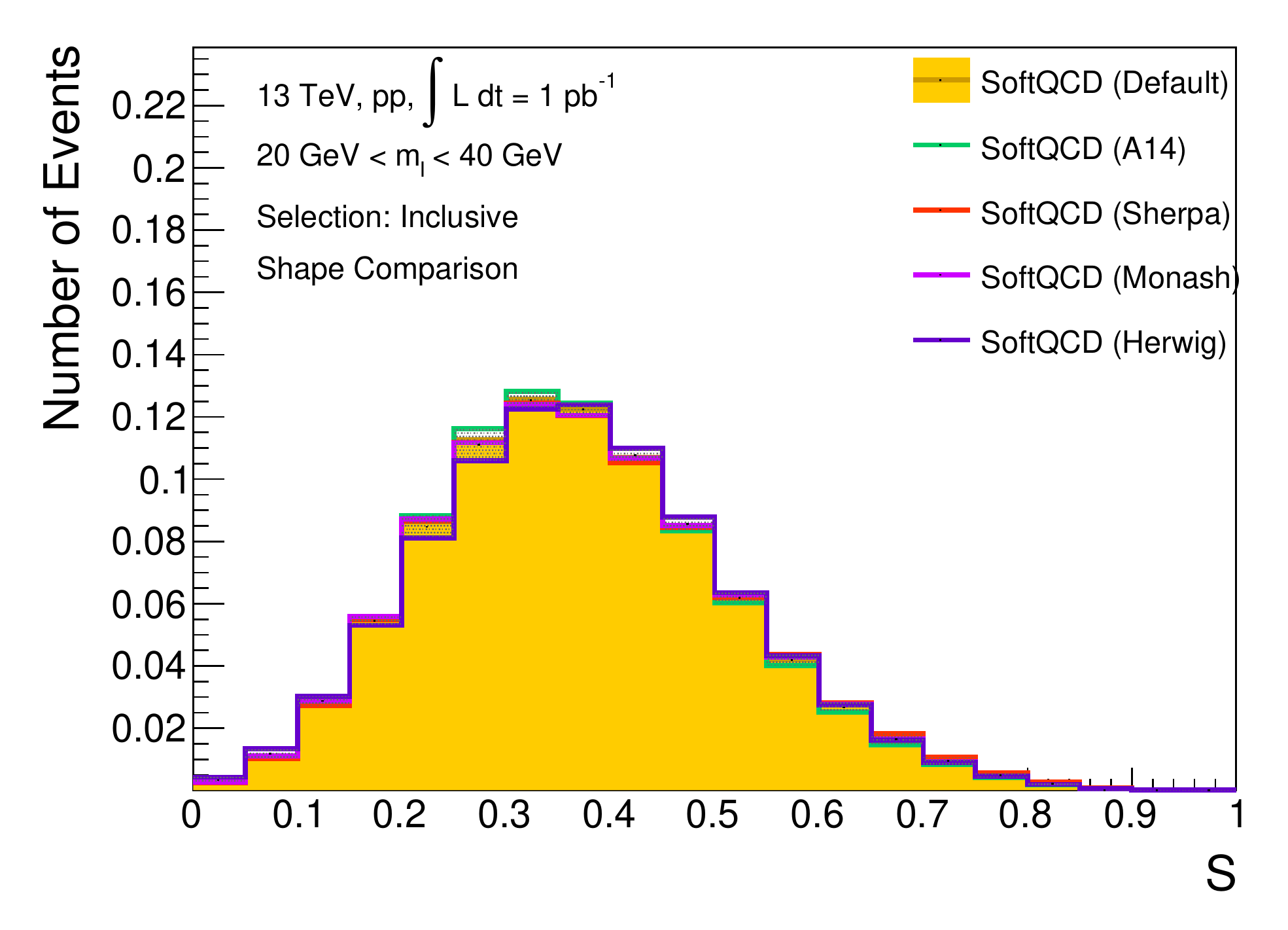}
\hspace{0.02cm}
\includegraphics[width=3.5cm]{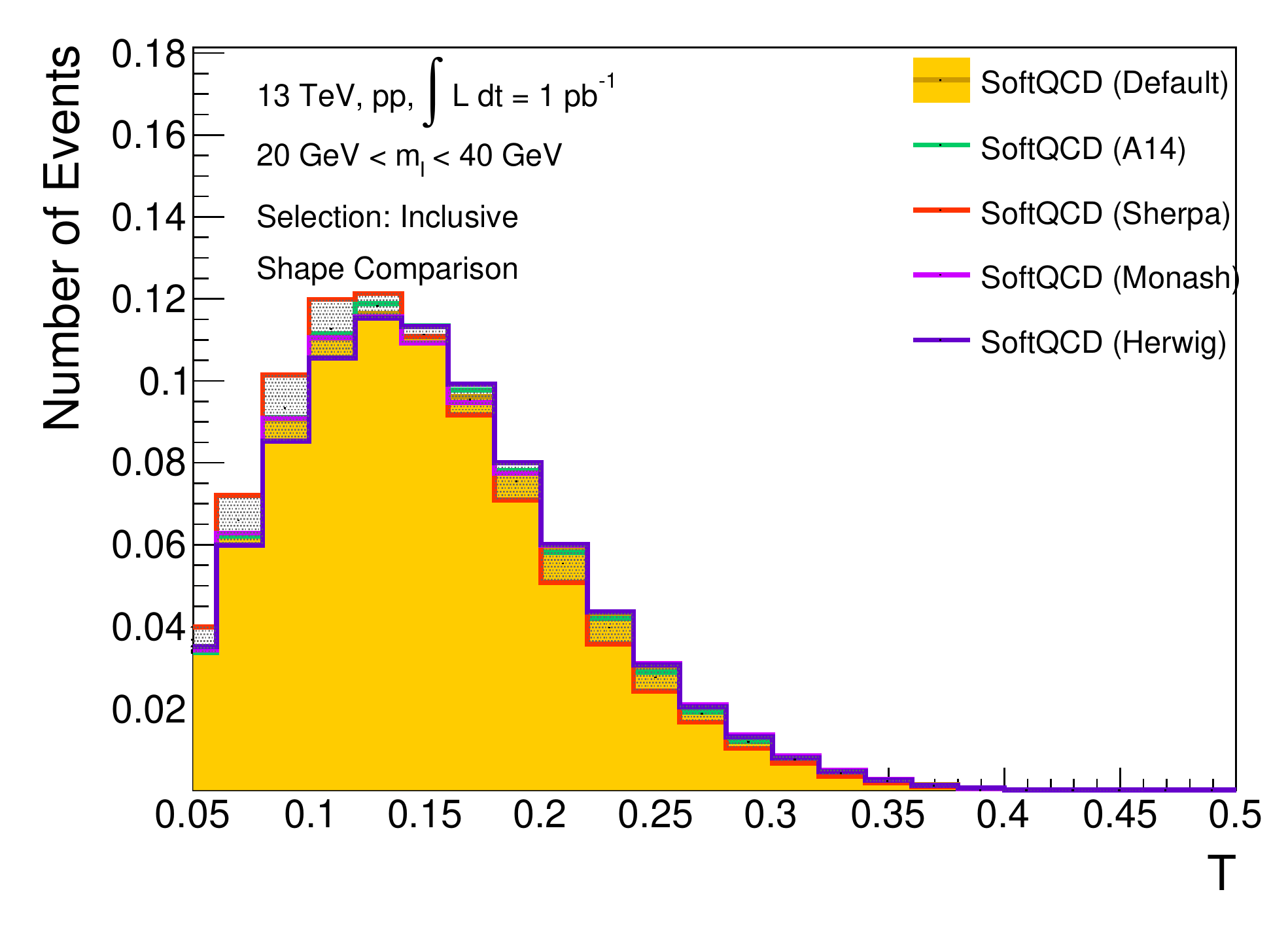} 
\hspace{0.02cm}
\includegraphics[width=3.5cm]{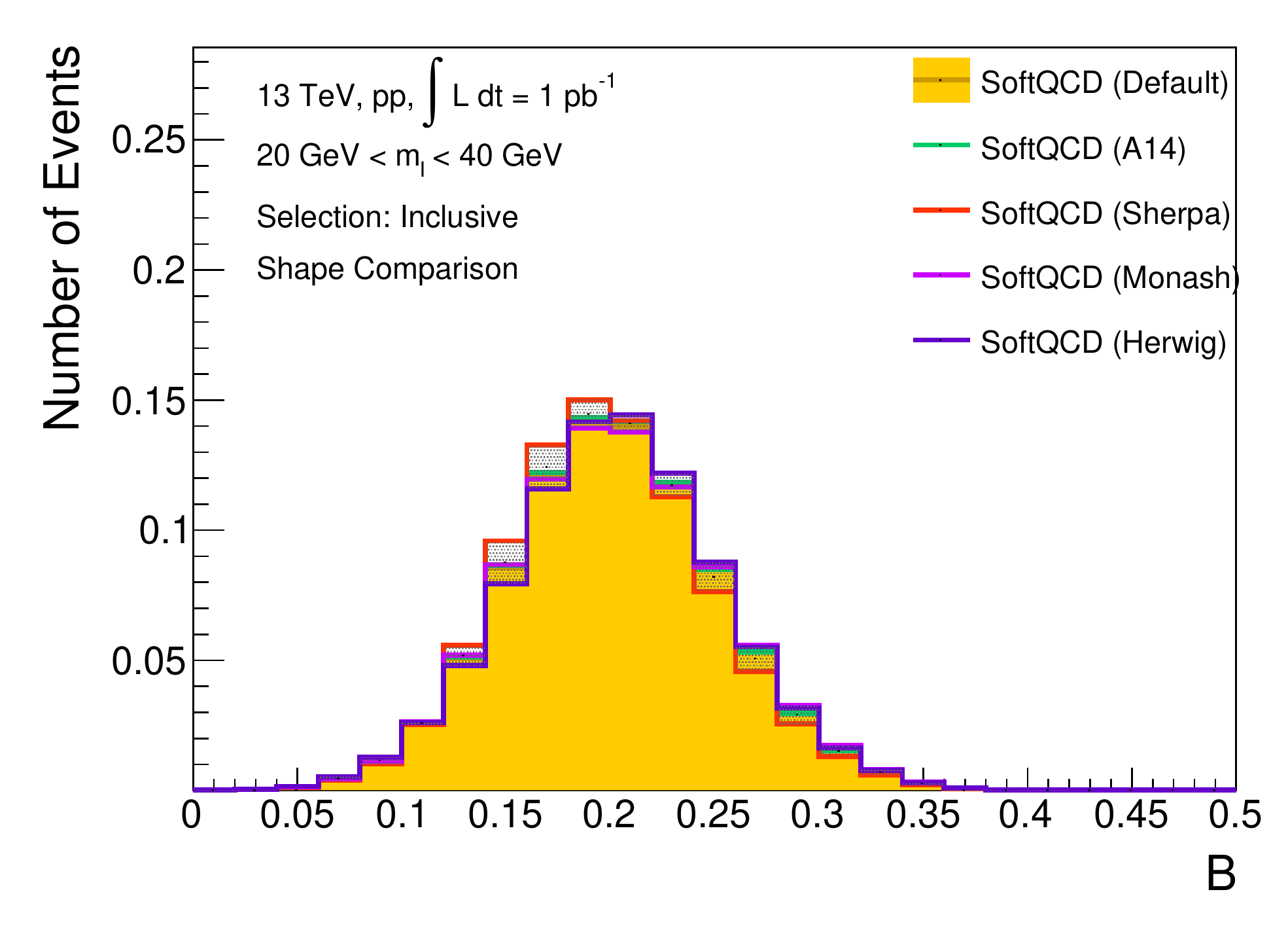}

\caption{\label{fig:ShapesSoftQCD} Normalized distributions for \softQCD processes with an invariant mass based on all tracks between 20 and 40 \GeV, predicted be different MC generators or generator settings: $\Ntrk$, $\mInst/\Ntrk$, $\NDisplaced$, $\Etatrk$, $\langle |\Etatrk| \rangle$, $\mathcal{S}$, ${\cal B}$, ${\cal T}$.}
\end{center}
\end{figure}

We also consider \softQCD production  in \textsc{Pythia8} using a different tune,  A14 \cite{ATL-PHYS-PUB-2014-021}, as well as the predictions of \softQCD processes from the \textsc{Herwig7} \cite{Bahr:2008pv} and the \textsc{Sherpa} \cite{Sherpa30, Bothmann:2019yzt} event generators.  The comparison of normalised distributions for the  observables of interest for Instanton processes  is shown in Figure \ref{fig:ShapesSoftQCD} for events with a reconstructed invariant mass based on reconstructed tracks between 20 and 40 \GeV{}. A good agreement is observed between the different predictions for most distributions, with the possible exception of the $\NDisplaced$ distribution, where the \textsc{Sherpa} predictions differs by about 20\% from the other generators. In the following, he maximal difference between the various \softQCD samples is taken as systematic uncertainty on the nominal \softQCD prediction from \textsc{Pythia8}. 

\subsection{Hard QCD and Further Processes \label{sec:hardQCD}}
High-\pT{} jet production processes can be predicted with high accuracy in perturbation theory. \textsc{Pythia8} with the \textsc{NNPDF23lo} PDF set was used to simulate \hardQCD multi-jet final states at a center of mass energy of 13~\TeV.
The transition between the \softQCD  and \hardQCD processes is not well defined. In our study, we use \softQCD samples for all events, which have no jet at particle level with a transverse momentum above 20~\GeV, while the simulation of \hardQCD processes is used for all other events. 

In addition to multi-jet final states,  the production of top-quark pairs and of $W$ and $Z$ bosons and di-boson processes can also lead to high multiplicity final states, in particular in their fully hadronic decay channels. These processes are also simulated with \textsc{Pythia8} at leading order in $\alpha_s$ with the \textsc{NNPDF23lo} PDF set. An uncertainty of 10\% on all predictions of multi-jet, $t\bar t$ and vector boson processes are assumed in the following, to account for  theoretical uncertainties from missing higher orders. However, it should be noted that it is possible to simulate these processes at higher perturbative accuracy. In addition,  a realistic data analysis could use the leptonic decay channels of  vector-boson and $t\bar t$ production to validate the theoretical predictions in dedicated control regions, thus reducing significantly the model uncertainties.

\subsection{Signal Samples}

\begin{figure}[thb]
\begin{center}
\includegraphics[width=7.3cm]{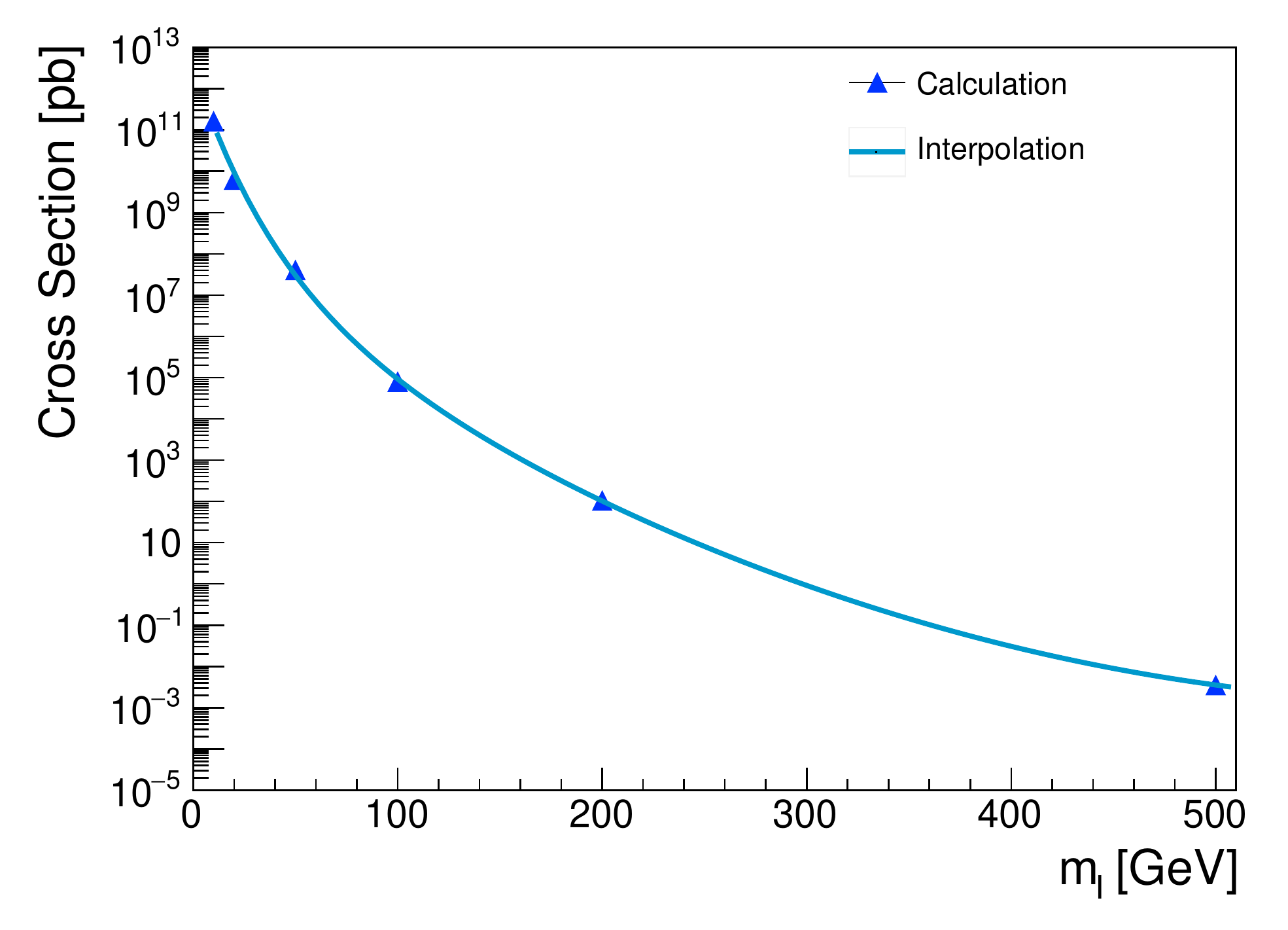} 
\includegraphics[width=7.3cm]{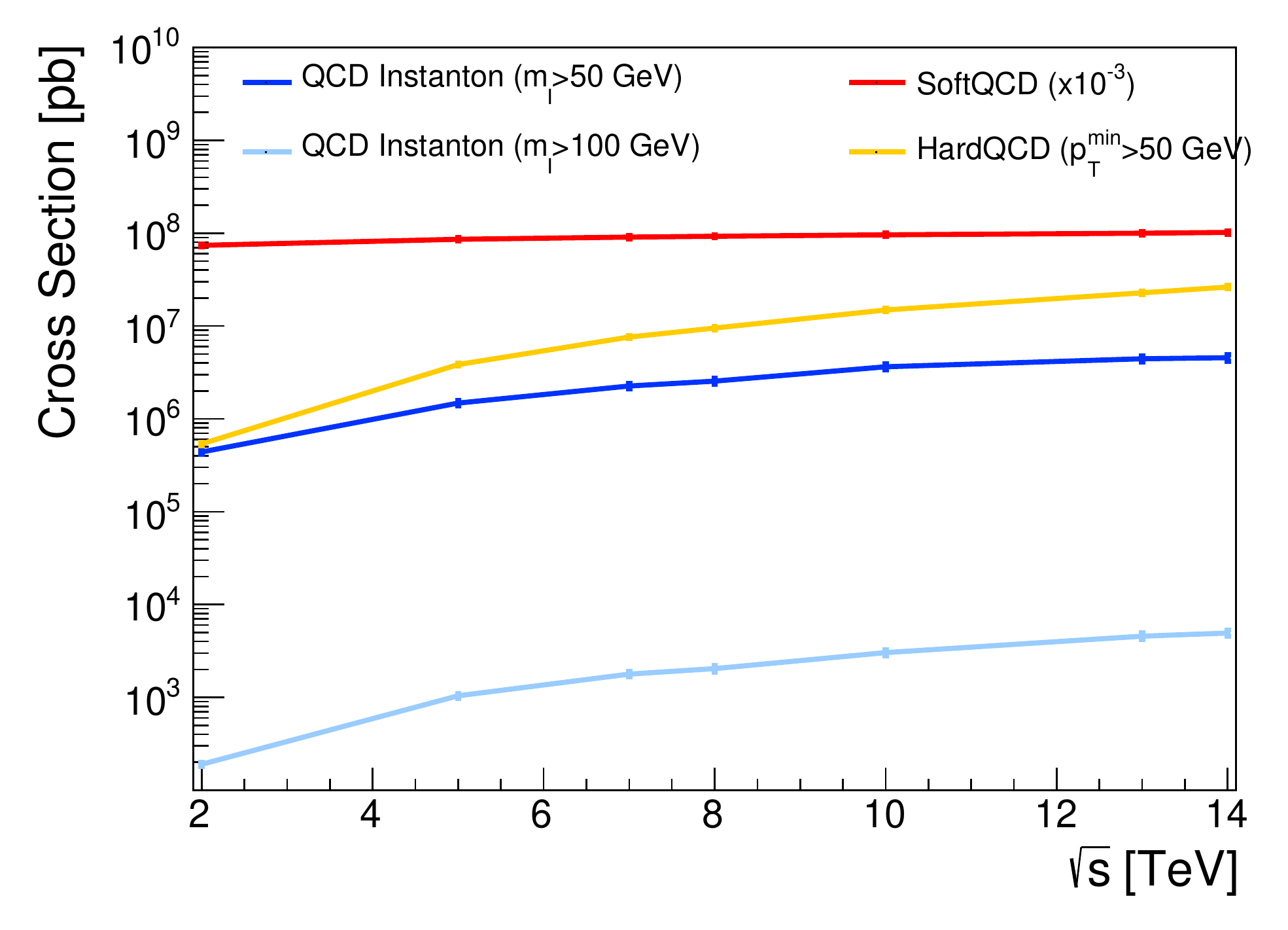} 
\caption{\label{fig:CrossSection} The production cross section of Instanton processes in proton proton collisions at $\sqrt{s}=13$~\TeV{} as a function of their mass (left) and a comparison of the cross section dependence on $\sqrt{s}$ between the Instanton signal and various background processes (right).}
\end{center}
\end{figure}

The Instanton signal samples have been produced with a modified version of the \textsc{Sherpa} event generator \cite{Sherpa30, Bothmann:2019yzt, Khoze:2019jta}. The predicted cross sections of Instanton induced processes for different Instanton masses $\mInst$  have been implemented based on the calculations in \cite{Khoze:2019jta} and are shown in Figure \ref{fig:CrossSection}. The figure also shows the predicted dependence of the Instanton cross section, for two different values of $\sqrt{\smin}$, on the center of mass energy of proton-proton collisions where also cross sections for soft- and hard-processes are shown in comparison. It is interesting to note that the Instanton cross sections exhibit a different dependence on $\sqrt{s}$ than \softQCD and \hardQCD processes.
 
\begin{figure}[htb]
\begin{center}
\includegraphics[width=2.8cm]{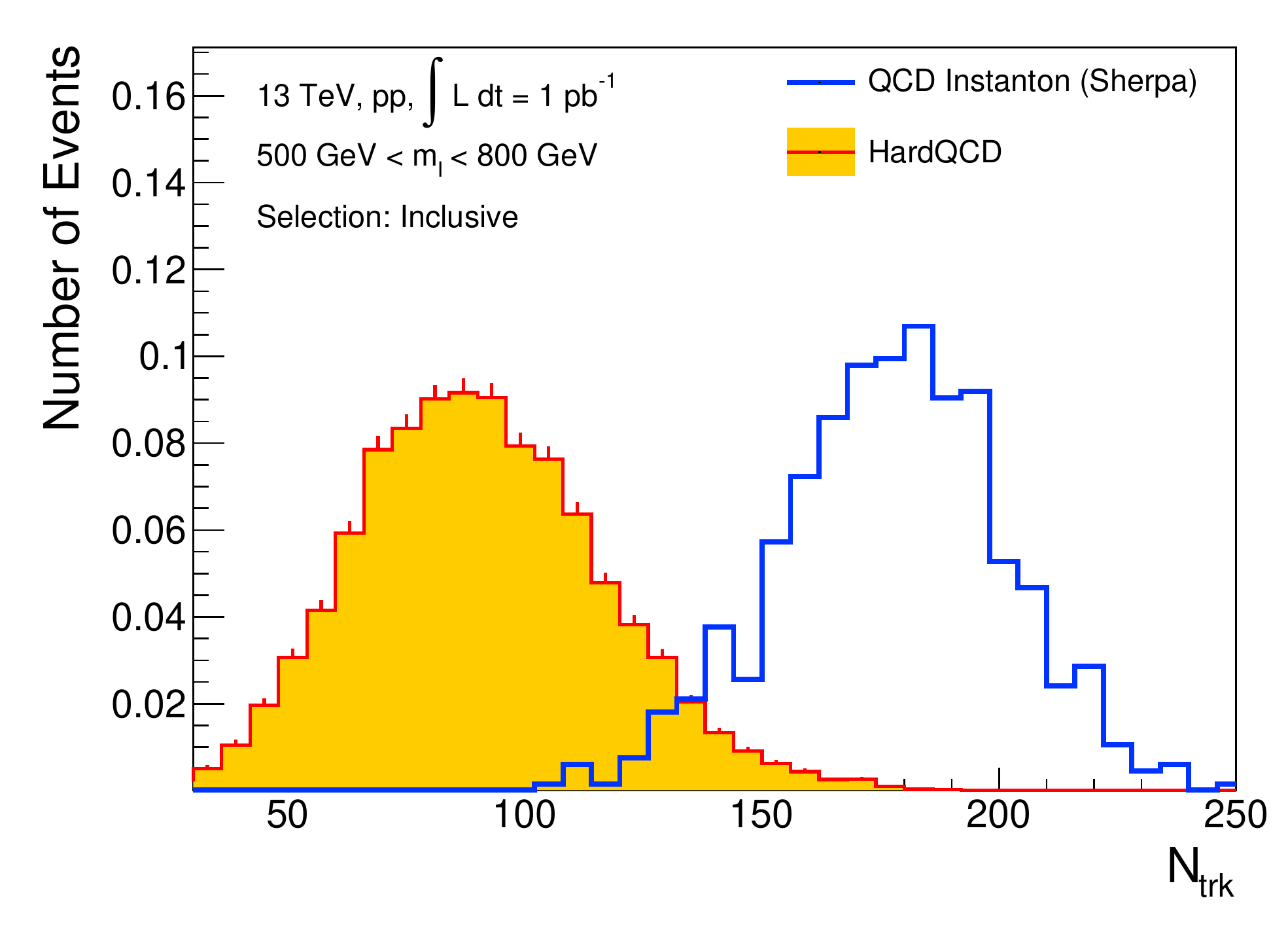} 
\hspace{0.02cm}
\includegraphics[width=2.8cm]{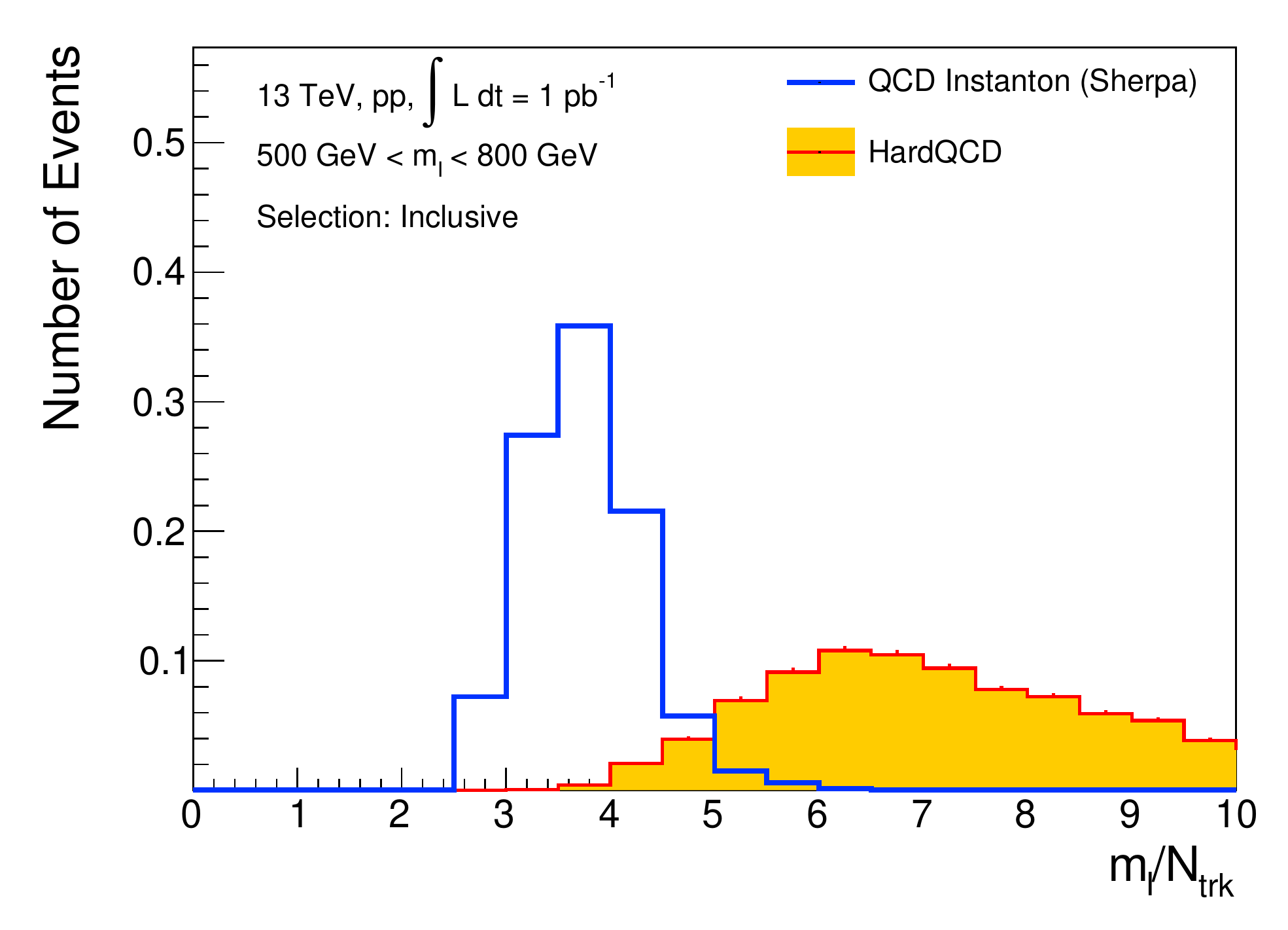}
\hspace{0.02cm}
\includegraphics[width=2.8cm]{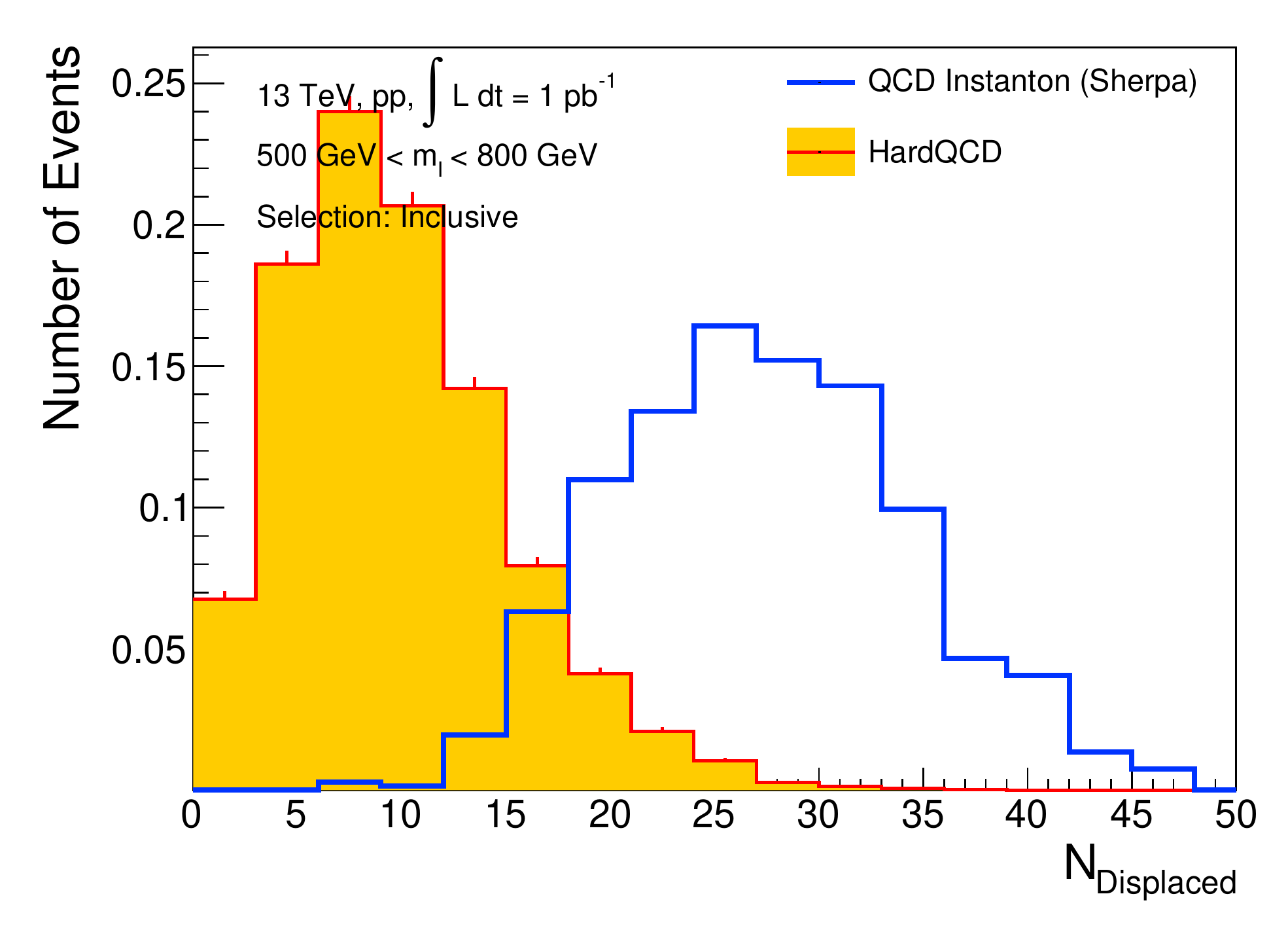} 
\hspace{0.02cm}
\includegraphics[width=2.8cm]{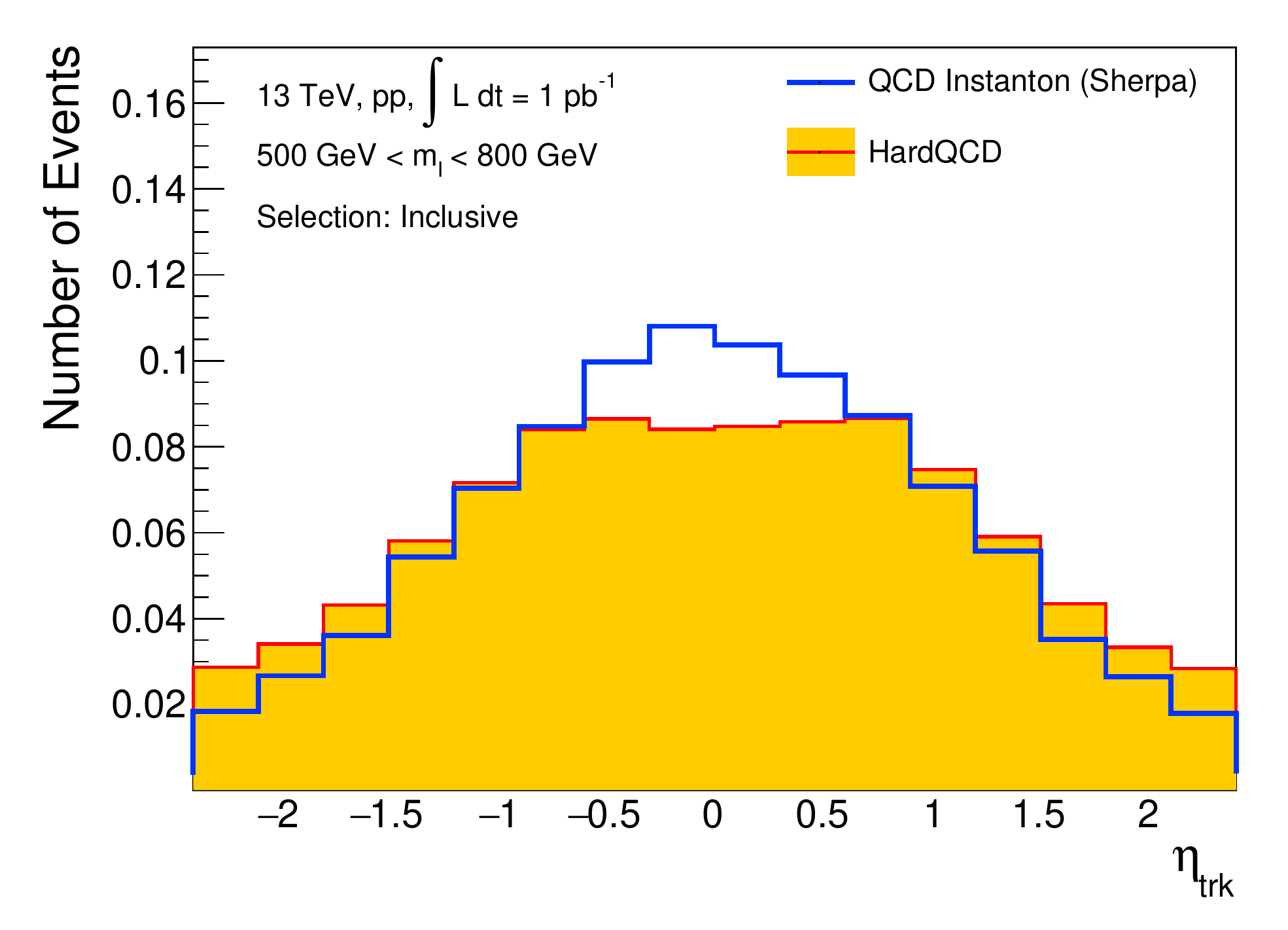}
\hspace{0.02cm}
\includegraphics[width=2.8cm]{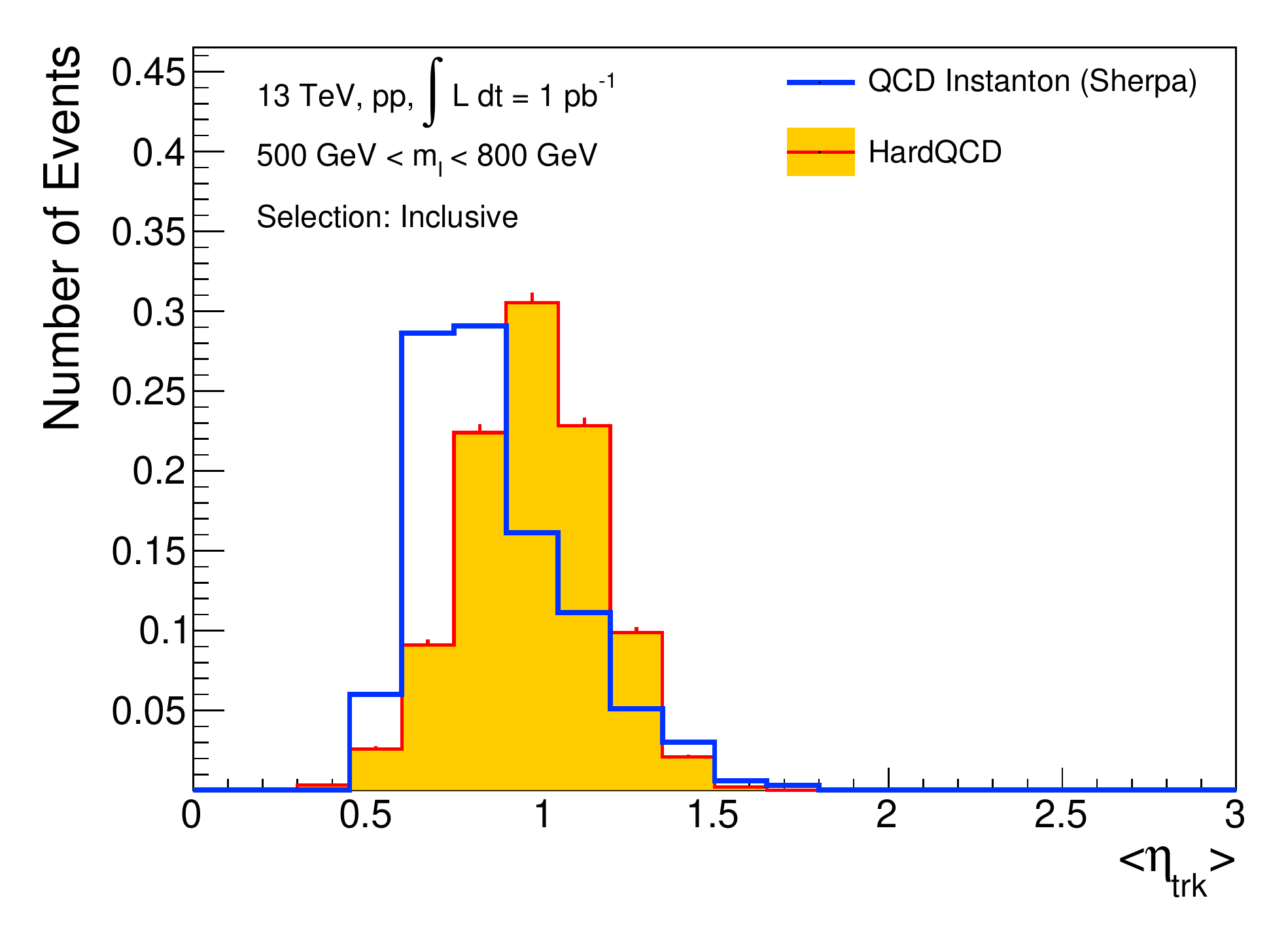}
\\
\includegraphics[width=2.8cm]{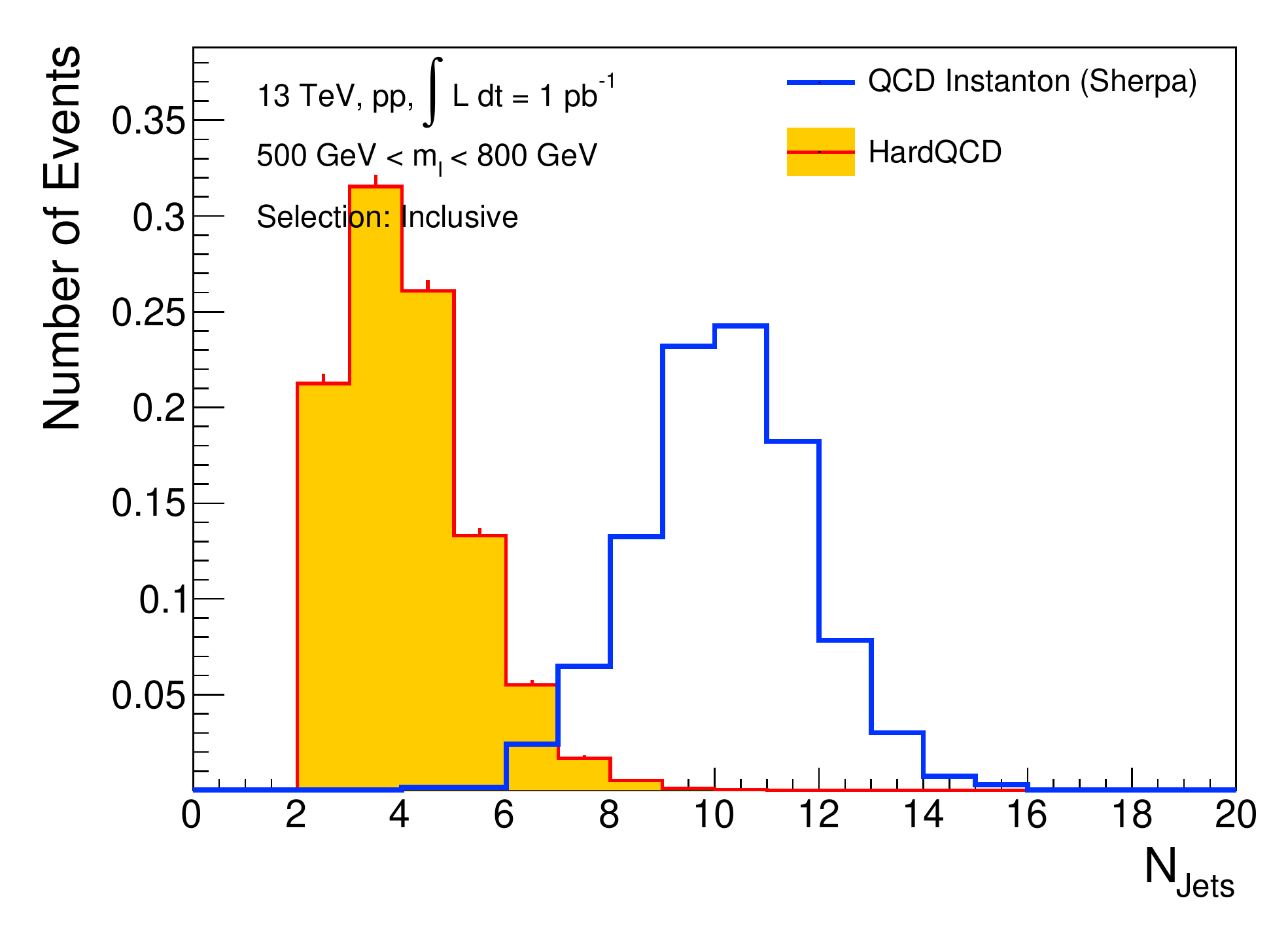} 
\hspace{0.02cm}
\includegraphics[width=2.8cm]{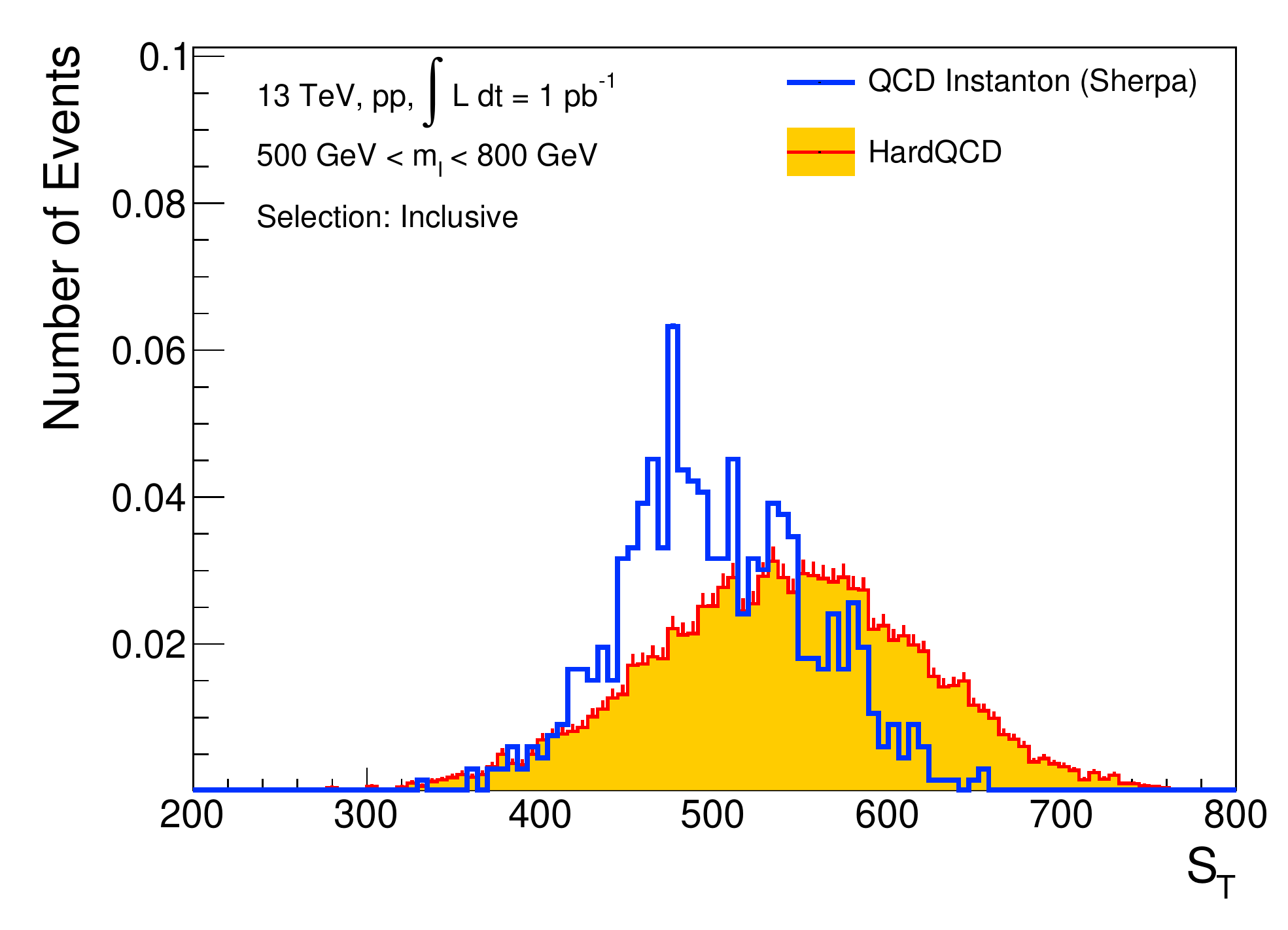}
\hspace{0.02cm}
\includegraphics[width=2.8cm]{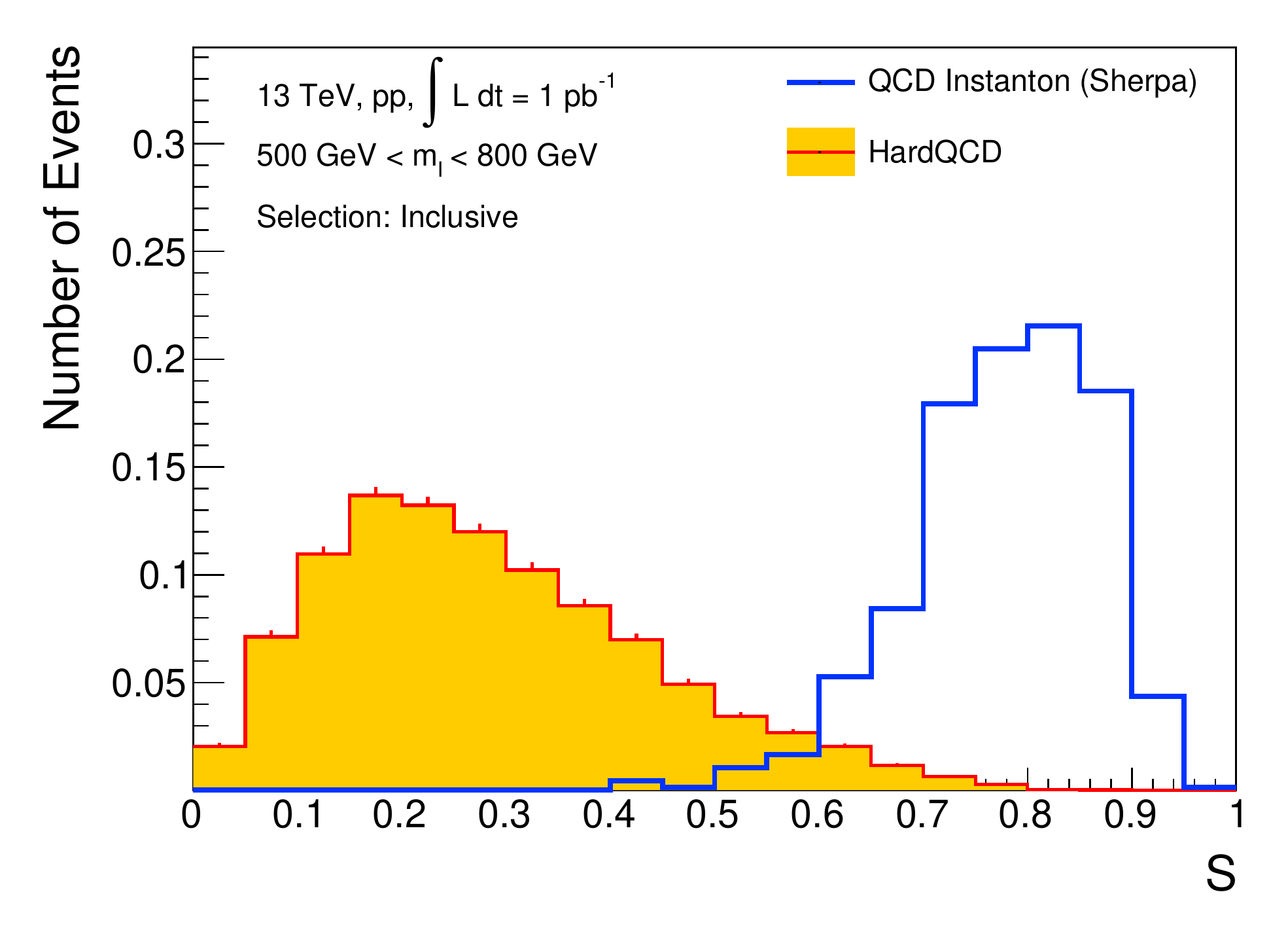} 
\hspace{0.02cm}
\includegraphics[width=2.8cm]{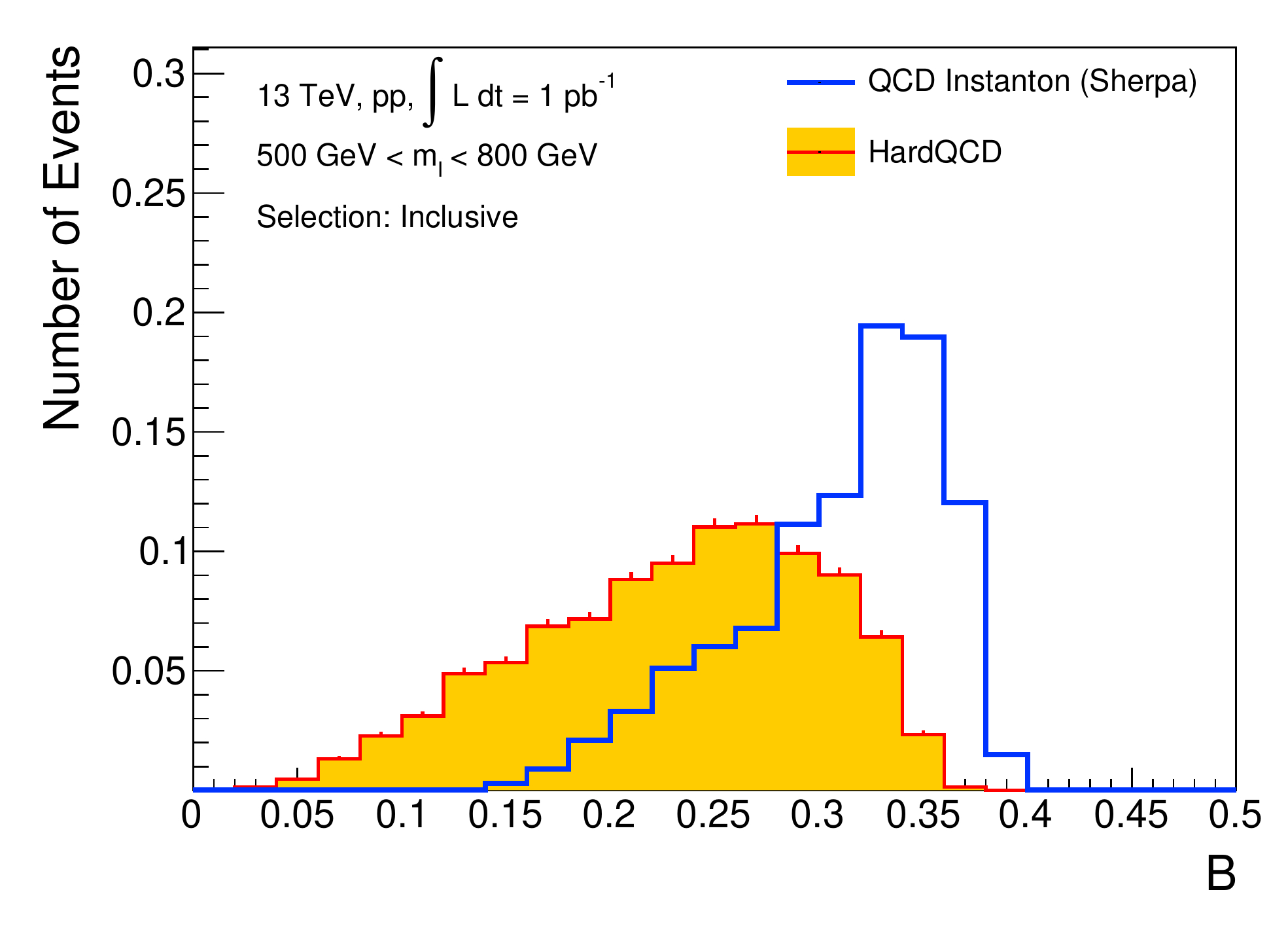}
\hspace{0.02cm}
\includegraphics[width=2.85cm]{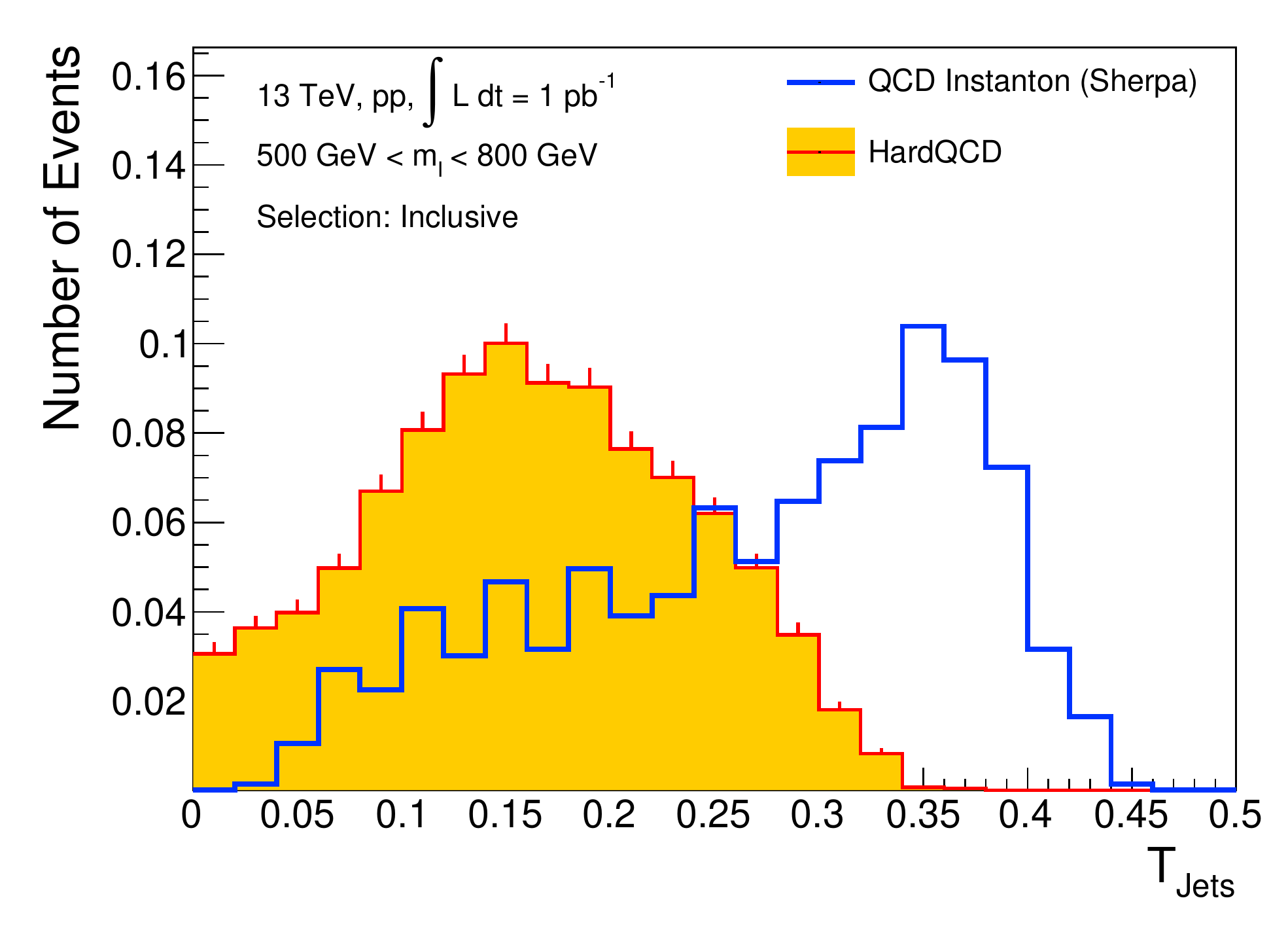}
\caption{\label{fig:ShapesInstanton} Normalized distributions of the Instanton signal sample, generated by Sherpa with Instanton masses between 500~\GeV{} and 800~\GeV{} as well as for \hardQCD processes in a similar mass range: $\Ntrk$, $\mInst/\Ntrk$, $\NDisplaced$, $\Etatrk$, $\langle |\Etatrk| \rangle$,$\NJets$, $S_T^{tracks}$ ,$\mathcal{S}$, ${\cal B}$, ${\cal T}$.}
\end{center}
\end{figure}

The decay of the Instanton pseudo particle in the Sherpa implementation proceeds as follows \cite{Khoze:2019jta}: first, the particle content of the final state is determined, where  quark--anti-quark pairs $q\bar q$, starting from the lowest mass, are added as long as the mass of the quark $m_q$ is smaller than a kinematics dependent threshold $\mu_q$, $m_q<\mu_q$ and as long as the combined mass of all pair-produced quarks is smaller than the Instanton mass. In a second step, the number of additional gluons is determined according to a Poissonian distribution with mean $\langle n_g\rangle$. The \textit{Rambo} algorithm~\cite{Kleiss:1985gy} is then used to distribute isotropically momenta to all decay products in the rest-frame of the Instanton pseudo particle and boosted back to the lab-frame. The subsequent showering and hadronization is based on the standard \textsc{Sherpa} implementation. The shapes of selected experimental observables for Instanton processes is shown in Figure \ref{fig:ShapesInstanton} for a mass range of 500~\GeV{} to 800~\GeV.  For comparison, also \hardQCD processes with a minimal energy $s'$ of 500~\GeV{} are shown. 

Several things have to be noted: first, the used implementation of the Instanton production shows significantly too little events for Instanton masses between 20 and 50 \GeV\ as well as some outliers at very high Instanton masses. To correct for this behaviour, all generated Instanton events are reweighted to the cross section prediction shown in Figure \ref{fig:CrossSection} based on the reconstructed Instanton mass on an event-by-event level. The potential bias which is introduced by this reweighting procedure is expected to be small compared to the theoretical uncertainties on the calculation. Secondly, it is not clear at which value of $\sqrt{\smin}$ the Instanton cross section calculation breaks down. Hence a search for Instantons over the full mass range is necessary. In total six Instanton signal samples have been produced, each covering an exclusive mass range between 20 and 600 \GeV. An overview is given in Table \ref{tab:samples}.

In order to validate the main properties of the Instanton decay in the \textsc{Sherpa} implementation, additional samples of the similar decay in the \textsc{Herwig7} generator \cite{Bellm:2019zci} have been produced, where isotropic decays of pseudo-particles with masses between 500~\GeV{} and 800~\GeV{} into $2\cdot N_f$-quarks and $n_g$ gluons have been simulated and reasonable agreement has been observed. 


\section{Search Strategies\label{Sec:Search}}

In contrast to most searches for new particles, no resonance behaviour is expected for Instanton induced processes,  rather a continuous, rapidly falling spectrum of invariant mass of all hadronic final state objects. This provides significant challenges in the search for  Instanton-induced processes. While sizeable cross sections are expected for small Instanton masses, the experimental signatures in this energy regime might be difficult to distinguish from soft QCD activity. At high luminosities, the  large amount of pile-up events further complicates such a search. In the high energy regime, the experimental signatures of Instanton-induced processes are strikin. their cross sections are however highly suppressed and hence difficult to observe in the first place.

\begin{figure}[htb]
\begin{center}
\includegraphics[width=7.3cm]{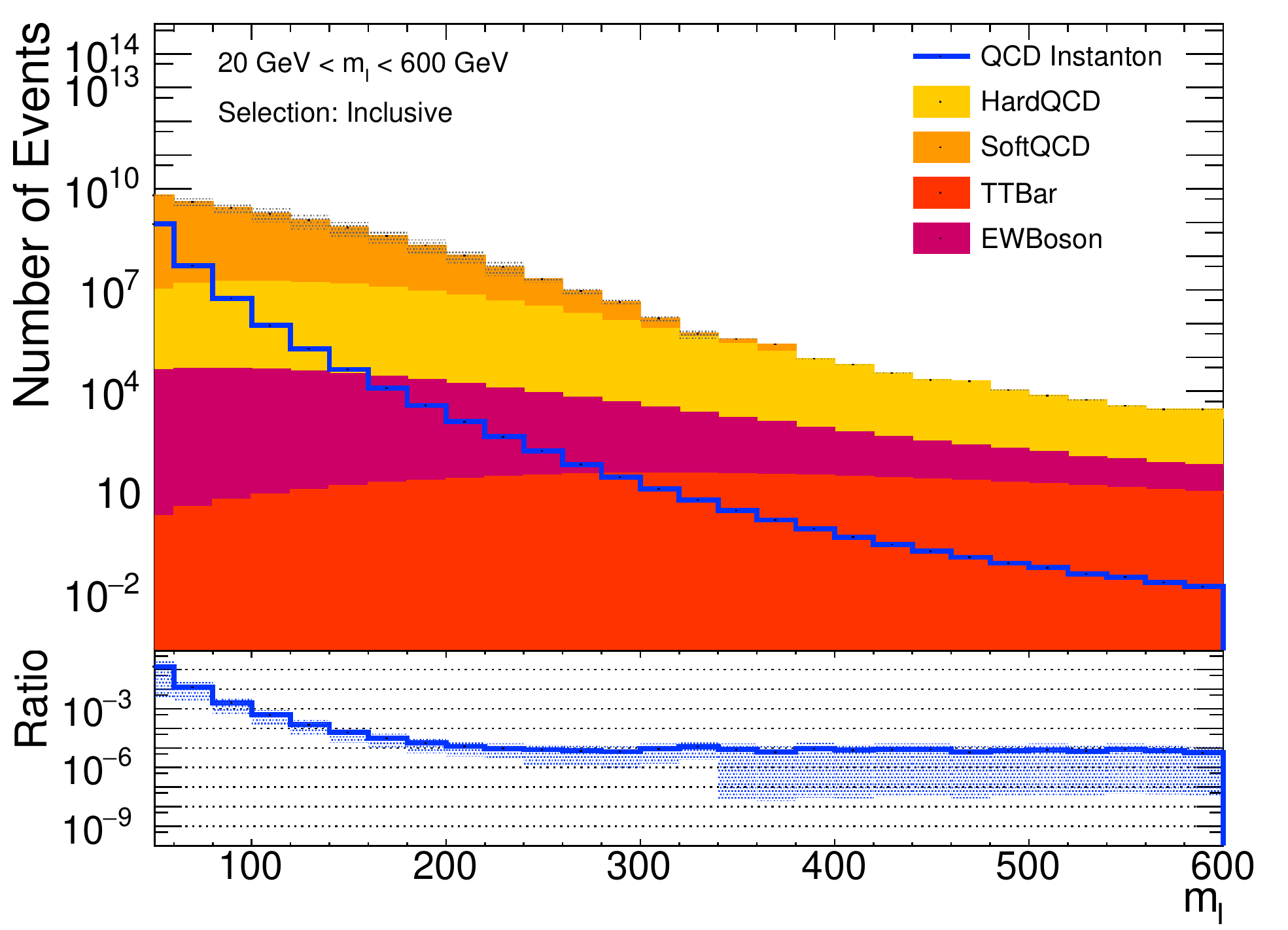} 
\hspace{0.1cm}
\includegraphics[width=7.3cm]{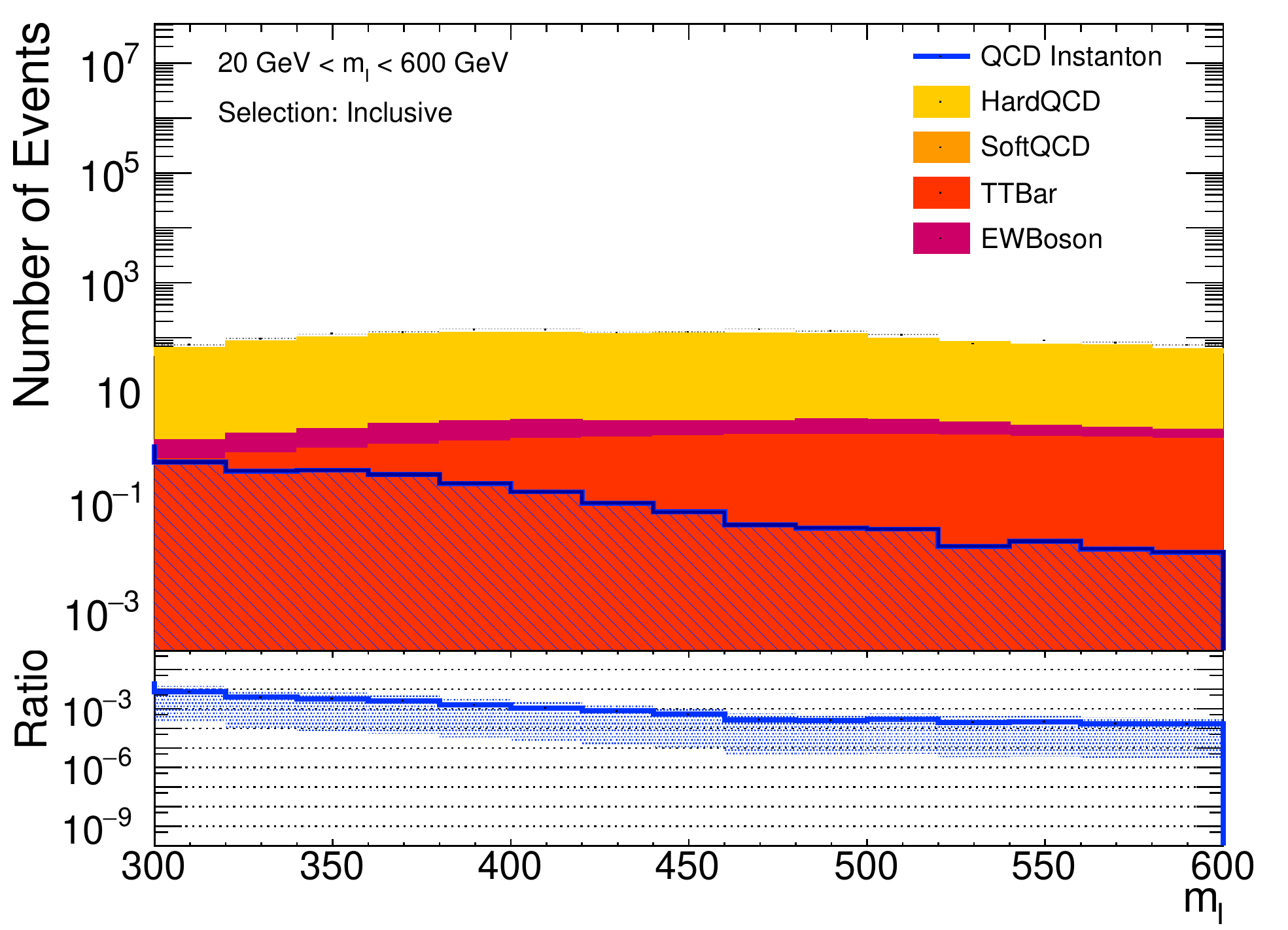}
\caption{\label{fig:SelectionLIM} Invariant mass distribution of all reconstructed tracks for standard model background processes and Instanton processes (left) as well as the same distribution for events with at least ten reconstructed particle jets with a $\pT>20$~\GeV{} (right). The events correspond to an integrated luminosity of $L=\int 1\,\inpb$ and the distributions from all processes except the Instanton process are stacked. The model uncertainties are indicated as bands. The lower plots show the signal over background ratio corresponding to the upper row.}
\end{center}
\end{figure}

The expected invariant mass distribution of reconstructed tracks is shown in Figure \ref{fig:SelectionLIM} for SM background  and Instanton processes, scaled to the expected event yields for an integrated luminosity of $\int L dt = 1\inpb$ . At low invariant masses, \softQCD processes are dominating, while for high invariant masses \hardQCD processes as well as top-quark pair and electroweak boson production becomes relevant.  The signal over background ratio falls rapidly with increasing mass,  suggesting a higher chance of observing of Instanton processes in the low mass regime.  As high multiplicity final state is  expected for the Instanton processes, the number of reconstructed  charged particle tracks or the number of reconstructed jets in the  event can be used as powerful discriminant against backgrounds.  Figure \ref{fig:SelectionLIM} shows the invariant mass distribution for signal and background processes when requiring at least eight reconstructed jets with a minimal momentum of 20~\GeV. The expected signal to background ratio increases by several orders of magnitude, remaining below $10^{-4}$ with the expectation of 1 Instanton process at a integrated luminosity of $\int L dt \approx 0.1\,fb^{-1}$, showing the challenge of   observing of Instanton processes for masses of several hundred~\GeV.

One possible avenue to  is to exploit the different  energy dependence of the production cross section of the Instanton processes and of the background SM processes. A simultaneous search at the different center-of-mass energies available at the LHC might help to strengthen a potential observation. 

In the following, we will present and discuss possible analysis strategies. Our analysis is performed in four different mass ranges: $20<\sqrt{s'}_{\rm min}<40$~\GeV{} and $40<\sqrt{s'}_{\rm min}<80$~\GeV{} for the \textit{low mass regime}, where \softQCD processes dominate, $200<\sqrt{s'}_{\rm min}<300$~\GeV{} for the \textit{medium mass regime} where \hardQCD processes dominate and $300<\sqrt{s'}_{\rm min}<500$~\GeV{} for the \textit{high mass regime}, where also  top-quark pair productions becomes relevant.

Different signal selections have been studies , optimised on the signal to background ratio. In addition, at least two control regions are defined for each mass range.  These regions are designed to have only a small signal contribution and can therefore be used for the validation of the modelling of background processes. In fact, the control regions can also be used for an ABCD-based background estimation technique, i.e.  to determine the background contribution in the signal regions in a fully data-driven way.

\clearpage
\newpage

\subsection{Very Low Instanton Masses: The Soft QCD Regime }

The very low Instanton mass regime is defined for two regions: the first requires the invariant mass of reconstructed tracks, $\mInst$, between 20~\GeV{} and 40~\GeV{}, the second between 40~\GeV{} and 80~\GeV. The average $\mInst$ values for Instantons in both regions are 24~\GeV{} and 46~\GeV, respectively.  A veto on 20~GeV jets at reconstruction level is applied for both mass ranges. This requirement is applied to keep these regions orthogonal to the regions where  \hardQCD processes dominate\footnote{Only \hardQCD events with at least one jet on particle level with $\pT>20$~\GeV, that has either not been reconstructed or reconstructed with a smaller value of $\pT$, pass the signal selection}.

\begin{figure}[htb]
\begin{center}
\includegraphics[width=7.3cm]{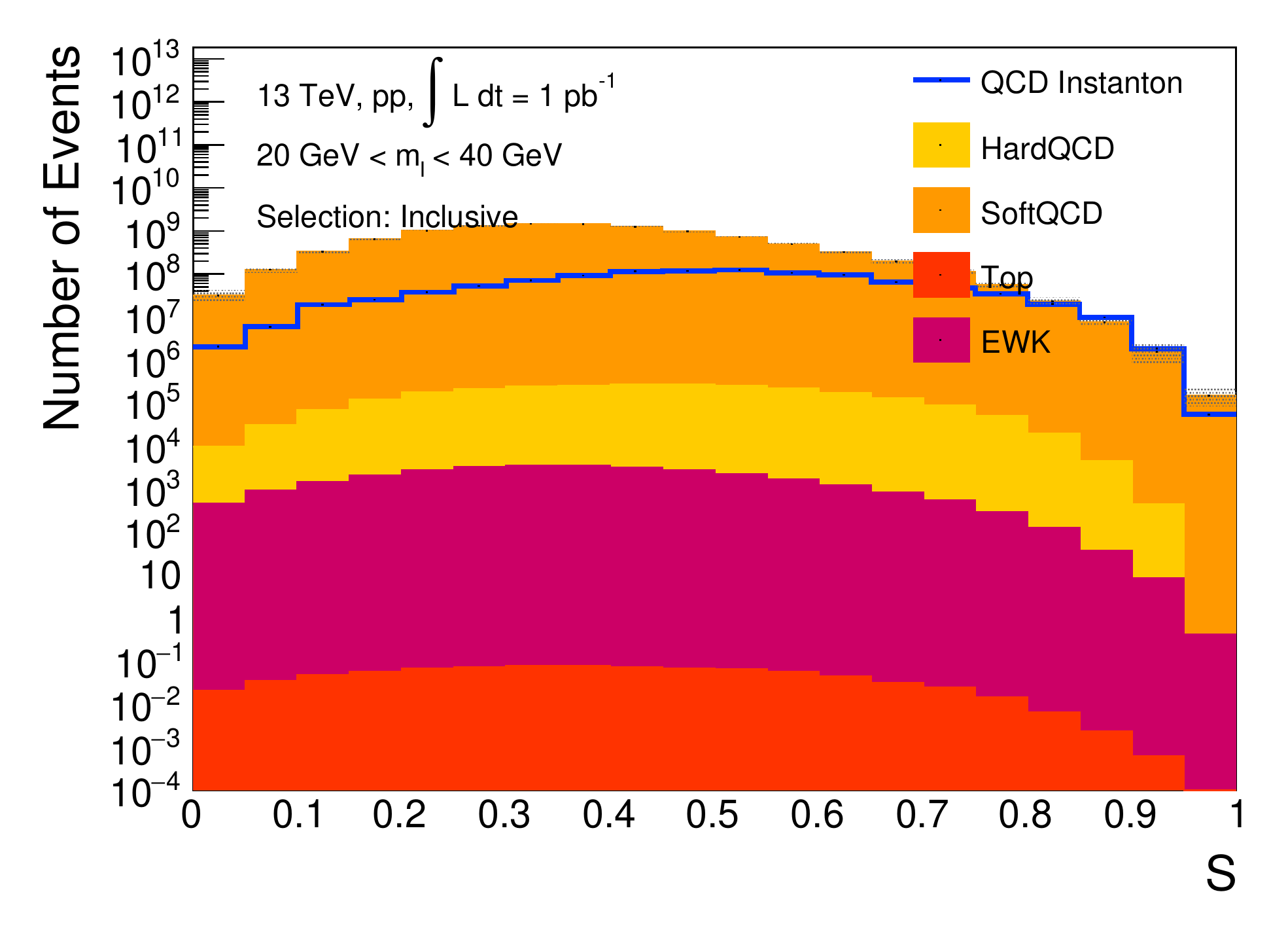} 
\hspace{0.1cm}
\includegraphics[width=7.3cm]{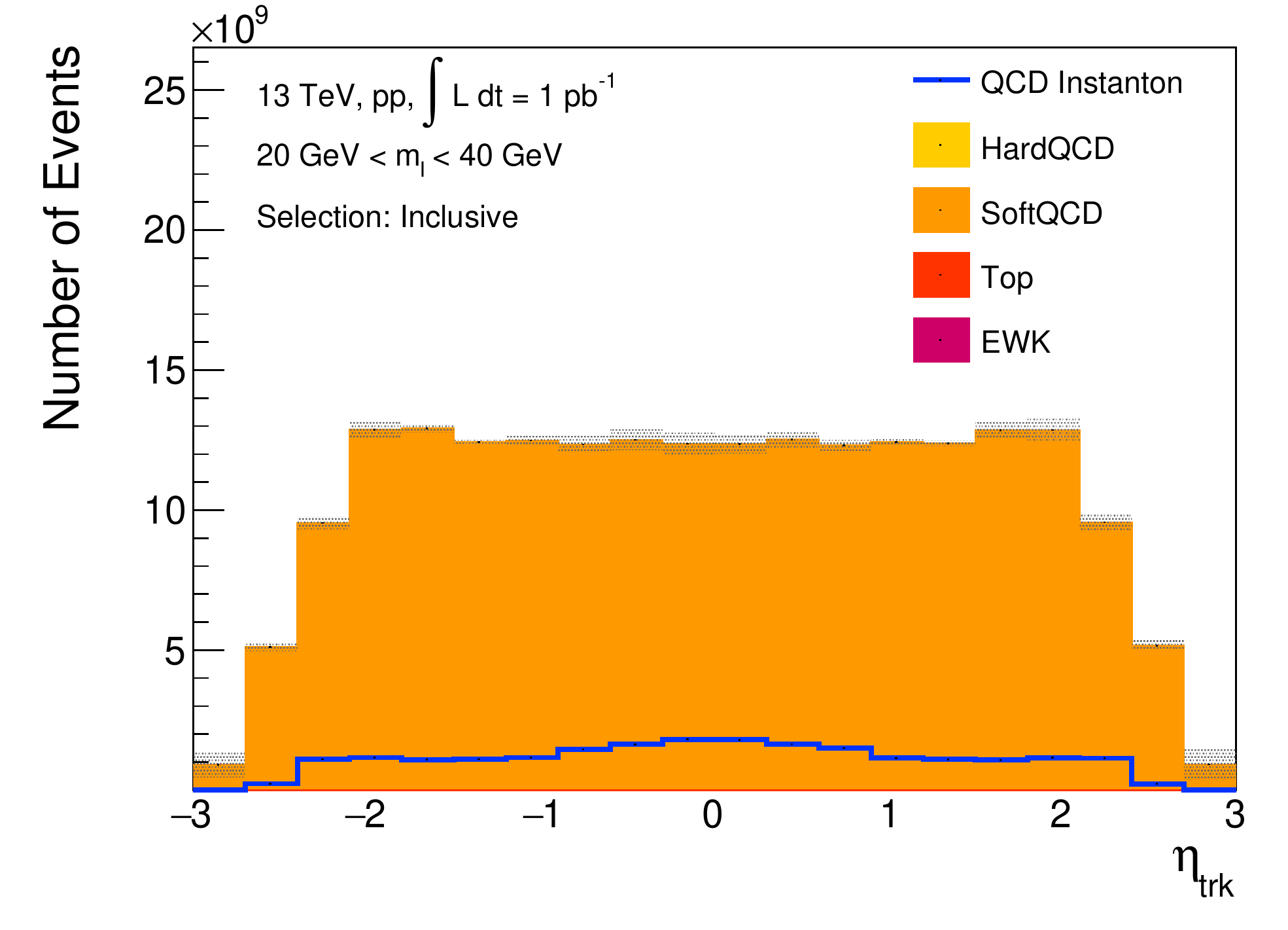}
\caption{\label{fig:LowMass1Overview} Predicted distributions of the event sphericity, $\cal{S}$ (left), and pseudo rapidity, $\eta$, of reconstructed charged particles (right) for various processes,normalised to the expected event yields for an integrated luminosity of $L=\int 1\,\inpb$. The invariant mass of all reconstructed tracks is required to be between 20~\GeV{} and 40~\GeV{} (very low mass regime). The distributions from all SM processes except for the Instanton are stacked. The model uncertainties are indicated as bands.}
\end{center}
\end{figure}

The lower mass region is discussed first: Figure \ref{fig:LowMass1Overview} shows the predicted distributions of the event sphericity, $\cal{S}$ and the pseudo rapidity, $\Etatrk$, of reconstructed charged particles for the various processes considered. The distributions are scaled to the expected event yields for an integrated luminosity of $L=\int 1\,\inpb$. While \softQCD processes dominate the background, the Instanton signal is enhanced at large values of the sphericity  and predicts a more central $\eta$ tracks distribution.

\begin{figure}[htb]
\begin{center}
\includegraphics[width=2.8cm]{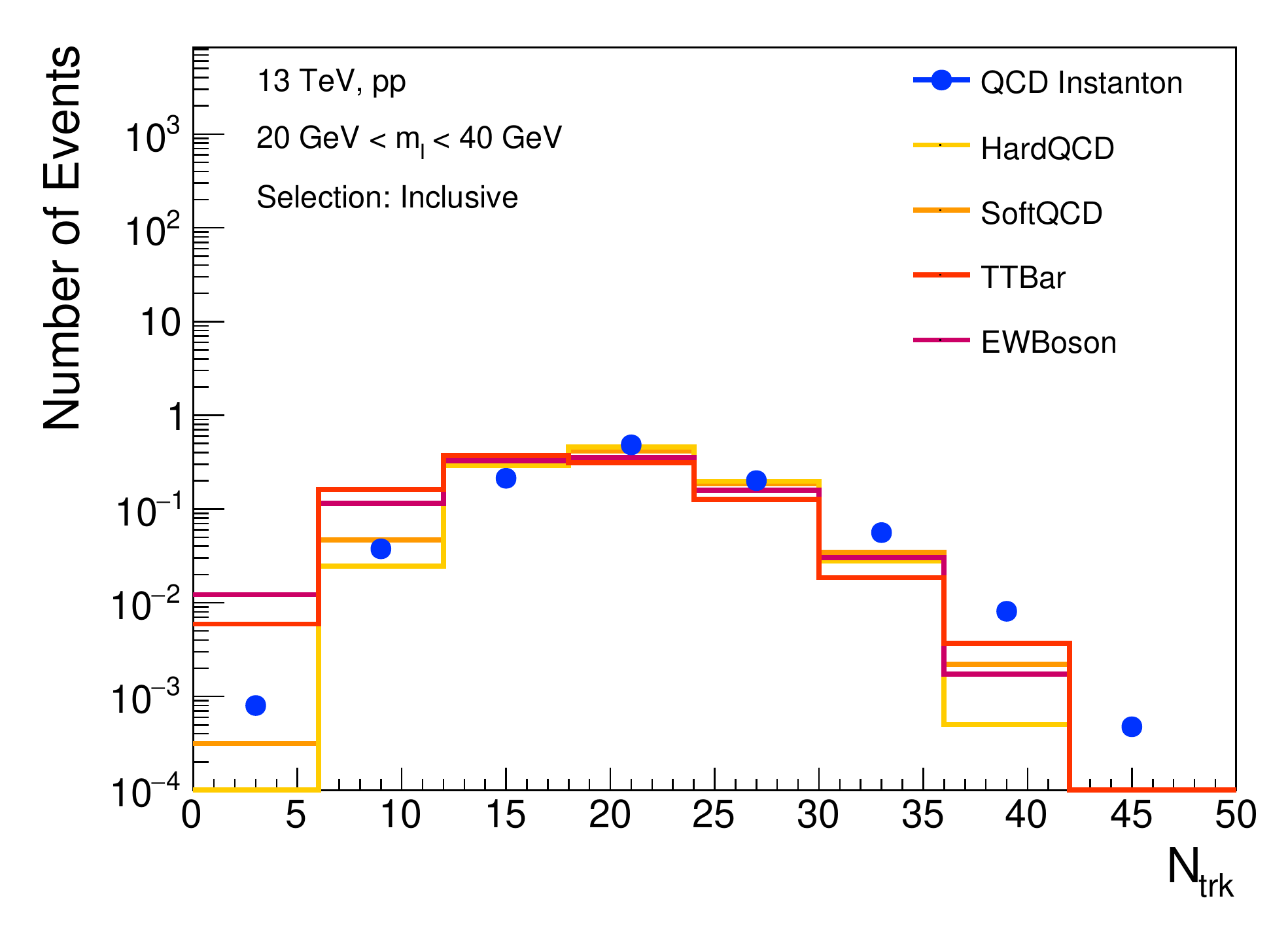} 
\hspace{0.02cm}
\includegraphics[width=2.8cm]{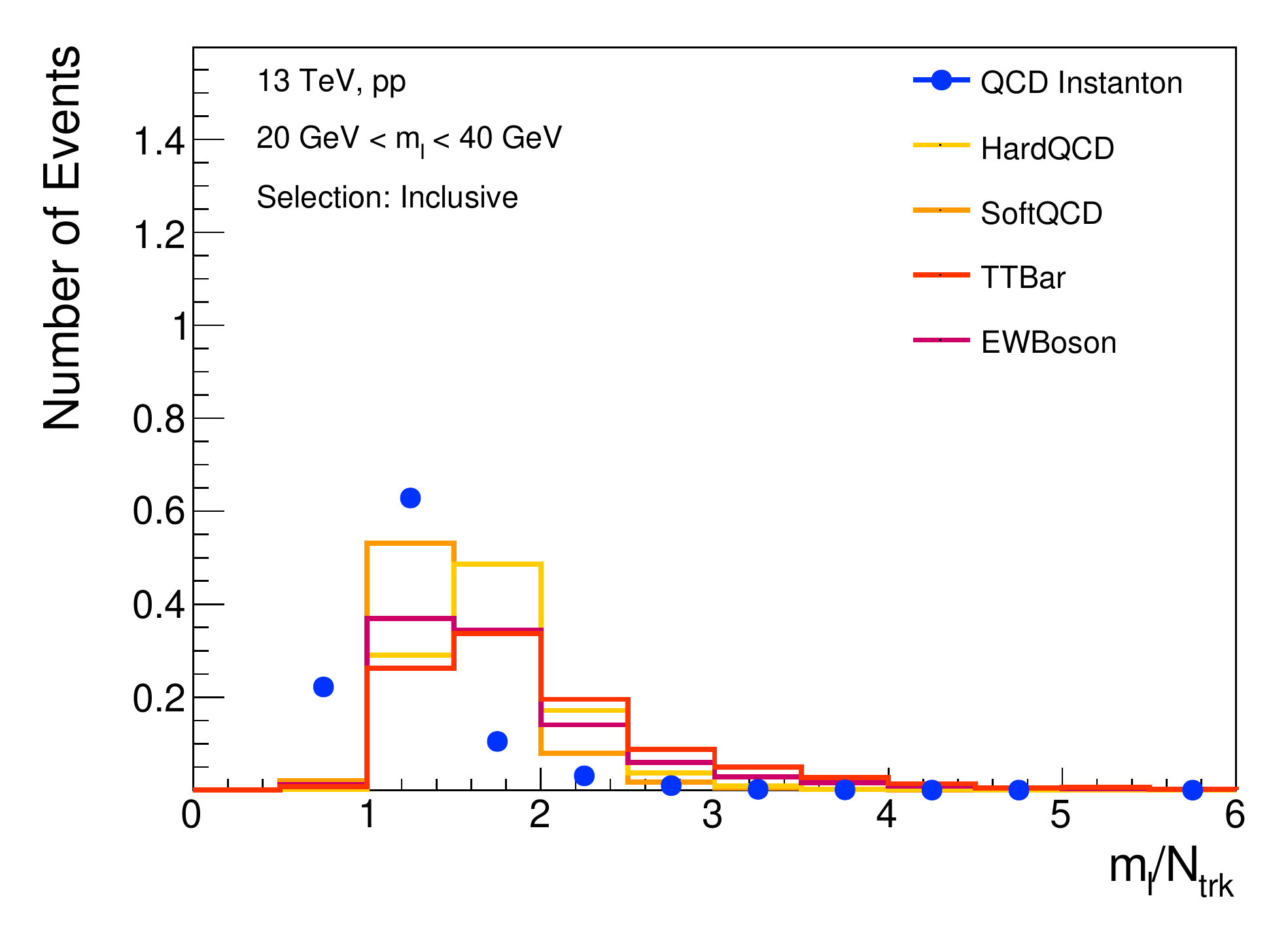}
\hspace{0.02cm}
\includegraphics[width=2.8cm]{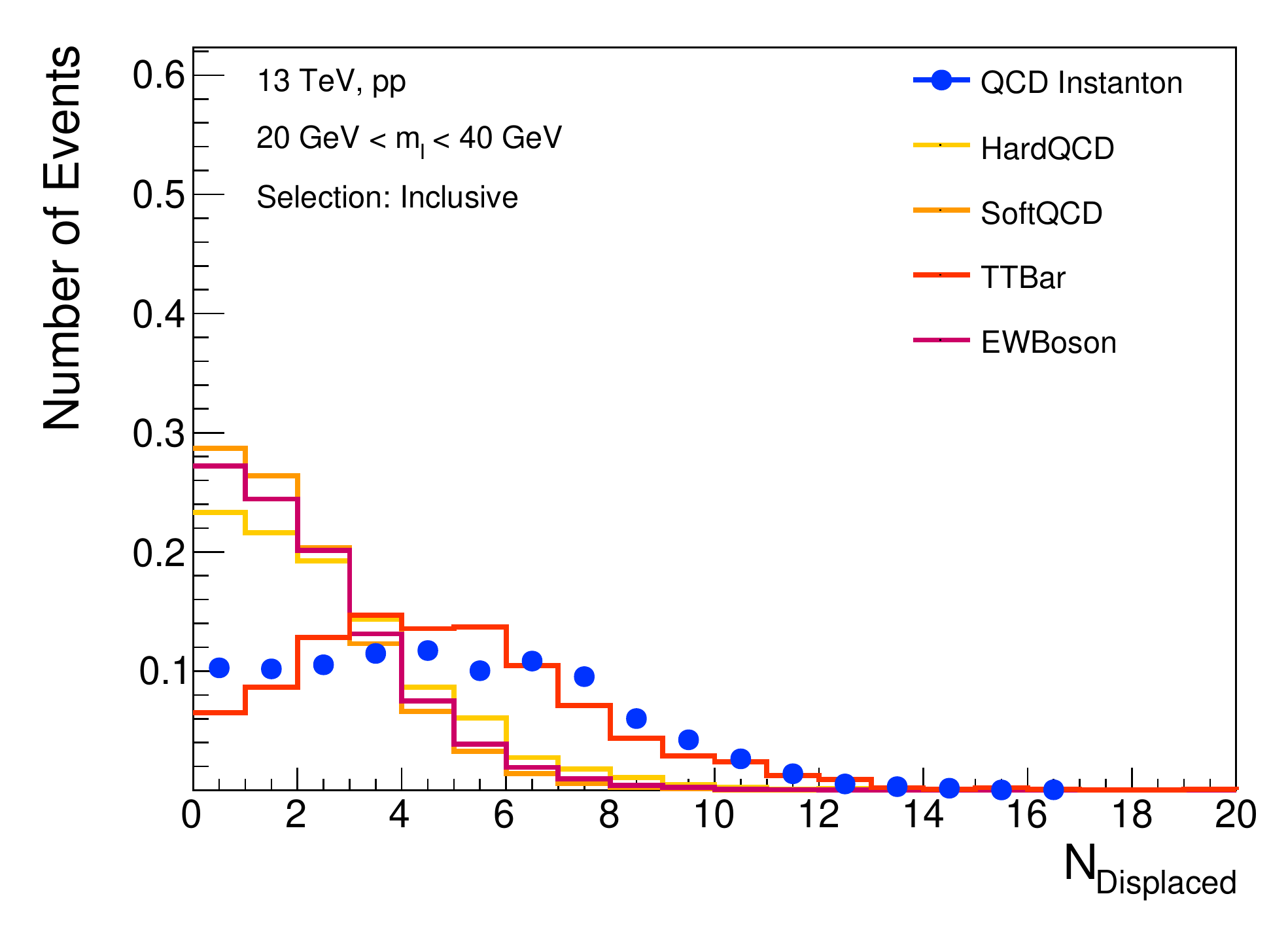} 
\hspace{0.02cm}
\includegraphics[width=2.8cm]{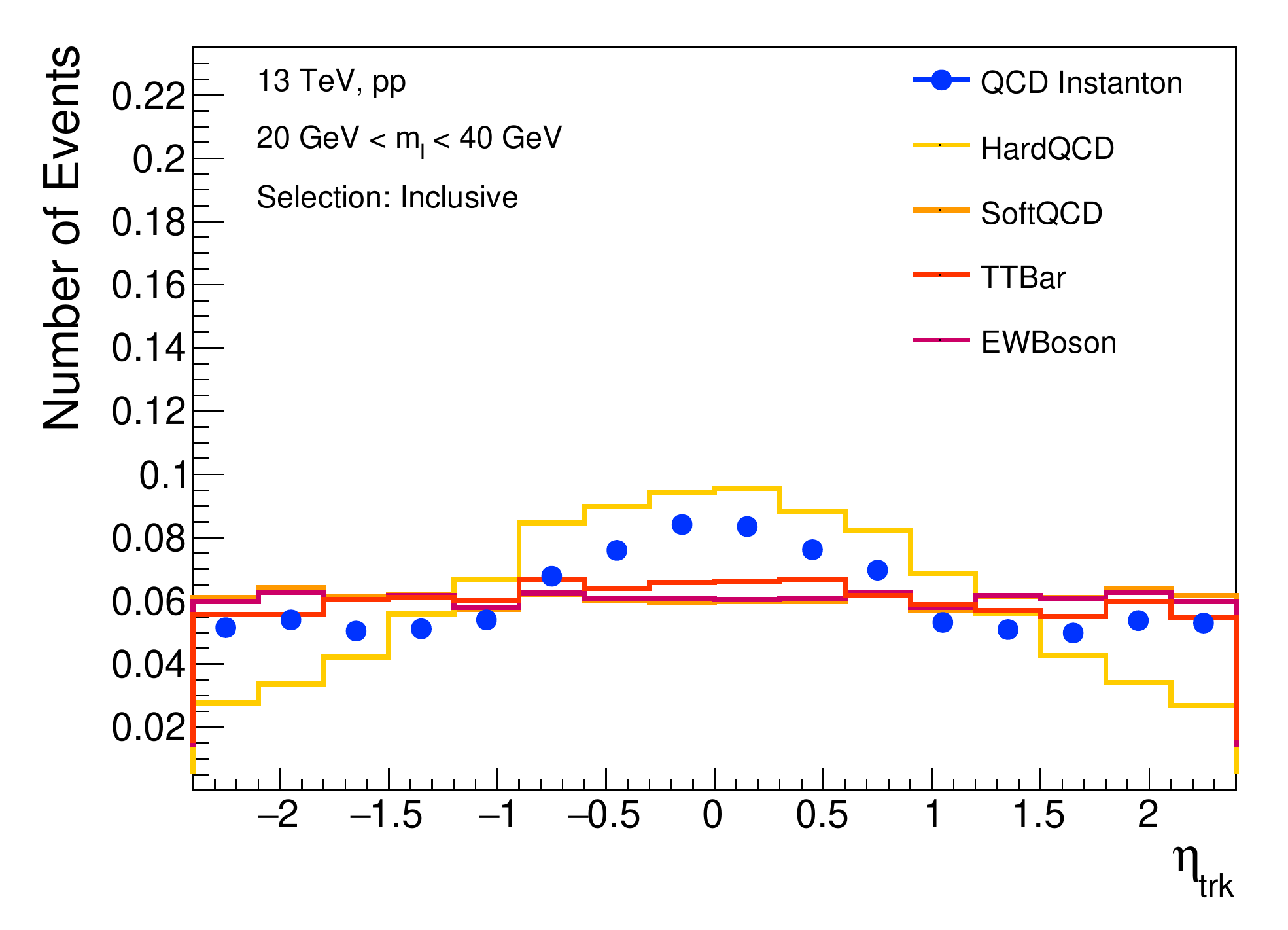}
\hspace{0.02cm}
\includegraphics[width=2.8cm]{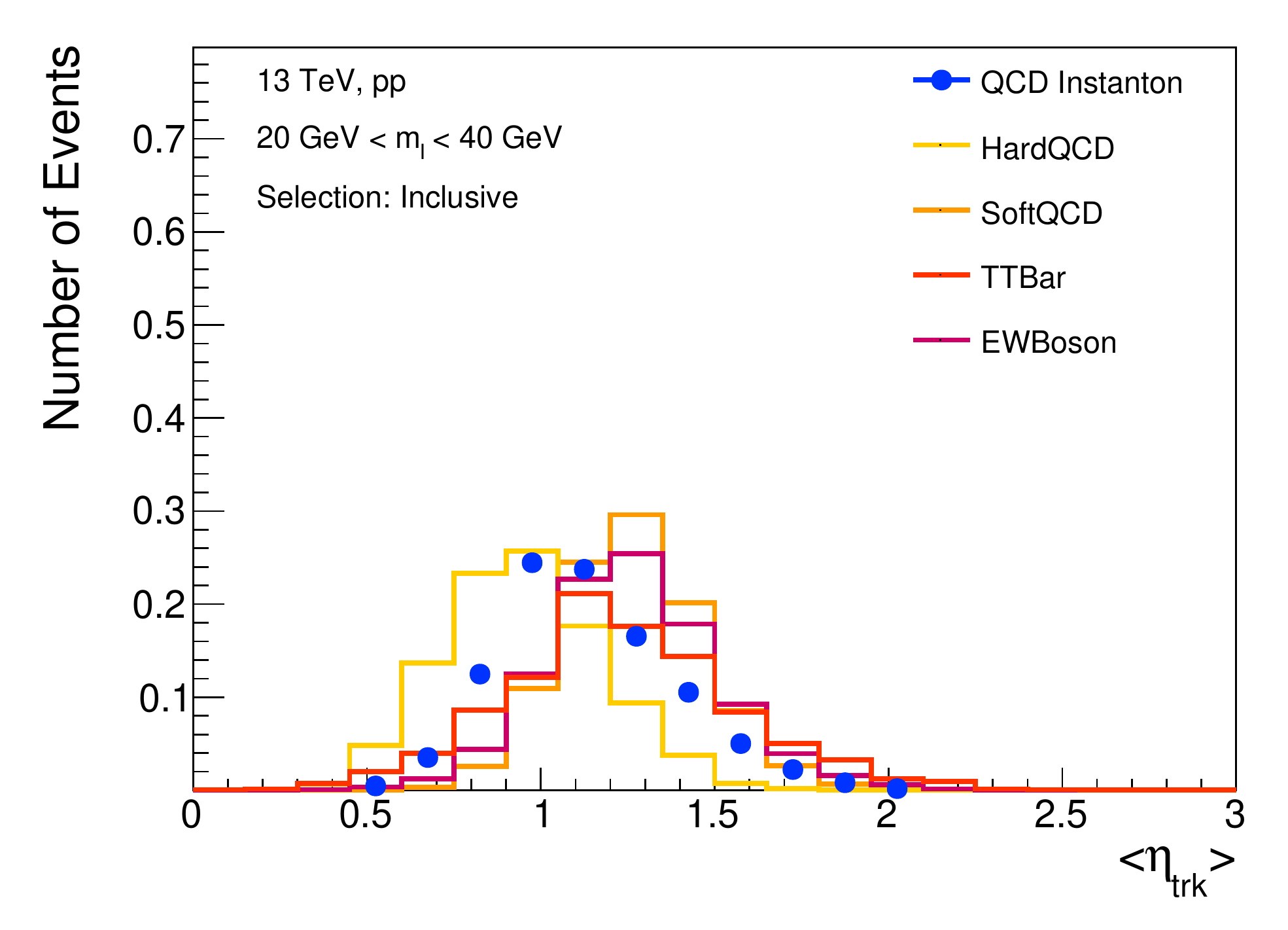}
\\
\includegraphics[width=2.8cm]{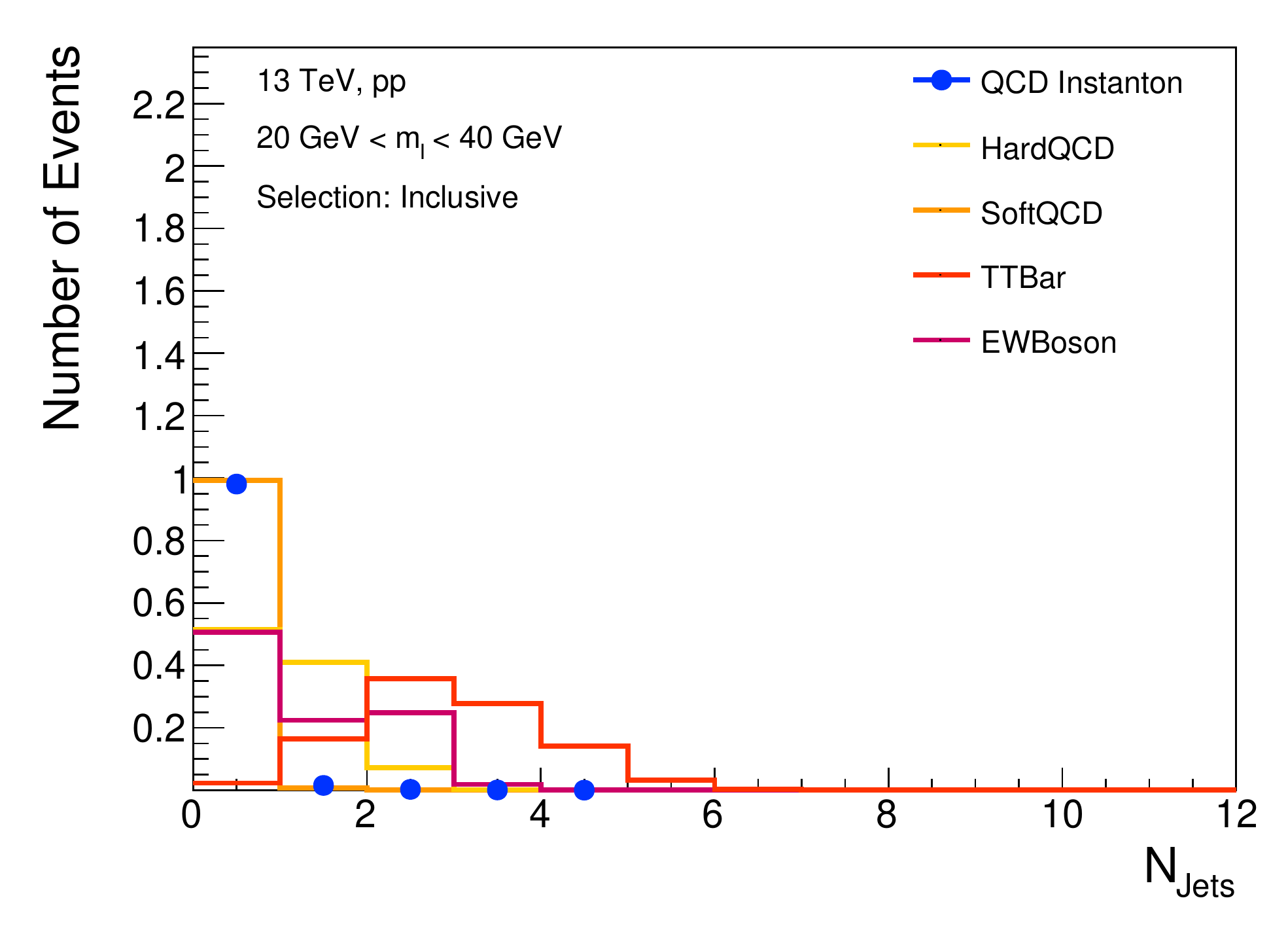} 
\hspace{0.02cm}
\includegraphics[width=2.8cm]{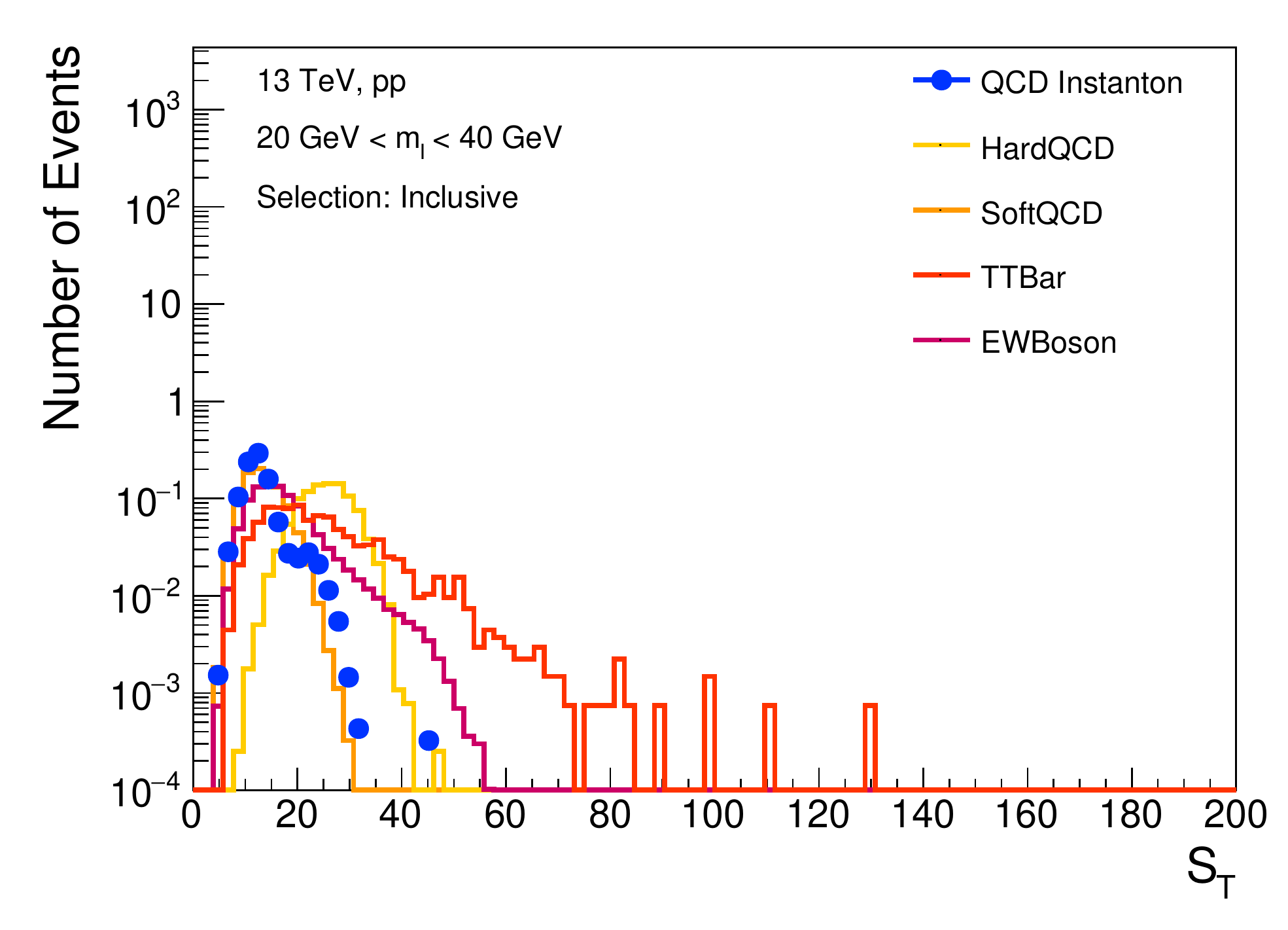}
\hspace{0.02cm}
\includegraphics[width=2.8cm]{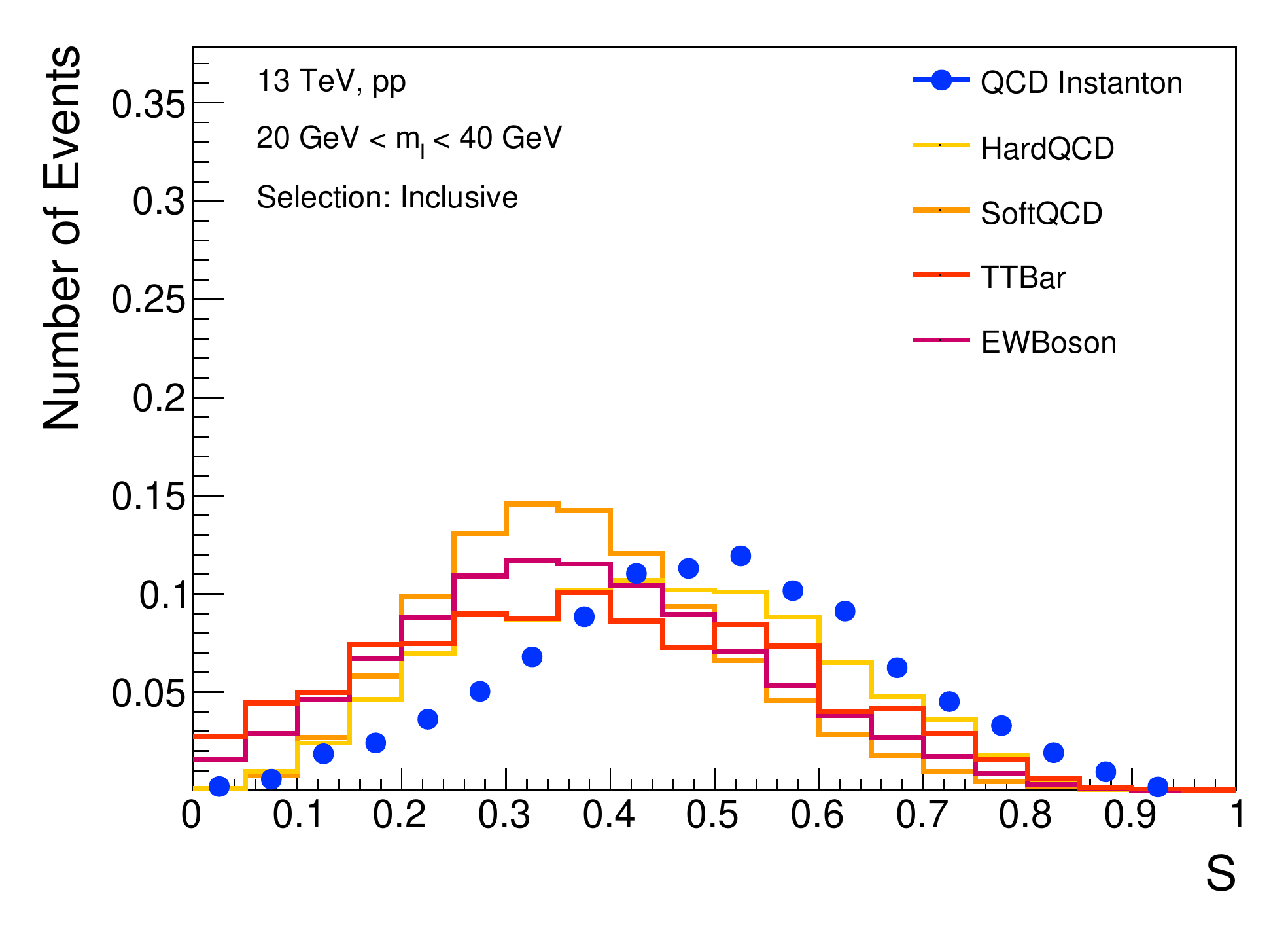} 
\hspace{0.02cm}
\includegraphics[width=2.8cm]{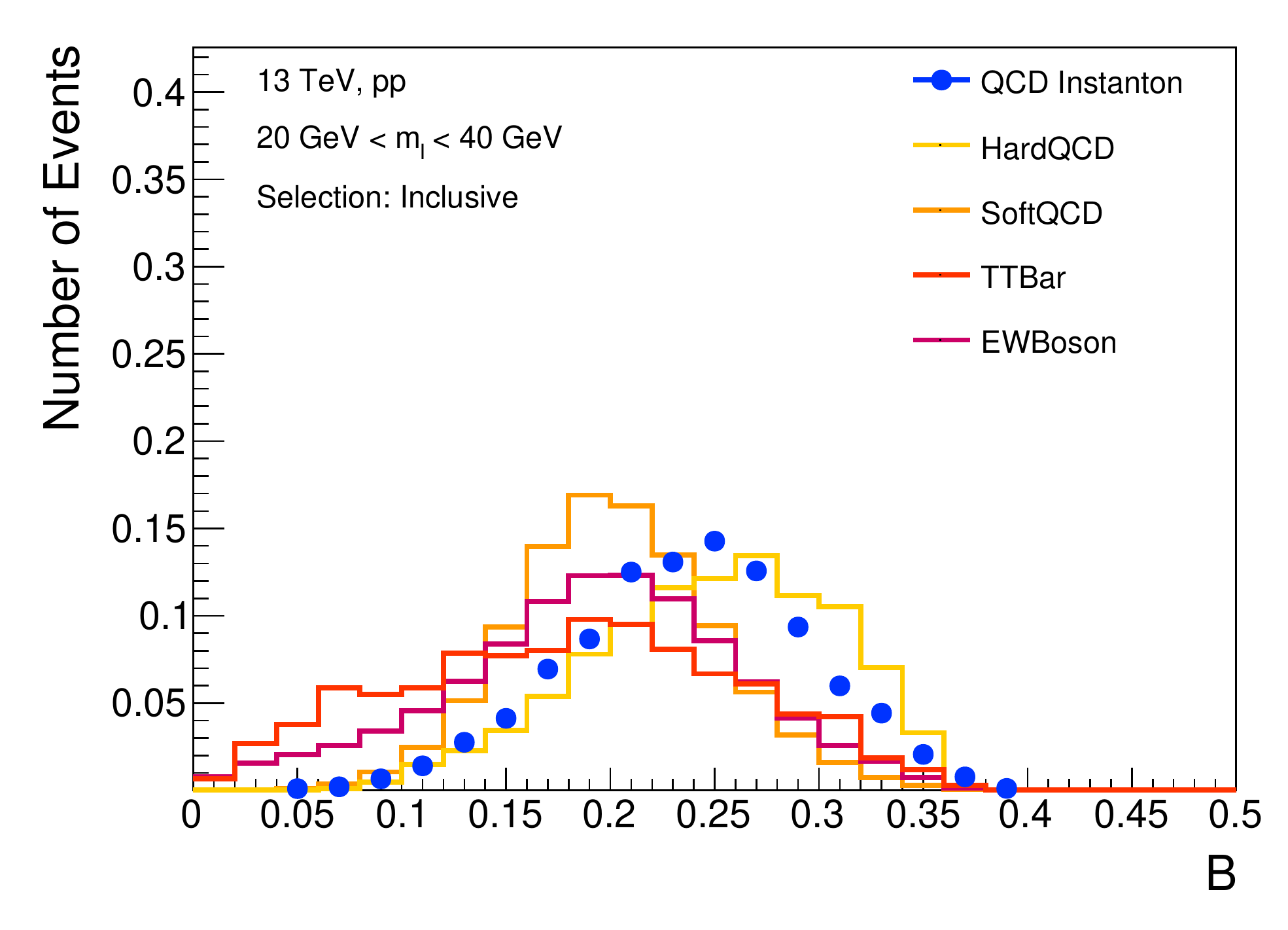}
\hspace{0.02cm}
\includegraphics[width=2.85cm]{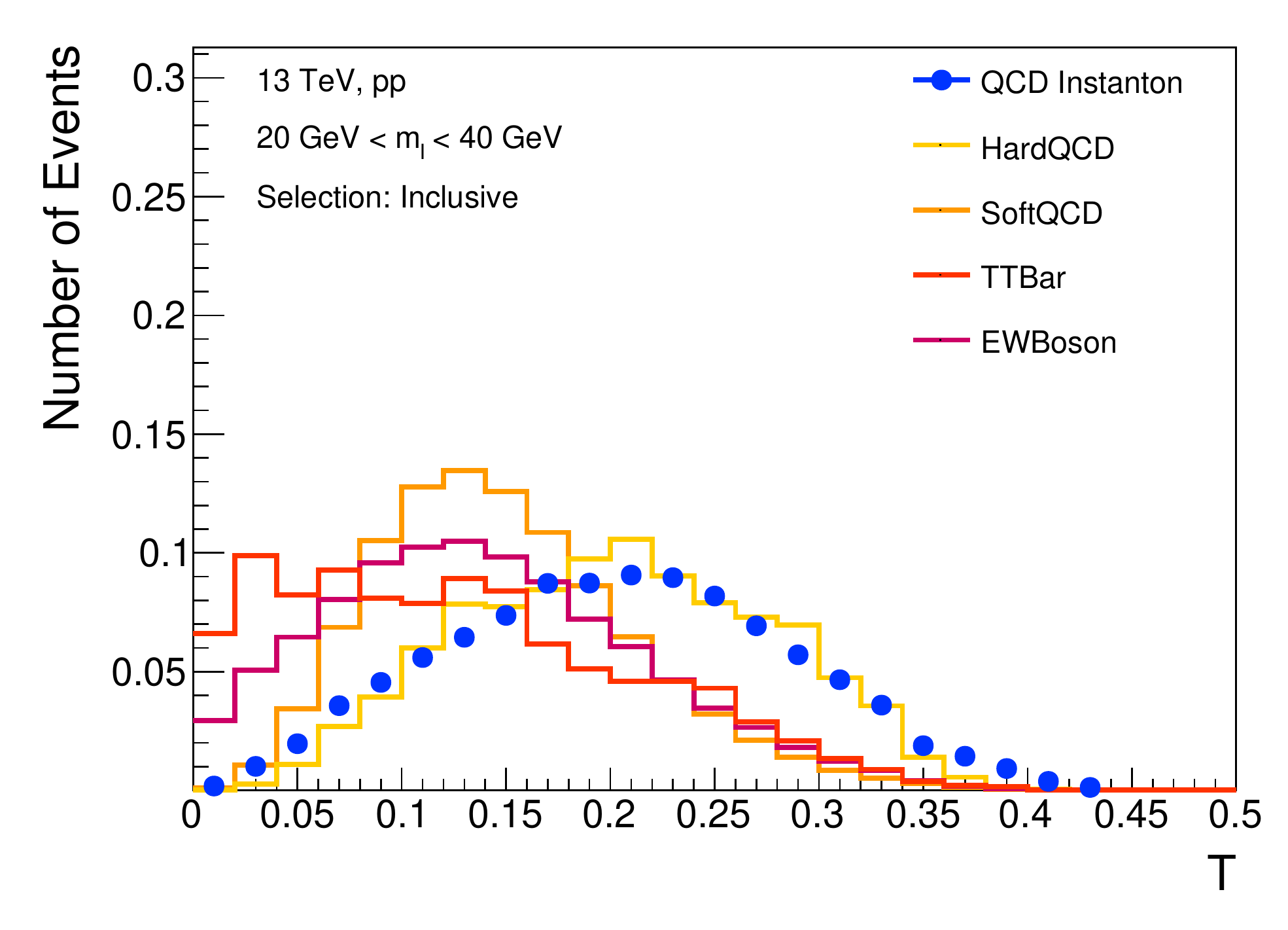}
\caption{\label{fig:ShapesVLIM1} Normalized distributions in the very low mass regime for various processes: $\Ntrk$, $\mInst/\Ntrk$, $\NDisplaced$, $\Etatrk$, $\langle |\Etatrk| \rangle$,$\NJets$, $S_T^{tracks}$ ,$\mathcal{S}$, ${\cal B}$, ${\cal T}$.}
\end{center}
\end{figure}

In order to determine possible observables that allow for a distinction between signal and background processes, it is illustrative to compare the shapes for various observables. An overview of eight relevant observables, previously introduced, is shown in Figure \ref{fig:ShapesVLIM1}. Instantons processes are expected to have larger track multiplicities and hence smaller values of $\mInst/\Ntrk$. As expected, the observables related to the event topology indicate more spherical events compared to the background processes. Highly interesting is the distribution of $\NDisplaced$, i.e. the number of tracks with a displaced origin, as it differs significantly for \textit{soft}- and \hardQCD processes and the signal process. This behavior might be explained by the fact that more heavy quarks in the final state of the Instanton decays are expected, which typically hadronize to long(er) lived mesons and hadrons. Based on these distributions, a few signal selection scenarios have been developed and are summarized in Table \ref{tab:SelectionVLIM1}. No requirements are made specifically for the event sphericity as well as the pseudo-rapidity of tracks. The idea behind this approach is, that these distributions could then be used in a combined fit of signal and background templates to data in order to extract a limit on the Instanton signal-strength. 

\begin{figure}[htb]
\begin{center}
\includegraphics[width=7.3cm]{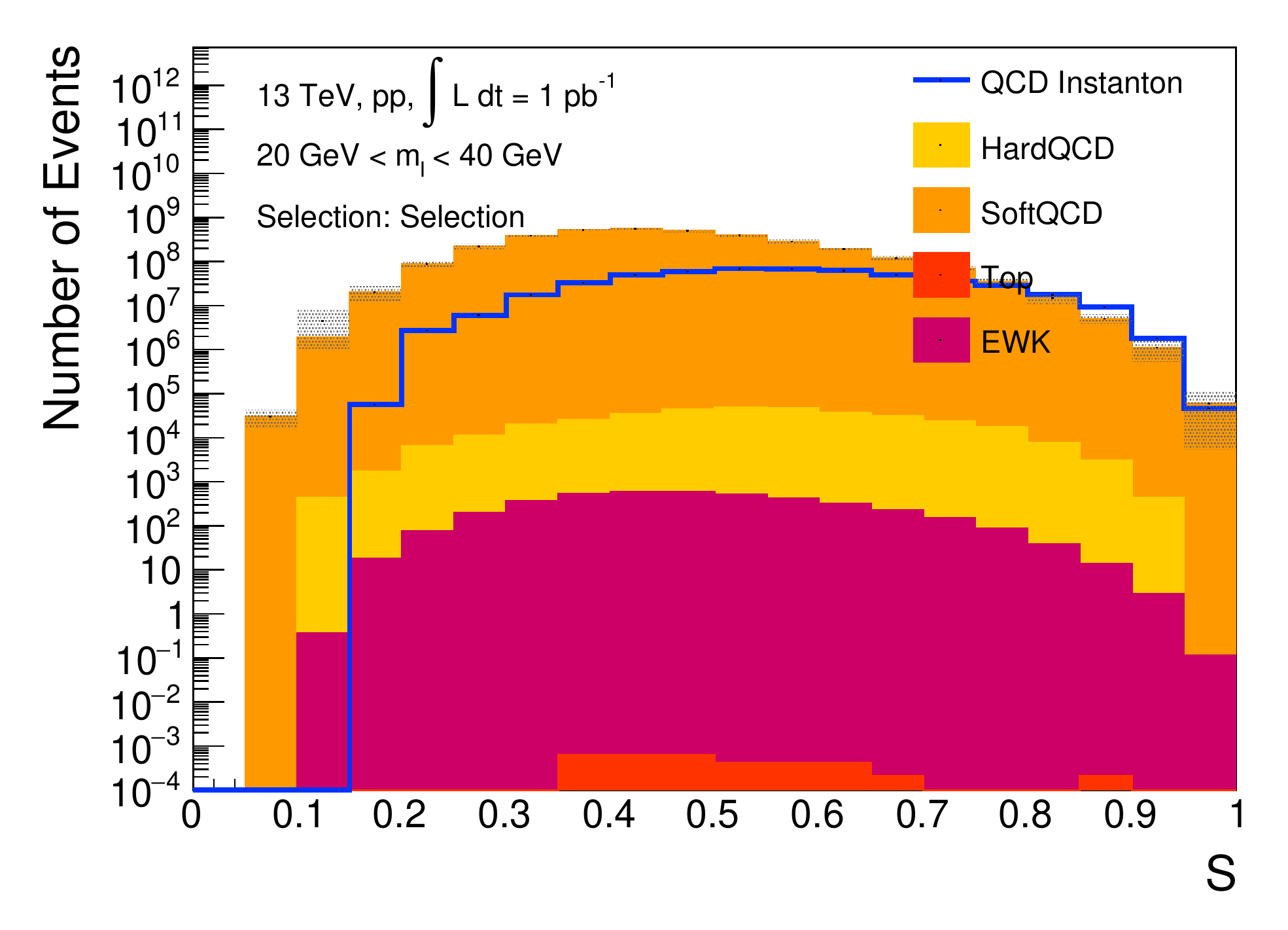} 
\hspace{0.1cm}
\includegraphics[width=7.3cm]{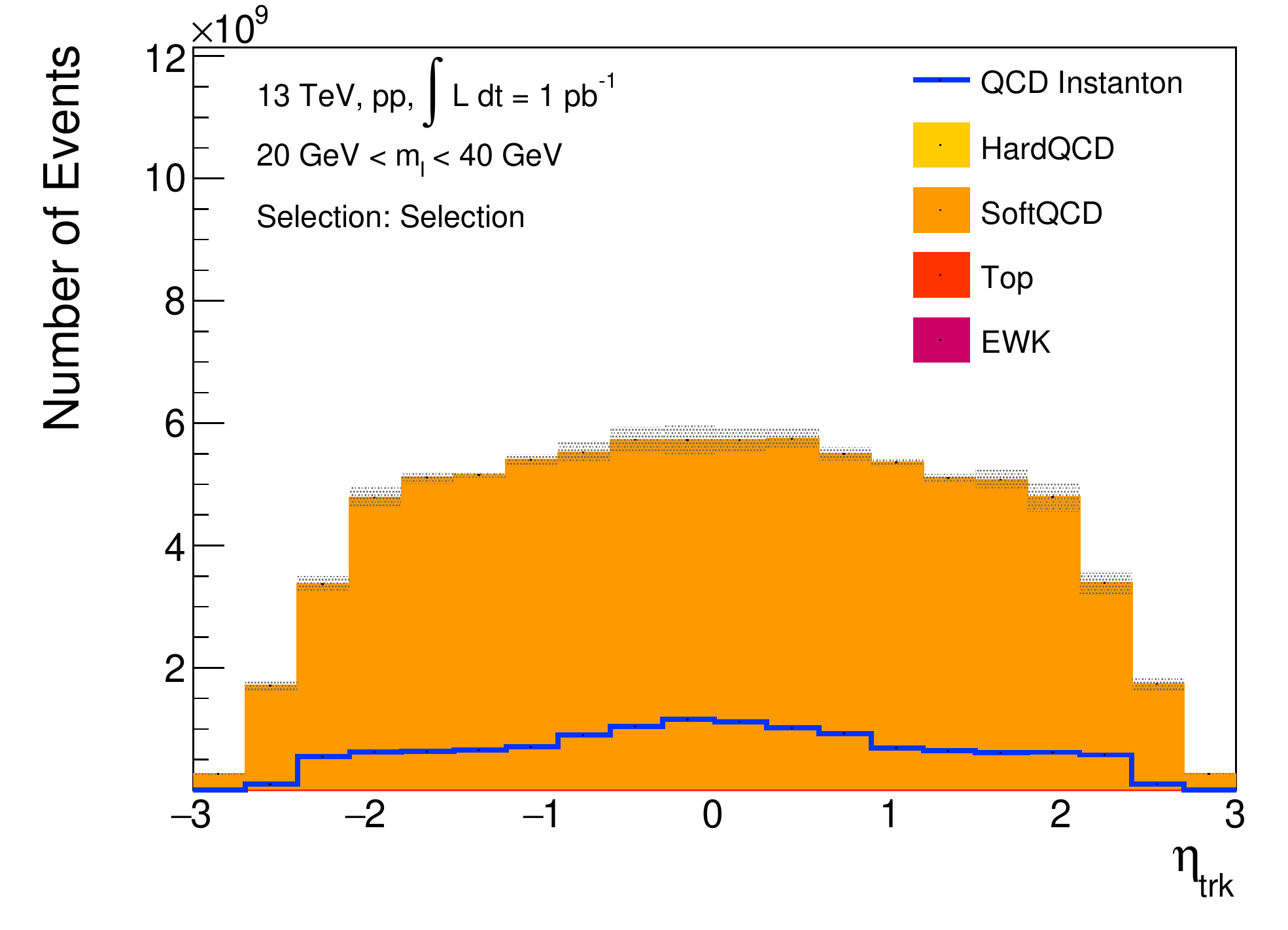}
\caption{\label{fig:SelectionVLIM1}Predicted distributions of the event sphericity (left) and pseudo rapidity, $\eta$, of reconstructed charged particles (right) for various processes, weighted by their predicted cross sections for an integrated luminosity of $L=\int 1\,\inpb$ after the nominal selection. The invariant mass of all reconstructed tracks is required to be between 20~\GeV{} and 40~\GeV{} (very low mass regime). The distributions from all processes except the Instanton process are stacked. The model uncertainties are indicated as bands.}
\end{center}
\end{figure}

\begin{figure}[htb]
\begin{center}
\includegraphics[width=7.3cm]{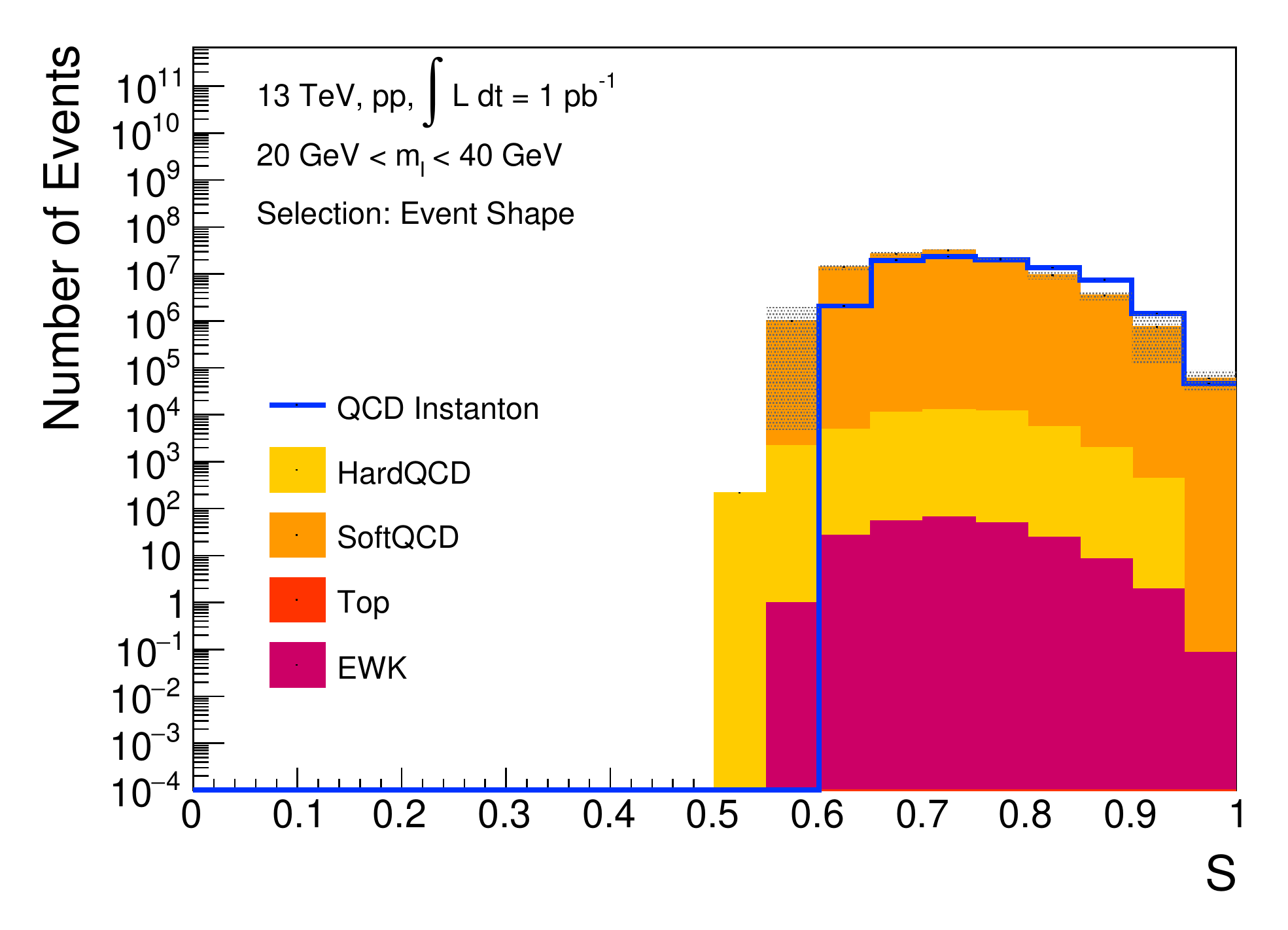} 
\hspace{0.1cm}
\includegraphics[width=7.3cm]{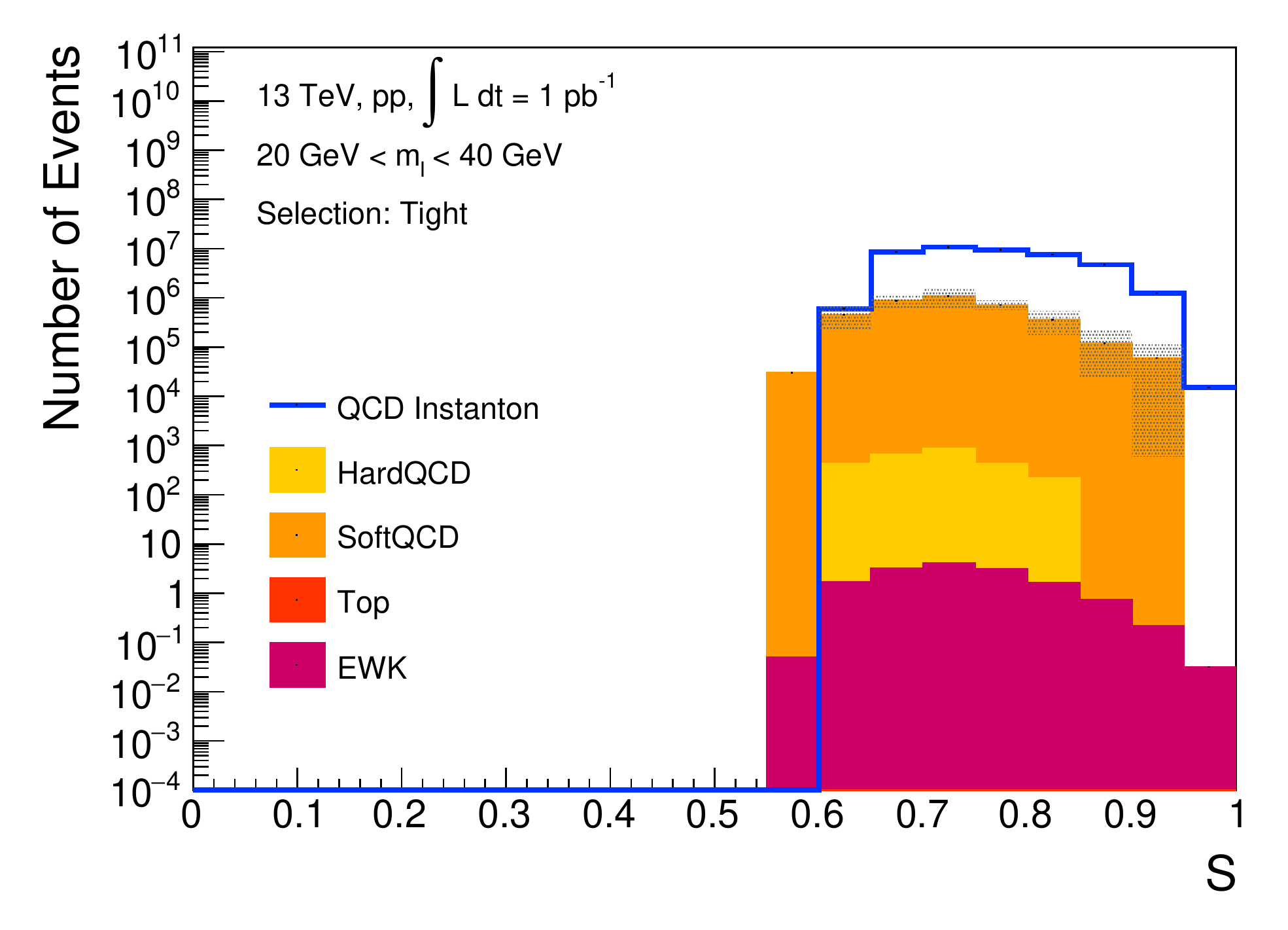}
\caption{\label{fig:SelectionEventTightVLIM} Predicted distributions of the event sphericity for various processes, weighted by their predicted cross sections for an integrated luminosity of $L=\int 1\,\inpb$ after the \textit{event-shape} based selected (left) and tight selection (right). The invariant mass of all reconstructed tracks is required to be between 20~\GeV{} and 40~\GeV{} (very low mass regime). The distributions from all processes except the Instanton process are stacked. The model uncertainties are indicated as bands.}
\end{center}
\end{figure}

The \textit{standard signal selection} applies requirements on the $\Ntrk$, $\mInst/\Ntrk$ and $\NJets$ distributions, where the latter is required to be 0 to reject \hardQCD processes. The resulting sphericity and track $\eta$ distributions for the signal and background processes is shown in Figure \ref{fig:SelectionVLIM1}.  A improvement by a factor of two in the signal over background ratio becomes visible after these selections are applied. In particular, the expected number of Instanton events becomes larger than the total SM background for event with sphericities  $\cal{S}$>0.85, which is used to define the signal region selections in the following.
The \textit{event-shape} signal selection adds in addition requirements on $\cal{B}$ and $\cal{T}$, hence affecting also the $\cal{S}$ distribution. The observables  after this selection are shown in Figure \ref{fig:SelectionEventTightVLIM}. The signal over background ratio improves further when also a requirement of $\NDisplaced>6$ is applied, which defines our \textit{tight signal selection} (Table \ref{tab:SelectionVLIM1}). The resulting sphericity distribution for the \textit{tight signal selection} is  shown in Figure \ref{fig:SelectionEventTightVLIM}. A very clean Instanton signal is expected for this tight selection.

\begin{table}[htb]
\footnotesize
\begin{center}
\begin{tabularx}{\textwidth}{l | l | l | l | c | c}
\hline
											&  \multicolumn{3}{c|}{Signal Region}		    	& \multicolumn{2}{c}{Control Region}	 \\
\hline
      											& Standard	& Event-		&	Tight	    	& A	& B \\
											& 			& Shape		&		    	& 	&	\\
\hline
Invariant mass of rec. tracks (Instanton Mass), $\mInst$   	& \multicolumn{5}{c}{$20 \GeV < \mInst < 40 \GeV$}					\\
\hline
\multicolumn{5}{c}{Selection Requirements}					\\
\hline
Number of rec. tracks, $\Ntrk$ 					& >20 			& >20   			& >20   			& >15 	& >20\\  
Number of rec. tracks/Instanton mass, $\mInst/\Ntrk$ 	& <1.5 			& <1.5   			& <1.5   			& >2.0 	& <1.5\\  
Number of Jets, $\NJets$                      				& =0 			& =0   			& =0   			& =0 	&=0\\
Broadening, $\BTracks$   					& 	 			& >0.3   			& >0.3   			& >0.3	& >0.3\\
Thrust, $\TTracks$   						& 	 			& >0.3   			& >0.3   			&  >0.3 	& >0.3\\
Number of displaced vertices, $\NDisplaced$   		& 	 			& 	   			& >6  			& 	 	&<4\\ 
\hline
\multicolumn{5}{c}{Expected Events for $\int L dt = 1\,\inpb$ in the Signal Region ($\cal S$ >0.85)}					\\
\hline
$N_{Signal}$ 									& $1.1\cdot 10^7$		& $8.9\cdot 10^6$		& $5.9\cdot 10^6$		& <1	 	&	$6.8\cdot 10^5$ \\
$N_{Background}$								& $6.2\cdot 10^6$		& $4.3\cdot 10^6$		& $1.8\cdot 10^5$		& $3\cdot 10^5$. &  $3.3\cdot 10^6$	\\
\hline
\end{tabularx}
\caption{Overview of the standard and tight signal selection as well as the definition of two control regions aiming at very low Instanton masses ($20 \GeV<\mInst<40 \GeV$)\label{tab:SelectionVLIM1}}
\end{center}
\end{table}

\begin{figure}[htb]
\begin{center}
\includegraphics[width=7.3cm]{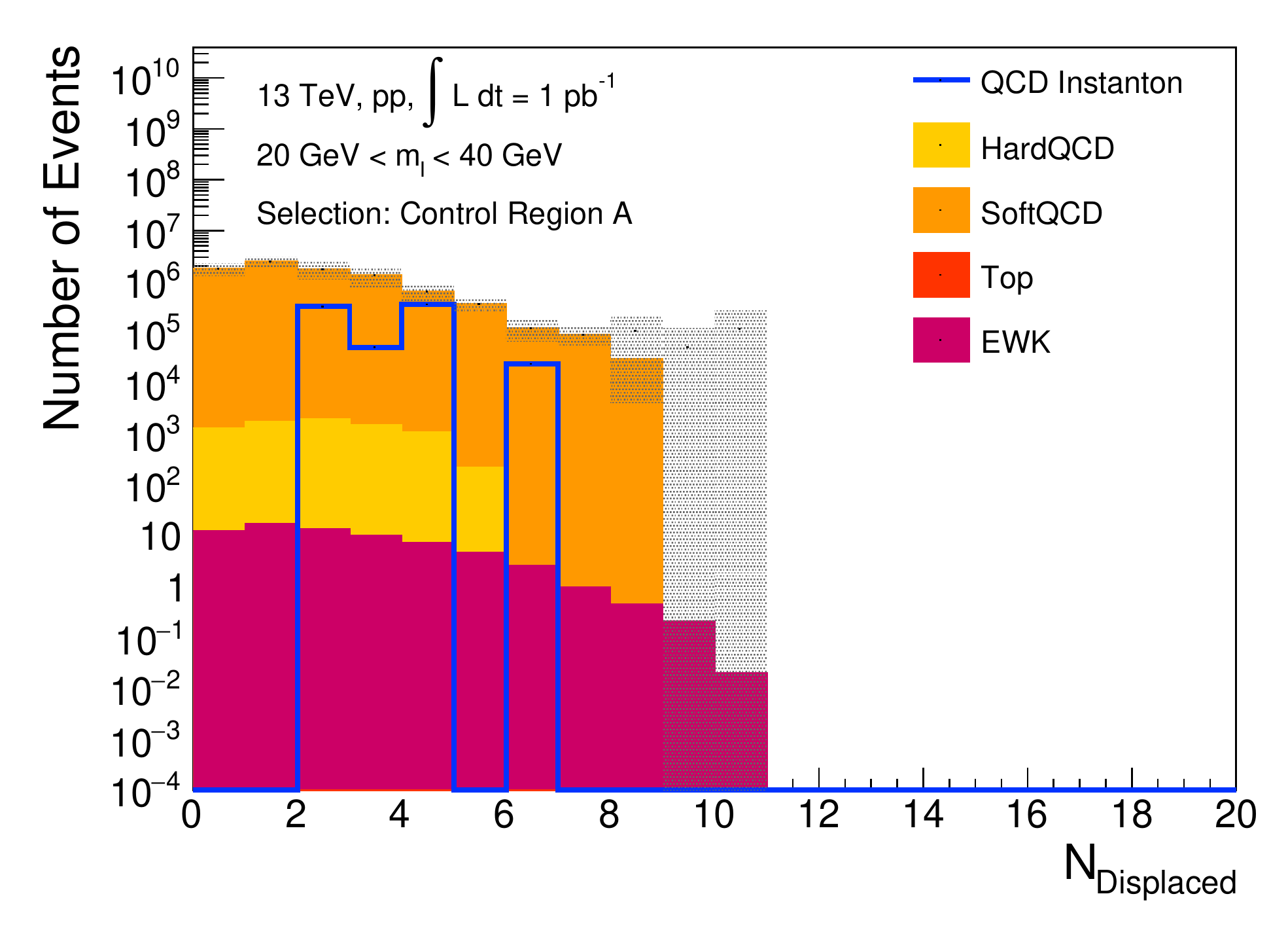}
\hspace{0.1cm}
\includegraphics[width=7.3cm]{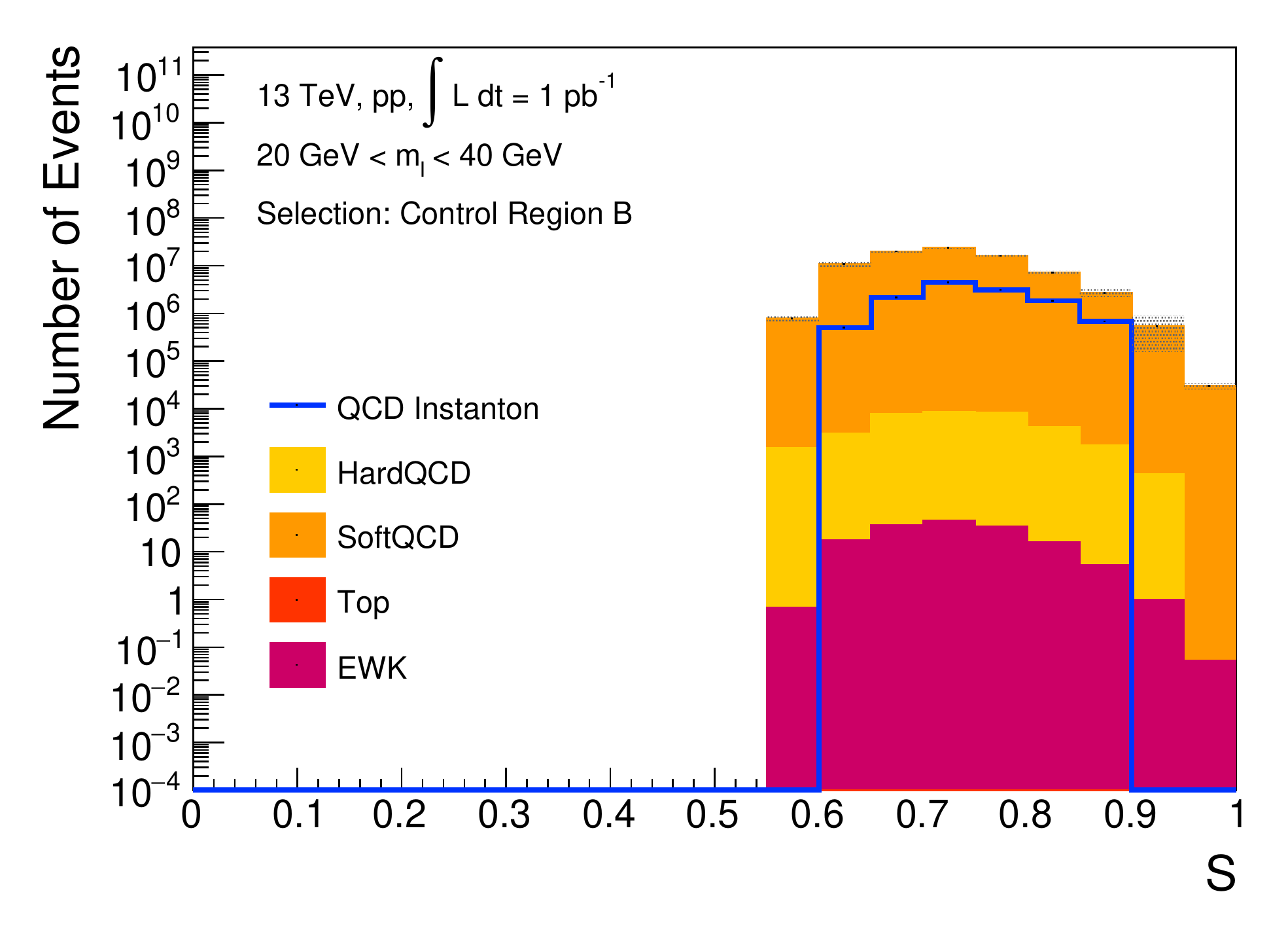} 
\caption{\label{fig:ControlVLIM1} Predicted distributions of the number of displaced tracks in the control region A (left) and the event sphericity in control region B (right), weighted by their predicted cross sections for an integrated luminosity of $L=\int 1\,\inpb$. The invariant mass of all reconstructed tracks is required to be between 20~\GeV{} and 40~\GeV{} (very low mass regime). The distributions from all processes except the Instanton process are stacked. The model uncertainties are indicated as bands.}
\end{center}
\end{figure}

Two possible definitions of control regions, called A and B, are summarized in Table \ref{tab:SelectionVLIM1}. Both exhibit a signal contamination smaller than 10\%. The $\NDisplaced$ distribution for control region A, as well as the sphericity distribution for control region B, are shown in Figure \ref{fig:ControlVLIM1}. 

We also study a higher range of Instanton masses, $40<\mInst<80$~\GeV{}, where we expect the cross section predictions for Instanton processes should be more reliable.
In this regime, the \softQCD background is still the dominant one in most regions of the phase space. Analogously to the previous case, three different signal selection scenarios and two control regions are defined, which is summarized in Table \ref{tab:SelectionVLIM2}. 

\begin{table}[htb]
\footnotesize
\begin{center}
\begin{tabularx}{\textwidth}{l | l | l | l | c | c}
\hline
											&  \multicolumn{3}{c|}{Signal Region}		    & \multicolumn{2}{c}{Control Region}	 \\
\hline
      											& Standard	& Event-		&	Tight	    	& A	& B \\
											& 			& Shape		&		    	& 	&	\\
\hline
Invariant mass of rec. tracks (Instanton Mass), $\mInst$   	& \multicolumn{5}{c}{$40 \GeV < \mInst < 80 \GeV$}					\\
\hline
\multicolumn{5}{c}{Selection Requirements}					\\
\hline
Number of rec. tracks, $\Ntrk$ 					& >30 			& >30   			& >30   			& <20 	& >30\\  
Number of rec. tracks/Instanton mass, $\mInst/\Ntrk$ 	& <2.0 			& <2.0   			& <2.0   			& >2.0 	& <2.0\\  
Number of Jets, $\NJets$                      				& =0 			& =0   			& =0   			& =0 	& =0\\
Broadening, $\BTracks$   					& 	 			& >0.3   			& >0.3   			&  >0.3 	&  >0.3 \\
Thrust, $\TTracks$   						& 	 			& >0.3   			& >0.3   			&  >0.3 	&  >0.3 \\
Number of displaced vertices, $\NDisplaced$   		& 	 			& 	   			& >12  			& 	 	& <8 \\ 
\hline
\multicolumn{5}{c}{Expected Events for $\int L dt = 1\,\inpb$ in the Signal Region ($\cal S$ >0.85)}					\\
\hline
$N_{Signal}$ 									& $4.8\cdot 10^5$	& $3.7\cdot 10^5$	& 52000			& 0	 	&	$1.3\cdot 10^5$ \\
$N_{Background}$								& $1.2\cdot 10^6$	& $8.7\cdot 10^5$	& 432			& 3000	 &  	$7.5\cdot 10^5$ \\
\hline
\end{tabularx}
\caption{Overview of the standard and tight signal selection as well as the definition of two control regions aiming at very low Instanton masses ($40 \GeV<\mInst<80 \GeV$)\label{tab:SelectionVLIM2}}
\end{center}
\end{table}

\begin{figure}[htb]
\begin{center}
\includegraphics[width=7.3cm]{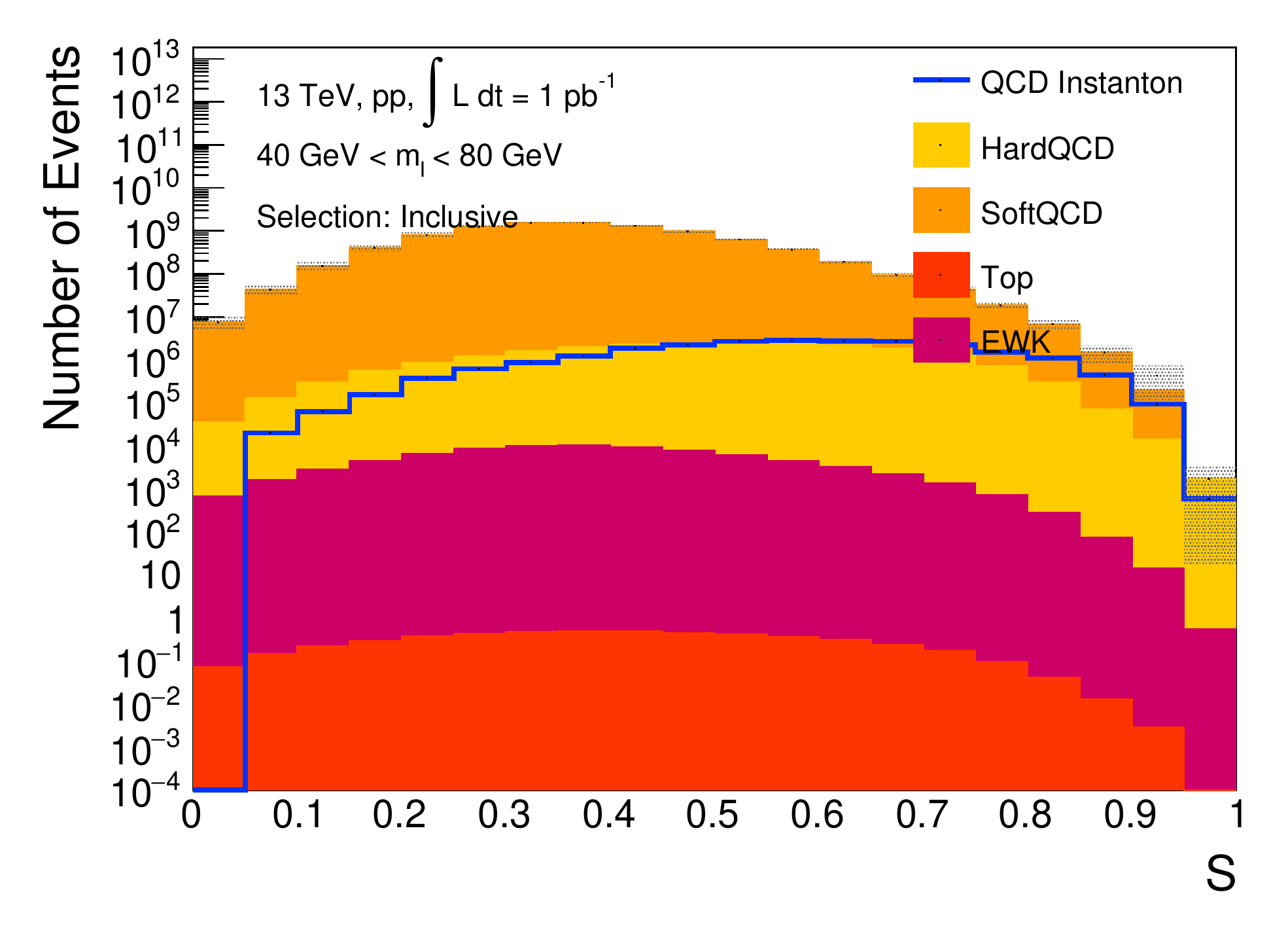} 
\hspace{0.1cm}
\includegraphics[width=7.3cm]{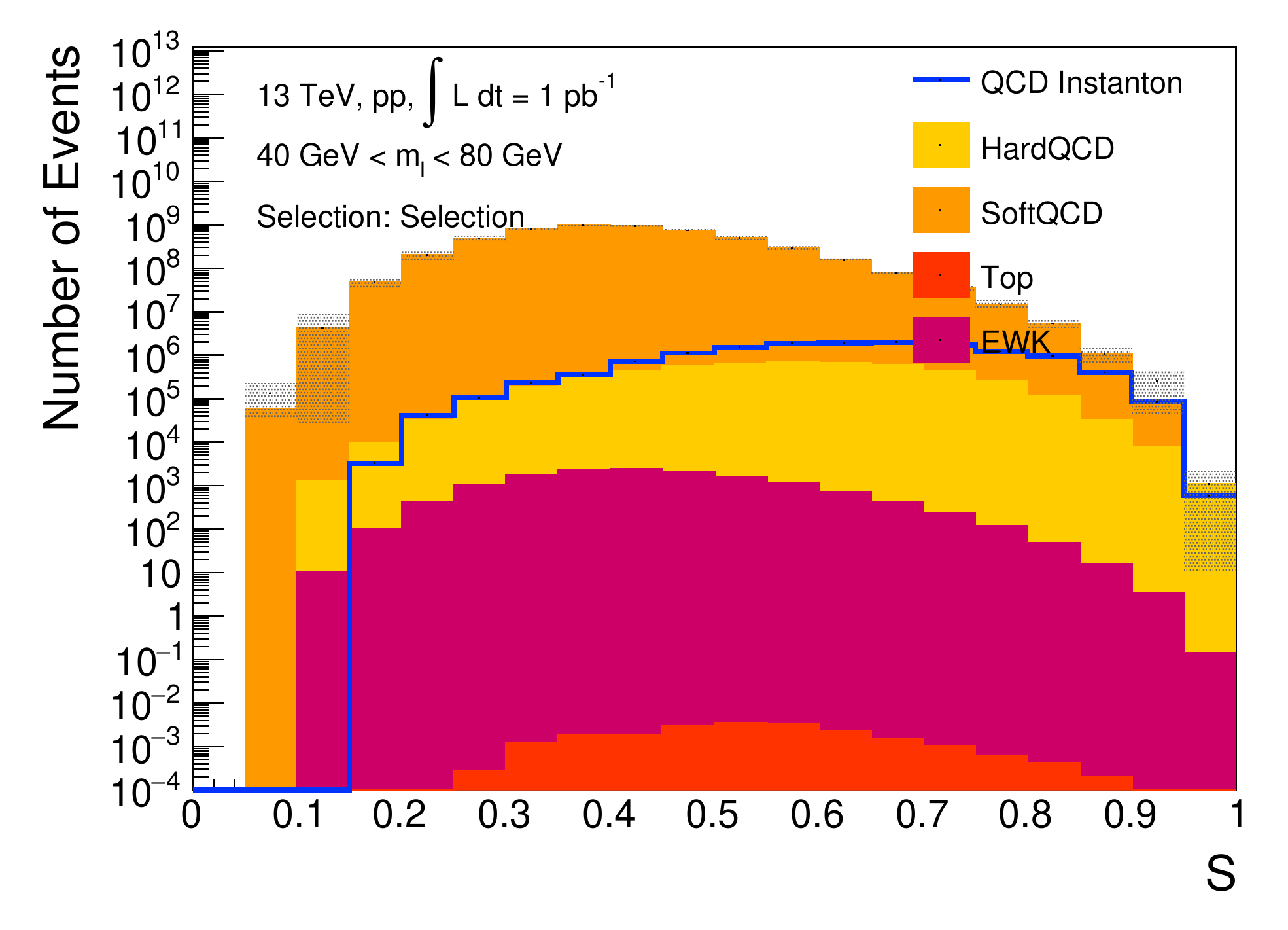}
\includegraphics[width=7.3cm]{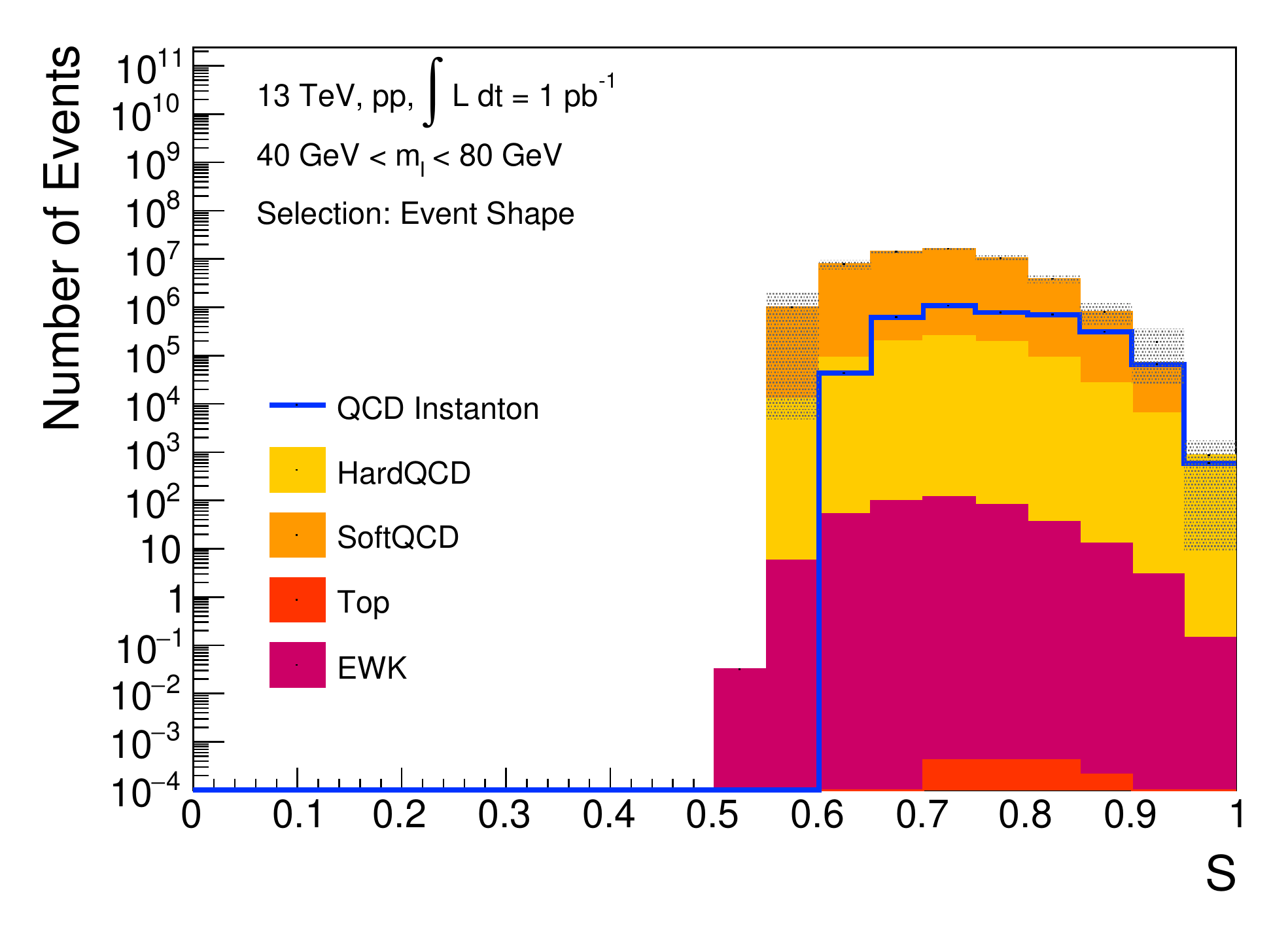} 
\hspace{0.1cm}
\includegraphics[width=7.3cm]{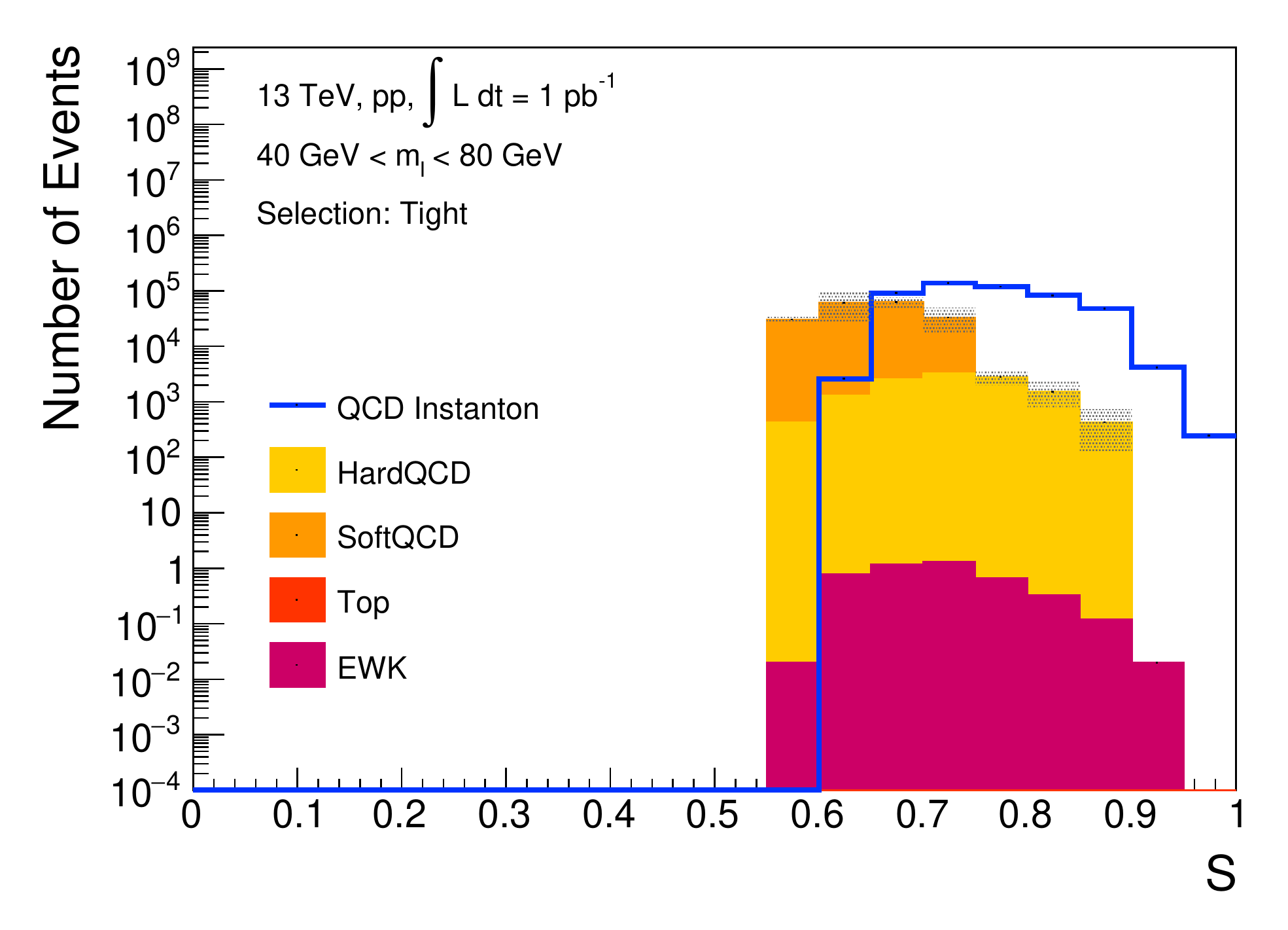}

\caption{\label{fig:SelectionVLIM2} Predicted distributions of the event sphericity for various processes, weighted by their predicted cross sections for an integrated luminosity of $L=\int 1\,\inpb$ for an inclusive selection (upper left), the nominal selection (upper right), the \textit{event-shape} based selection (lower left) and the tight selection (lower right). The invariant mass of all reconstructed tracks is required to be between 40~\GeV{} and 80~\GeV. The distributions from all processes except the Instanton process are stacked. The model uncertainties are indicated as bands.}
\end{center}
\end{figure}

The sphericity distribution for the signal and background processes for an inclusive selection\footnote{inclusive is defined here as only applying selection cuts on the reconstructed invariant mass \mInst of the Instanton} in this mass range, as well as the three signal region definitions (\textit{standard}, \textit{event-shape}, \textit{tight}) are shown in Figure \ref{fig:SelectionVLIM2}, together with the expected signal and background events in the signal region in Table \ref{tab:SelectionVLIM2}.  The \textit{standard} and \textit{event-shape} selections yield signal to background ratios below 1 and are dominated by \softQCD processes. The \textit{tight} selection significantly enhances the signal and would allow for a clear observation. In this selection the dominant background contribution in the signal region comes from \hardQCD processes. We note however that the negligible contribution from \softQCD could just be a consequence of  the limited  statistics employed in this study, and it is possible that the actual background from \softQCD processes is larger.The interplay between \softQCD and \hardQCD processes could be experimentally studied, by applying various requirements on the number of reconstructed jets and thus define different control regions. 

\clearpage
\newpage

\subsection{Medium Instanton Masses: The Hard QCD Regime}

\begin{figure}[htb]
\begin{center}
\includegraphics[width=7.3cm]{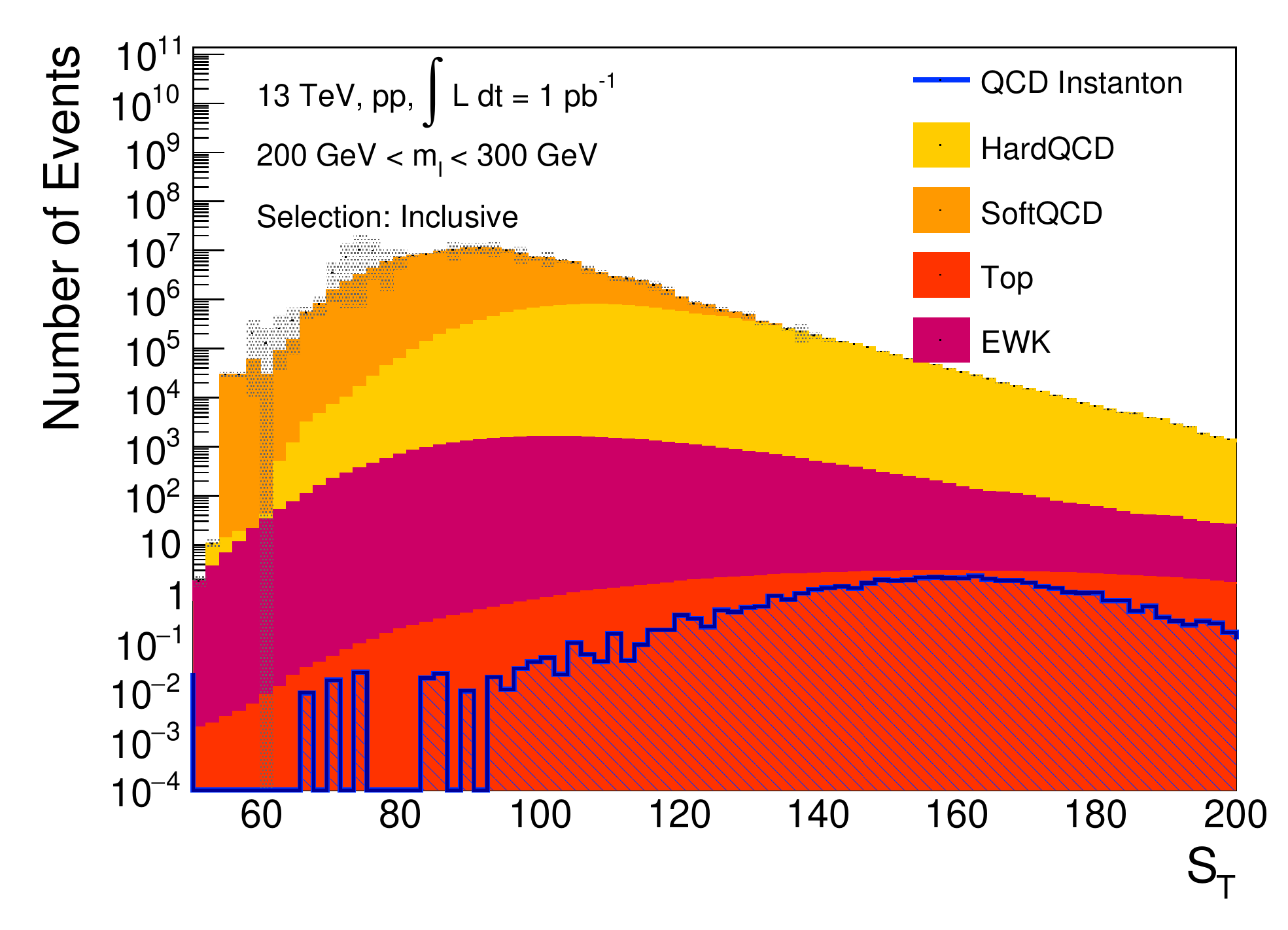} 
\hspace{0.1cm}
\includegraphics[width=7.3cm]{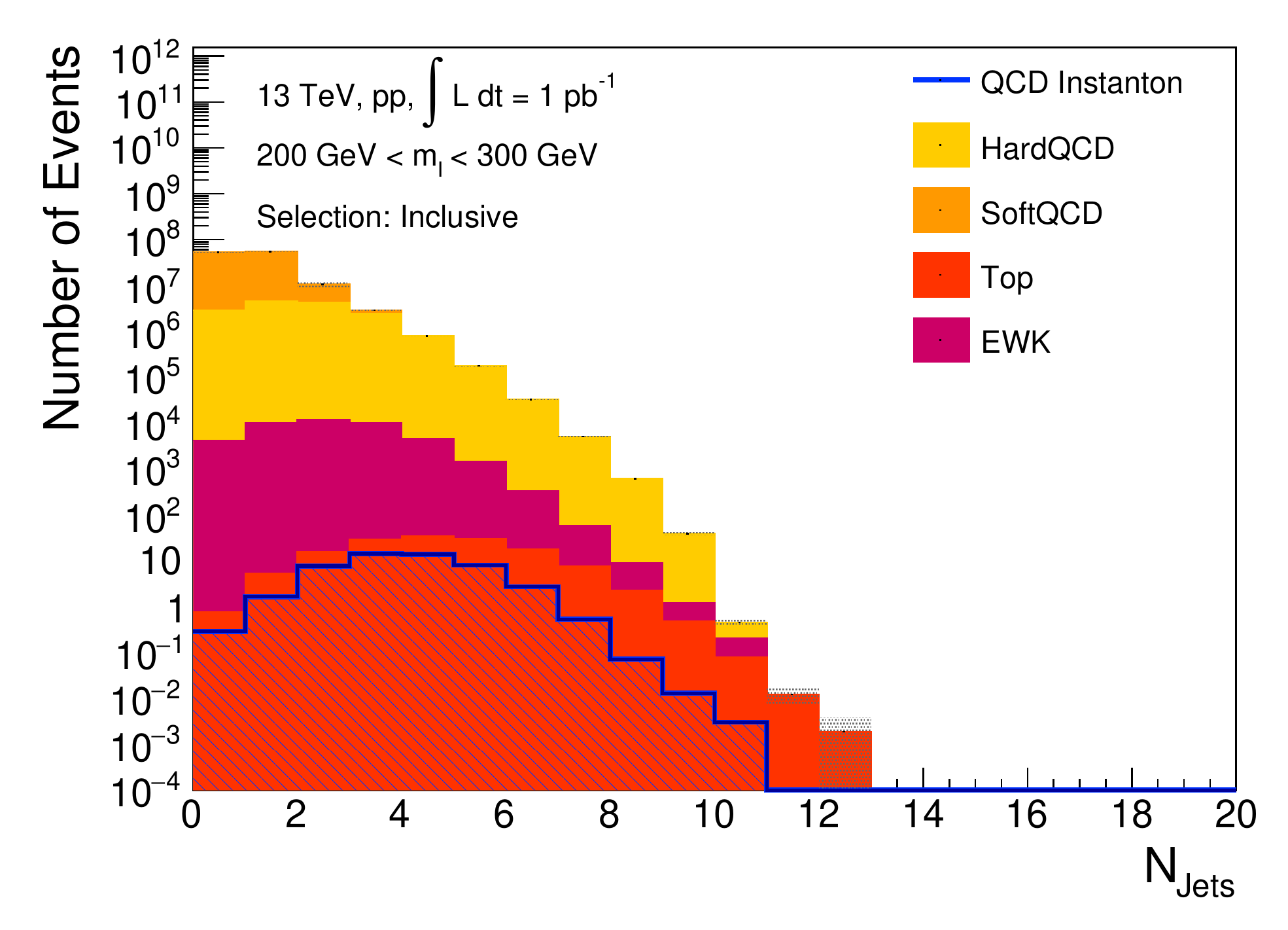}
\caption{\label{fig:OverviewLIM} Predicted distributions of the ST (left) and number of reconstructed jets (right) for various processes, weighted by their predicted cross sections for an integrated luminosity of $L=\int 1\,\inpb$. The invariant mass of all reconstructed tracks is required to be between 200~\GeV{} and 300~\GeV{} (medium mass regime). The distributions from all processes except the Instanton process are stacked. The model uncertainties are indicated as bands.}
\end{center}
\end{figure}

As one explores  higher Instanton invariant masses, one enters the regime of perturbative QCD, and the background prediction becomes less uncertain. For this, we studied medium mass range of $200<\mInst<300$~\GeV{} with an average Instanton mass of 220~\GeV{} was studied. The $S_T^{tracks}$ distribution and the number of reconstructed jets with $\pT>20$~\GeV{} in this mass range is shown in Figure \ref{fig:OverviewLIM} for the signal and background processes. 
The Instanton processes are expected to peak for $3\leq N_{jet} \leq 6$, and one can see events with $S_T>150$~\GeV~have only a negligible contribution form \softQCD processes.

A \textit{standard signal selection} imposes requirements on $\Ntrk$, $\mInst/\Ntrk$ and the $\NJets$ distribution.  The \textit{event-shape} selection applies in addition a minimum requirement on $\cal{B}$ and $\cal{T}$. Similarly to the low mass scenarios, a requirement on $\NDisplaced$ is made for the \textit{tight signal} selection. All cuts for the signal selection as well as the definitions of the two control regions are summarized in Table \ref{tab:SelectionVMedium}. The  \textit{event-shape} selection yields 1 signal event and approximately 10 background events for an integrated luminosity of $\int L dt = 1\,\inpb$. For the \textit{tight  selection} 0.5 signal events and 0.6 background events are expected.  An observation would therefore only be possible  with an integrated luminosity of about $\int L dt \sim 10\,\inpb$. 

In this context the ability of the LHC experiments to trigger on these event topologies becomes relevant. For the studies of \softQCD processes at the LHC, special triggers are used, which record collision events even with limited activity in the detector, i.e. are nearly free of any bias towards a certain physics signature. Given the enormous rates of such \textit{minimum bias triggers}, only a small fraction of these events can be actually stored on tape. However, the published \softQCD analysis at the LHC indicate that sufficient statistics has been already collected to allow for Instanton searches in the low mass regime. In the medium and high Instanton mass regime the available statistics is a larger challenge, as the required integrated luminosity increases significantly. Typically, multi-jet triggers require minimal transverse jet energies of 50~\GeV{} or more and hence are of no use. It might be therefore important to develop new trigger strategies for the upcoming LHC runs to able to record sufficient statistics, as it was already pointed out previously in \cite{Khoze:2020tpp}.

\begin{table}[tb] 
\footnotesize
\begin{center}
\begin{tabularx}{\textwidth}{l | l | l | l | c | c}
\hline
											&  \multicolumn{3}{c|}{Signal Region}		    & \multicolumn{2}{c}{Control Region}	 \\
\hline
      											& Standard	& Event-		&	Tight	    	& A	& B \\
											& 			& Shape		&		    	& 	&	\\
\hline
Invariant mass of rec. tracks (Instanton Mass), $\mInst$   	& \multicolumn{4}{c}{$200 \GeV < \mInst < 300 \GeV$}					\\
\hline
\multicolumn{5}{c}{Selection Requirements}					\\
\hline
Number of rec. tracks, $\Ntrk$ 					& >80 			& >80   			& >80   			& >80	& >80 \\  
Number of rec. tracks/Instanton mass, $\mInst/\Ntrk$ 	& <3.0 			& <3.0   			& <3.0   			& >3.0 	& <3.0\\  
Number of Jets, $\NJets$                      				& 3-6 			& 3-6   			& 3-6   			& 3-6 	& 3-6\\
Broadening, $\BTracks$   					& 	 			& >0.3   			& >0.3   			&  >0.3	 &  >0.3 \\
Thrust, $\TTracks$   						& 	 			& >0.3   			& >0.3   			&  >0.3	&  >0.3 \\
Number of displaced vertices, $\NDisplaced$   		& 	 			& 	   			& >15  			& 	 	& <10 \\ 
\hline
\multicolumn{5}{c}{Results}					\\
\hline
\multicolumn{5}{c}{Expected Events for $\int L dt = 1\,\inpb$ in the Signal Region ($\cal S$ >0.85)}					\\
\hline
$N_{Signal}$ 									& 5.6				& 1.0		& 0.54		& 0.04	 	&	0.21	 \\
$N_{Background}$								& 1900			& 9.6		& 0.64		& 200	 	&  	1100 \\
\hline
\end{tabularx}
\caption{Overview of the standard and tight signal selection as well as the definition of two control regions aiming at very low Instanton masses ($200 \GeV<\mInst<300 \GeV$)\label{tab:SelectionVMedium}}
\end{center}
\end{table}

\begin{figure}[hbt]
\begin{center}
\includegraphics[width=7.2cm]{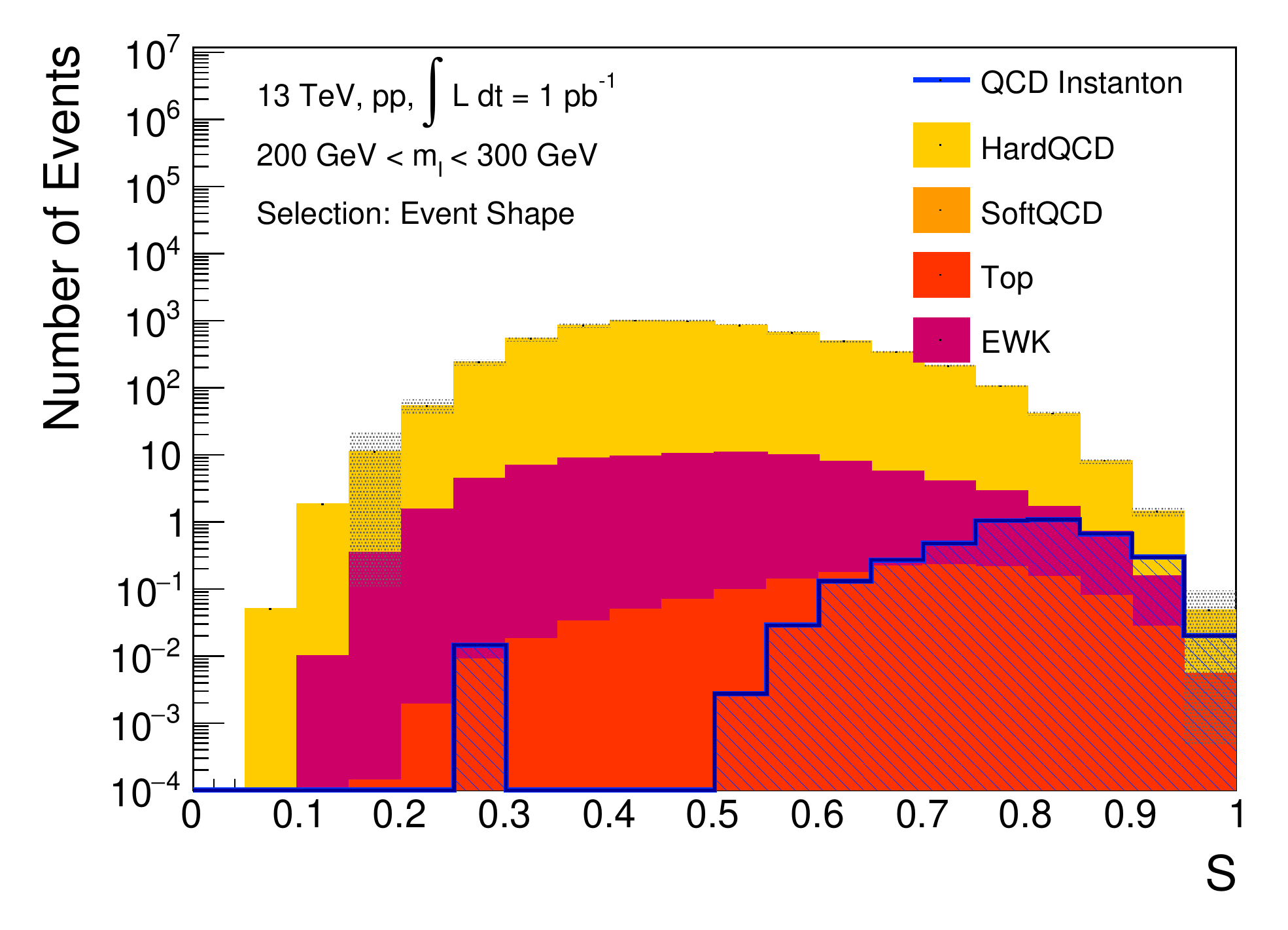}
\hspace{0.1cm}
\includegraphics[width=7.2cm]{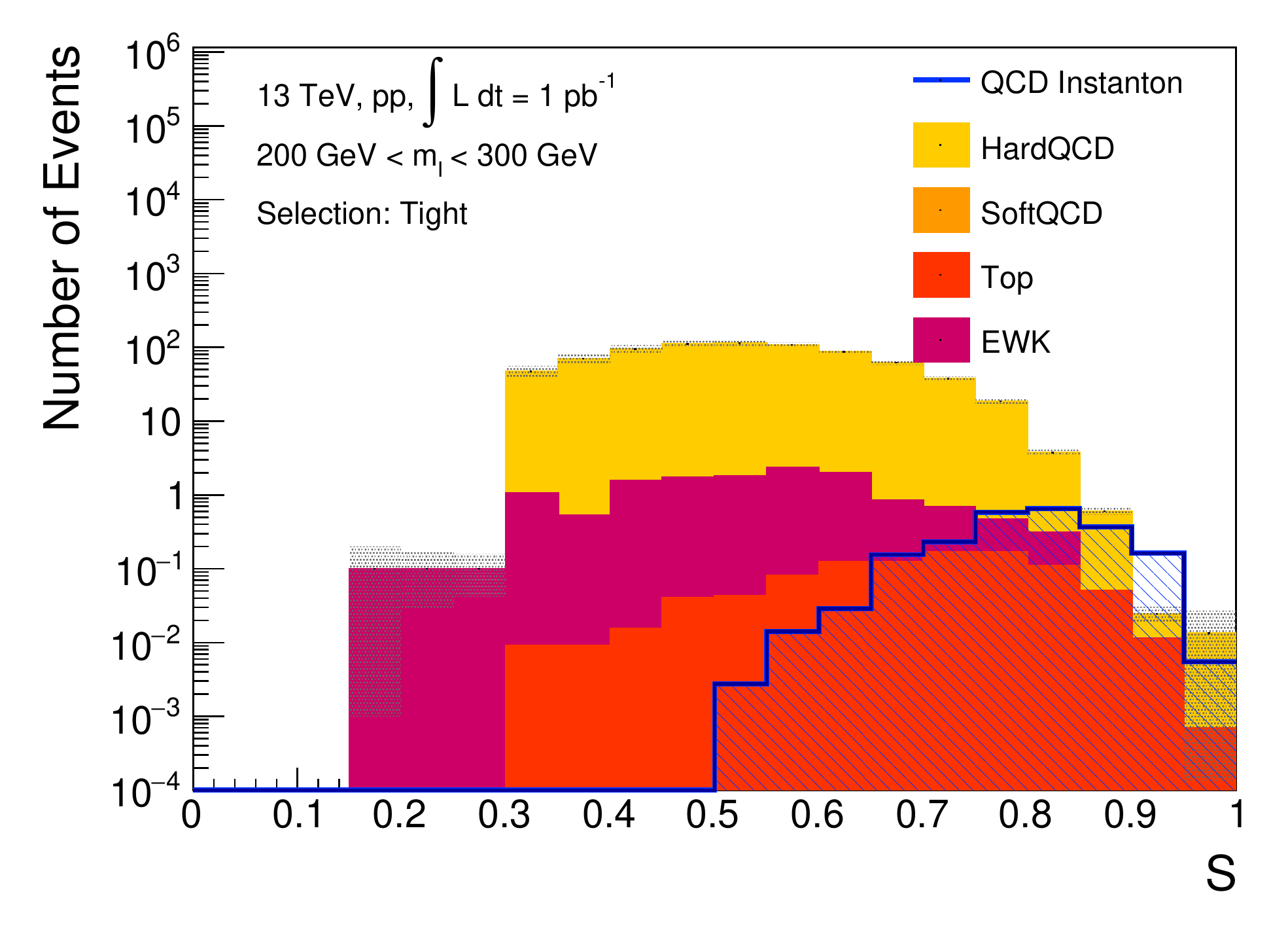}
\caption{\label{fig:SelectionVLIMSpher} Predicted distributions of $\cal{S}$ for an integrated luminosity of $L=\int 1\,\inpb$ for 
the \textit{event-shape} based selection (left) and the \textit{tight} selection (right). The invariant mass of all reconstructed tracks is required to be between 200~\GeV{} and 300~\GeV{} (medium mass regime)}
\end{center}
\end{figure}

\begin{figure}[tbh]
\begin{center}
\includegraphics[width=7.3cm]{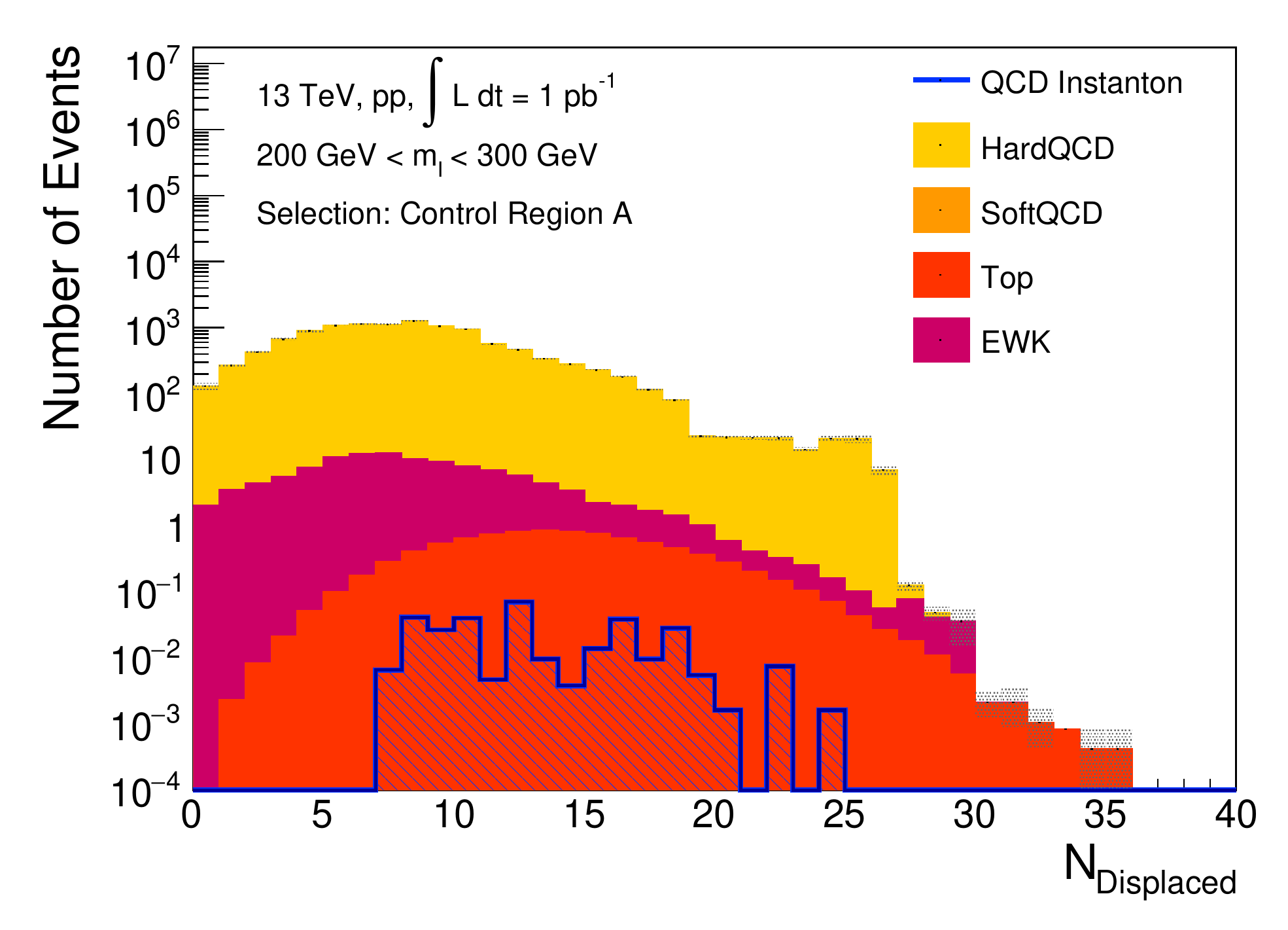}
\hspace{0.1cm}
\includegraphics[width=7.3cm]{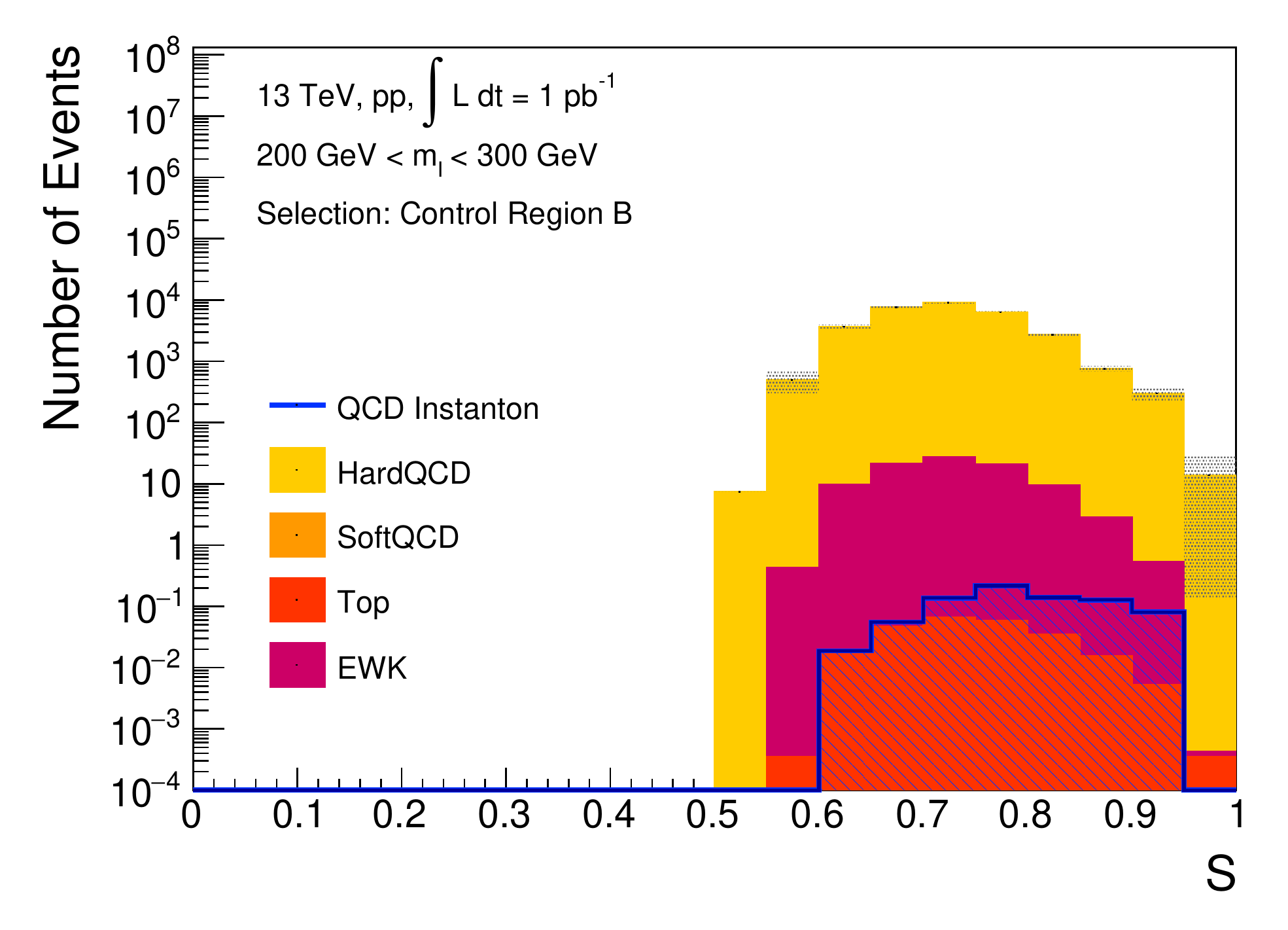} 
\caption{\label{fig:ControlVLIM} Predicted distributions of the number of displaced tracks in the control region A (left) and the event sphericity in control region B (right), weighted by their predicted cross sections for an integrated luminosity of $L=\int 1\,\inpb$. The invariant mass of all reconstructed tracks is required to be between 200~\GeV{} and 300~\GeV{} (medium mass regime). The distributions from all processes except the Instanton process are stacked. The model uncertainties are indicated as bands.}
\end{center}
\end{figure}

The predicted distributions of the event sphericity for the signal and background processes, scaled to the expected event yields for an integrated luminosity of $L=\int 1\,\inpb$ is shown in Figure \ref{fig:SelectionVLIMSpher} for the \textit{event-shape} and the \textit{tight} selection. The dominant background are multijet events from \hardQCD processes. Figure \ref{fig:ControlVLIM} shows the predicted number of displaced tracks in the control region A as well as the event sphericity in control region B, which again can be used to validate the modelling of the background processes.

\subsection{High Instanton Masses: The Top Quark Regime}

The \hardQCD multi-jet background is also dominant in the $300<\mInst<500$~\GeV{} mass ranges. This can be seen in Figure \ref{fig:OverviewLIM}, which shows the event sphericity  and the number of reconstructed particle jets with $\pT>20$~\GeV for signal and background events. However, it might be more interesting to apply a dedicated event selection that promotes another process as dominant background, which does not suffer from the same model uncertainties as the background in the medium mass regime. Hence we focus our selection here on top-quark pair events. While it is obvious to enhance the top-quark (as well as the Instanton) contribution by requiring reconstructed jets which are tagged to stem from $b$-hadrons, this was not required within this study. Our results on the expected signal over background ratio are therefore conservative. 

\begin{figure}[htb]
\begin{center}
\includegraphics[width=7.3cm]{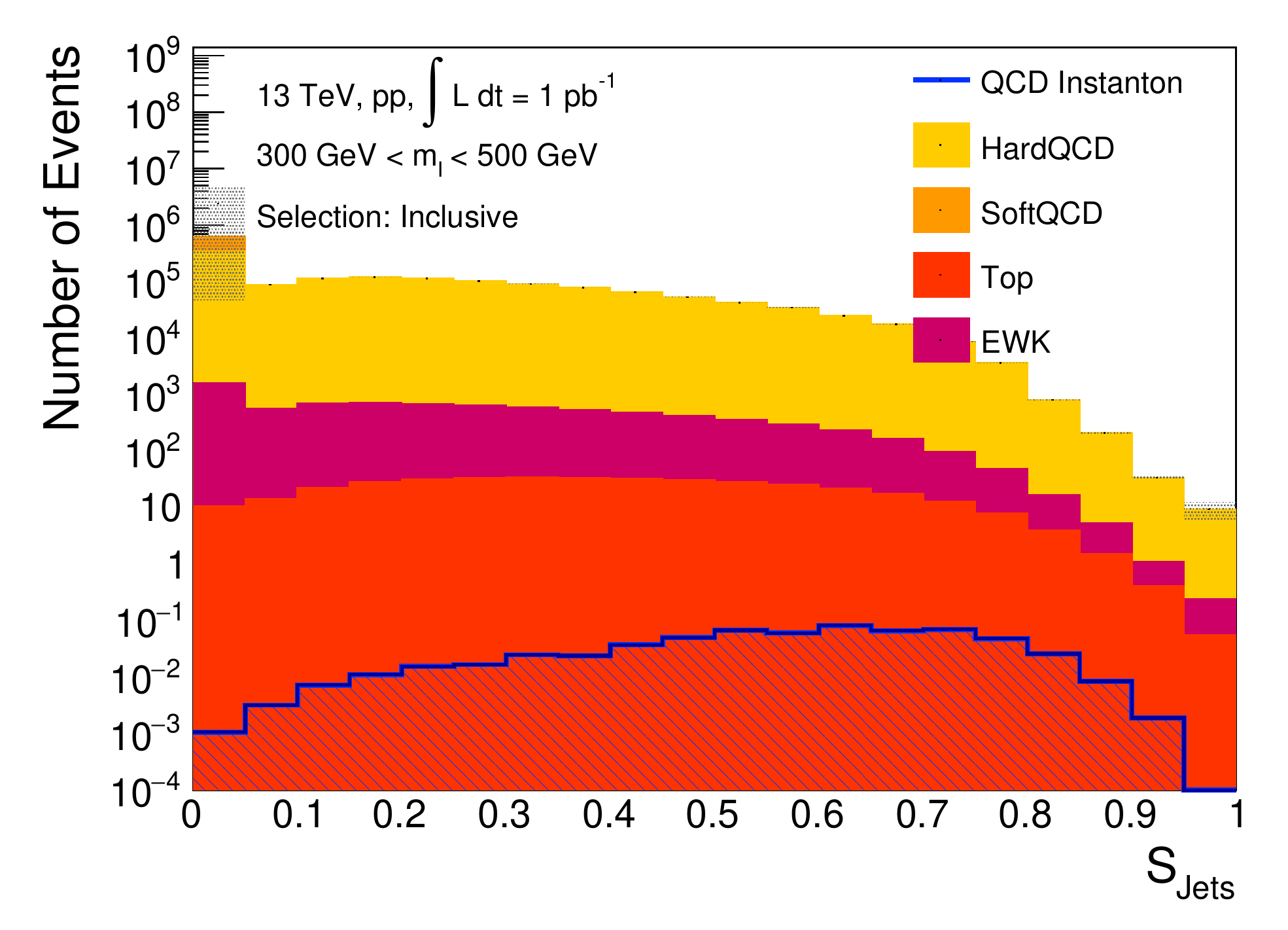} 
\hspace{0.1cm}
\includegraphics[width=7.3cm]{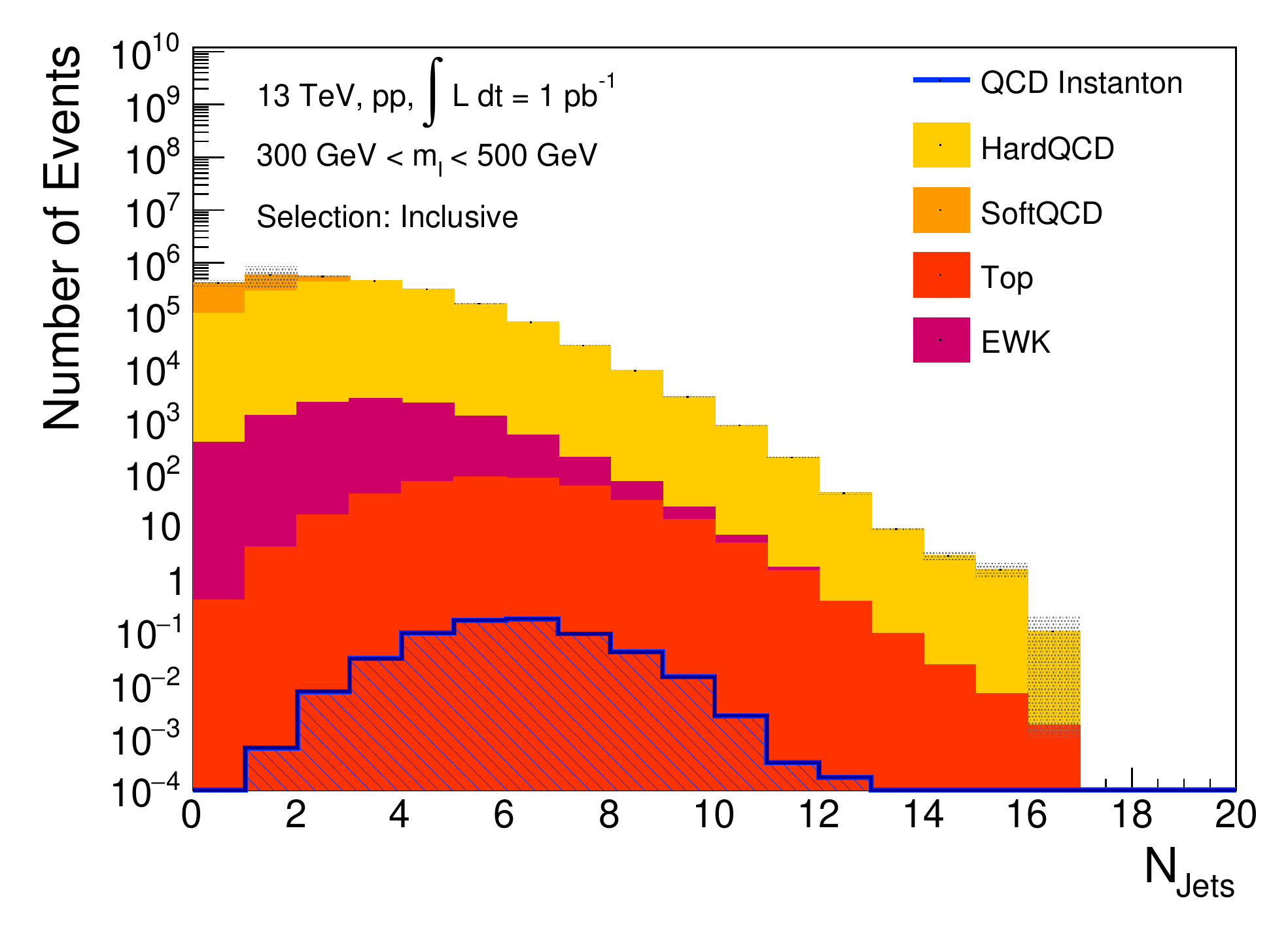}
\caption{\label{fig:OverviewLIM1} Predicted distributions of the event sphericity (left) and number of reconstructed jets (right) for various processes, weighted by their predicted cross sections for an integrated luminosity of $L=\int 1\,\inpb$. The invariant mass of all reconstructed tracks is required to be between 300~\GeV{} and 500~\GeV{} (high mass regime). The distributions from all processes except the Instanton process are stacked. The model uncertainties are indicated as bands.}
\end{center}
\end{figure}

\begin{figure}[htb]
\begin{center}
\includegraphics[width=7.3cm]{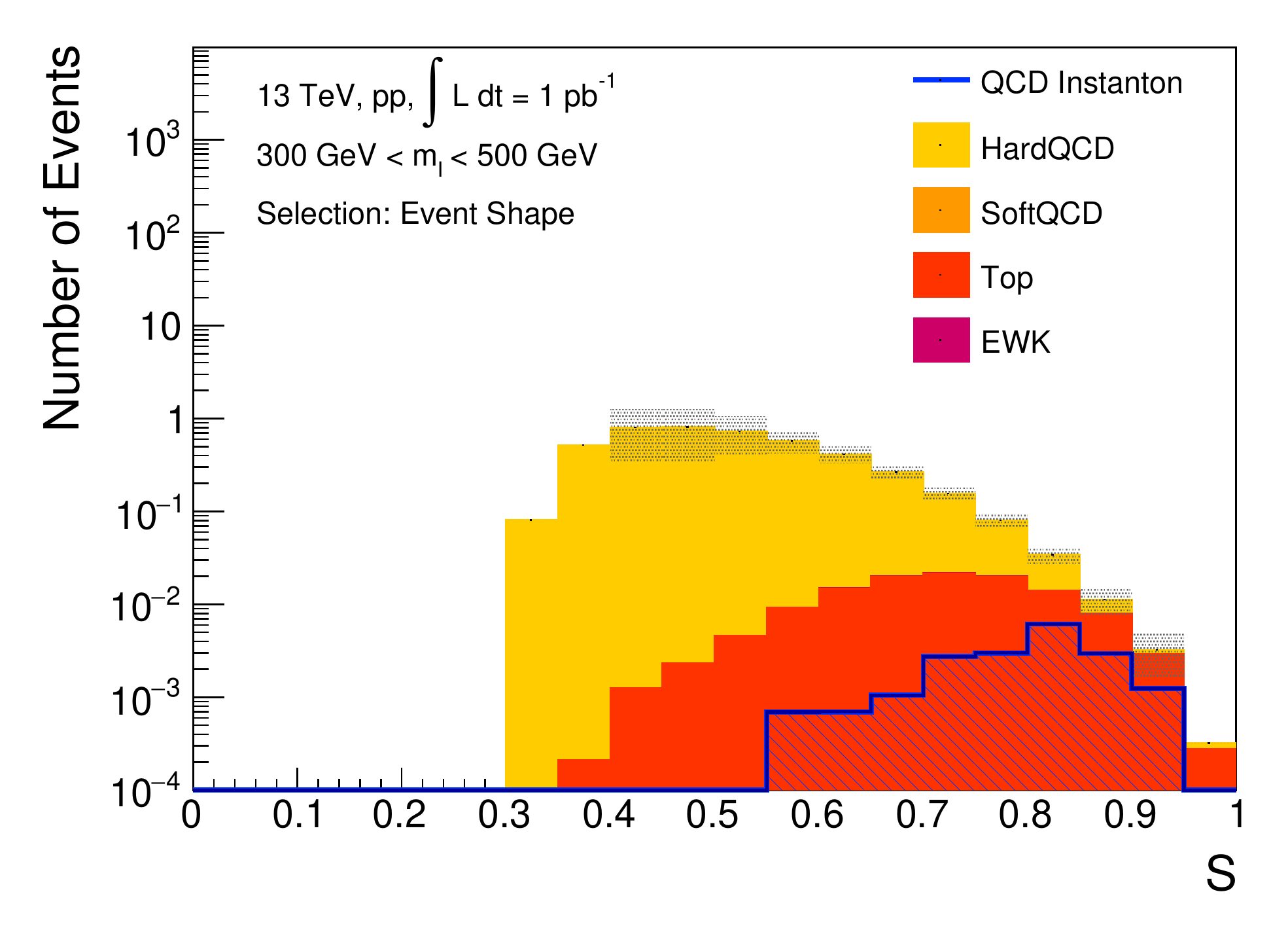}
\hspace{0.1cm}
\includegraphics[width=7.3cm]{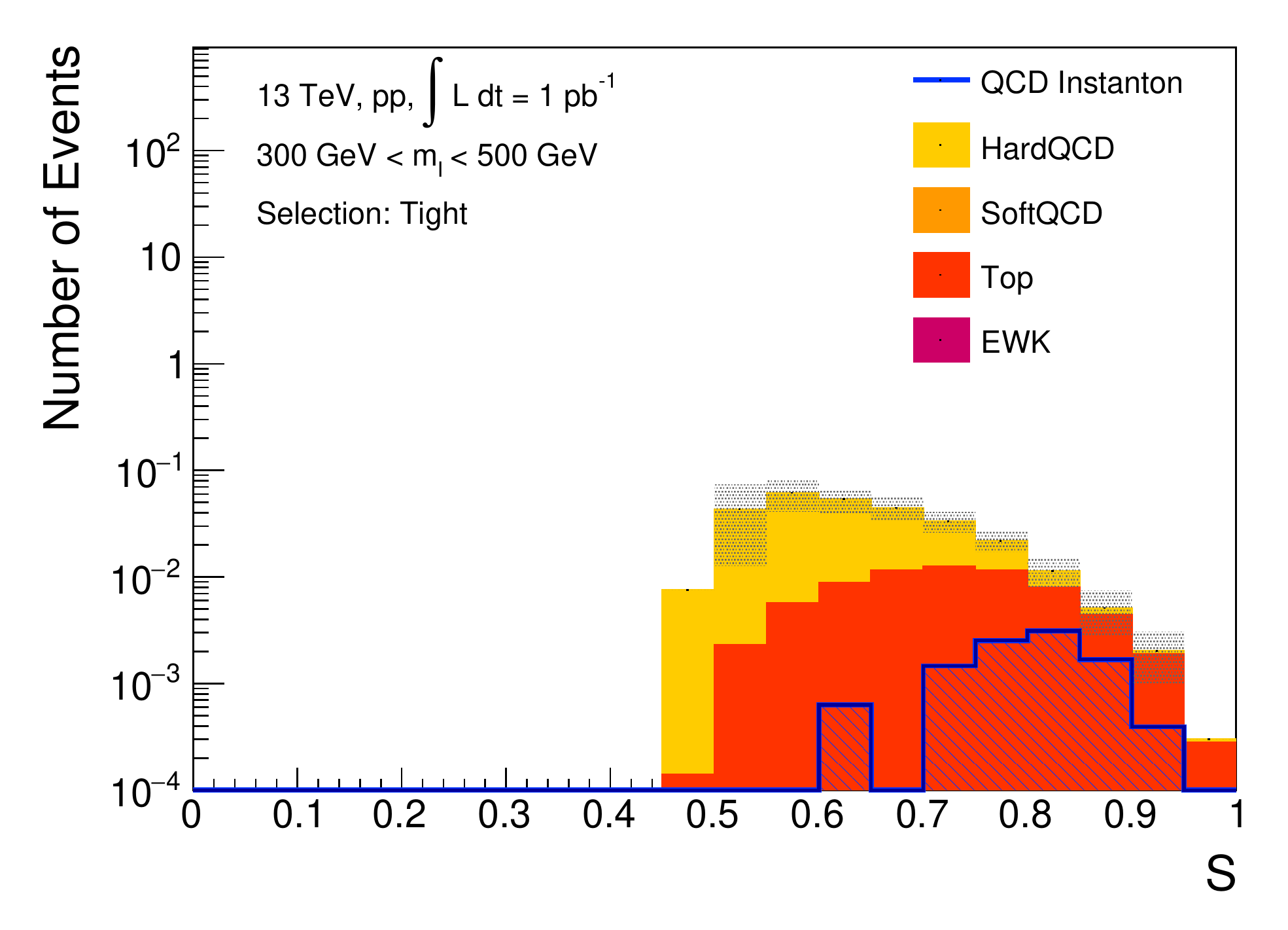}
\caption{\label{fig:SelectionVLhigh} Predicted distributions of the event sphericity for various processes, weighted by their predicted cross sections for an integrated luminosity of $L=\int 1$pb$^{-1}$ for 
the \textit{event-shape} based selection (left) and the \textit{tight} selection (right). The invariant mass of all reconstructed tracks is required to be between 300~\GeV{} and 500~\GeV{} (high mass regime). The distributions from all processes except the Instanton process are stacked. The model uncertainties are indicated as bands.}
\end{center}
\end{figure}

The large track multiplicity in  events within $300<\mInst<500$~\GeV{} does not allow anymore for a clean separation between signal and background processes. Hence no signal selection cut involving $\Ntrk$ is applied. The \textit{standard} signal selection requires therefore only $\mInst/\Ntrk<3.0$ as well as more than seven reconstructed jets with $\pT>20$~\GeV. While the latter requirement is not optimal for the enhancement of the overall signal over background ratio, it allows to enhance the top-quark background contribution. The \textit{event-shape} and \textit{tight} signal selections follow the same lines as for the lower mass ranges, i.e. impose cuts on the event topology as well as on $\NDisplaced$. A summary of all signal selection criteria is given in Table \ref{tab:SelectionHigh}. The corresponding event sphericity distributions for the \textit{event-shape} and the \textit{tight} selection is shown in Figure \ref{fig:SelectionVLhigh}, where the top-quark background starts to dominate large values of $\cal{S}$. A further enhancement of the top-quark contribution can be achieved by an additional $b$-tagging requirement without impacting significantly the signal yield. 

\begin{table}[htb]
\footnotesize
\begin{center}
\begin{tabularx}{\textwidth}{l | l | l | l | c | c| c}
\hline
											&  \multicolumn{3}{c|}{Signal Region}		    & \multicolumn{3}{c}{Control Region}	 \\
\hline
      											& Standard	& Event-		&	Tight	    	& A	& B & C\\
											& 			& Shape		&		    	& 	&	& \\
\hline
Invariant mass of rec. tracks $\mInst$   	& \multicolumn{4}{c}{$300 \GeV < \mInst < 500 \GeV$}					\\
\hline
\multicolumn{5}{c}{Selection Requirements}					\\
\hline
Number of rec. tracks, $\Ntrk$ 					& -				& -	   			& -	   			& 		& 		&\\  
Number of rec. tracks/inst. mass, $\mInst/\Ntrk$ 		& <3.0 			& <3.0   			& <3.0   			& >3.0 	& <3.0	&	<3.0\\  
Number of Jets, $\NJets$                      				& >7 			& >7   			& >7   			& >7 	&>7		&>5\\
Broadening, $\BTracks$   					& 	 			& >0.3   			& >0.3   			&  >0.3	& >0.3	&>0.3\\
Thrust, $\TTracks$   						& 	 			& >0.3   			& >0.3   			&  >0.3	& >0.3	&>0.3\\
Number of displaced vertices, $\NDisplaced$   		& 	 			& 	   			& >20  			& 	 	&	<15	&>20\\ 
Identified Leptons						   		& -	 			& -	   			& -  				& - 	 	& -		&1\\ 
\hline
\multicolumn{5}{c}{Expected Events for $\int L dt = 1\,\inpb$ in the Signal Region ($\cal S$ >0.85)}					\\
\hline
$N_{Signal}$ 									& 0.007			& 0.004			& 0.0021			&	0.002	& 0.0003	& -\\
$N_{Background}$								& 0.204			& 0.015			& 0.0074			&	33	 	& 0.0022	& 0.002\\
\hline
\end{tabularx}
\caption{Overview of the standard and tight signal selection as well as the definition of two control regions aiming at very low Instanton masses ($300 \GeV<\mInst<500 \GeV$)\label{tab:SelectionHigh}}
\end{center}
\end{table}

The advantage of this signal selection relies on the experimentally well understood top-quark pair production. In addition to similar control regions as in the low mass ranges, also a top-quark specific selection can be envisioned, e.g. by requiring one additional reconstructed lepton (electron or muon) in the event, which stems from the leptonic decay of one top-quark (see definition of Control Region C in Table \ref{tab:SelectionHigh}). This ensures large experimental constrains on the background uncertainties in the signal region. However, the signal over background ratio is only in the order of 30\% and only 2 signal events are expected for an integrated luminosity of $\int L dt = 1$\,fb$^{-1}$, implying an observation with a $5\sigma$ significance would require an integrated luminosity of more than 80\,fb$^{-1}$ based on pure statistical considerations. Such large luminosities might require either dedicated developed triggers for high multiplicity jet events during the Run-3 of the LHC, or a long data-taking period with pre-scaled jet triggers during the high luminosity phase of the LHC.

\section{Limits on Instanton Processes in Proton-Proton Collisions\label{Sec:Limit}}

\subsection{How to mimic QCD Instanton Signatures}

As shown in the previous sections, the most promising mass range for the observation of Instanton induced processes at the LHC is below 100~\GeV, where the \softQCD background contribution dominates. It is therefore crucial to understand if the \softQCD phenomenological models have enough freedom to mimic the QCD Instanton signatures. A first indication that this might not be easily achieved comes from the observation that the \softQCD predictions from the Pythia, Sherpa and the Herwig7 generators, which implement different models, are remarkably consistent for the observables considered in this study. As the \softQCD  model in event generators contain many parameters  optimised using data, it remains remains however possible that with a suitable parameter choice the \softQCD predictions can be made more similar to the Instanton.

\begin{figure}[th]
\centering
\begin{minipage}{7.1cm}
\centering
\includegraphics[width=1.0\linewidth]{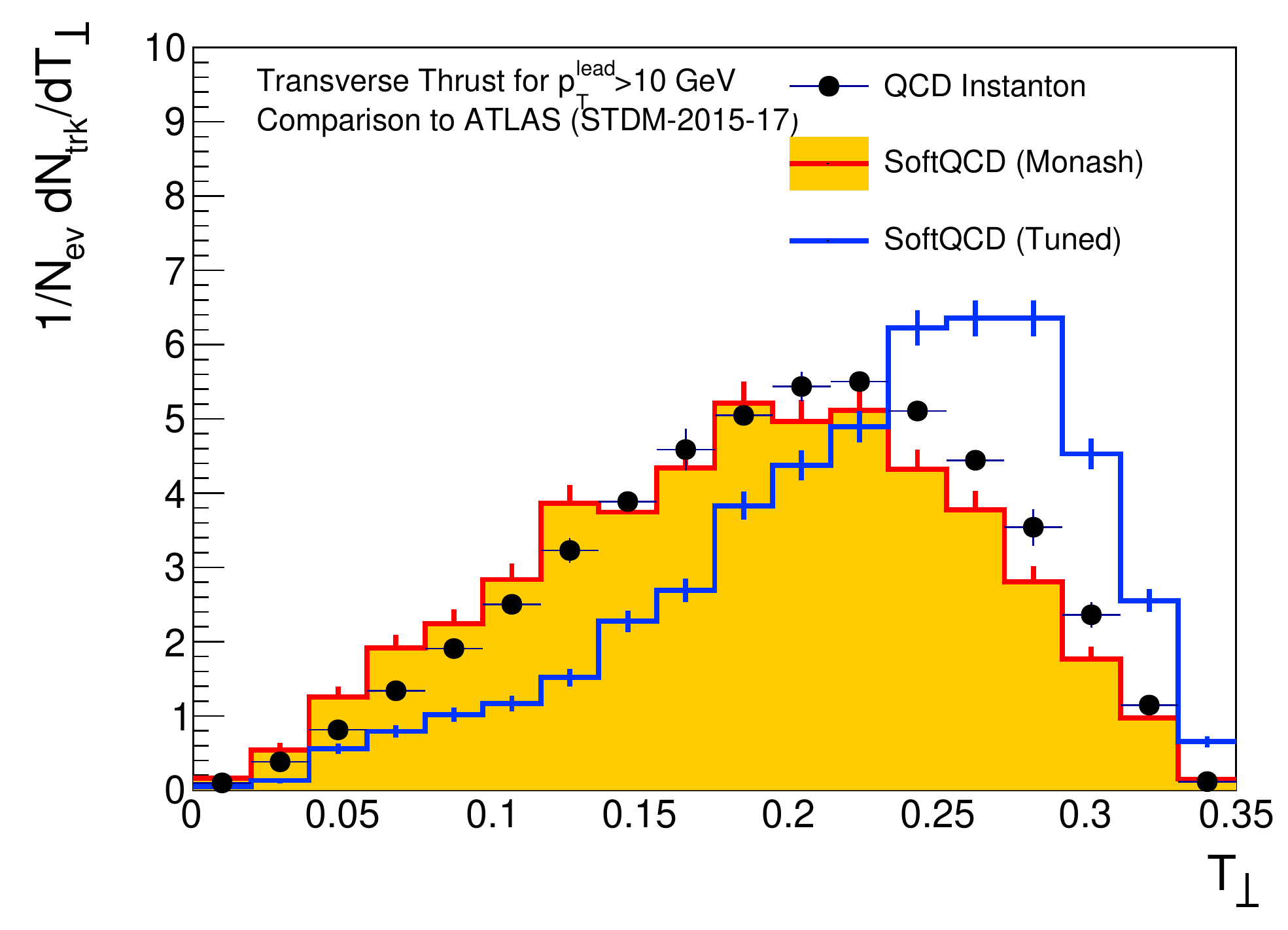}
\caption{Predicted distribution of the event thrust of the \textsc{Monash} \softQCD tune of \textsc{Pythia8} as well as a modified version with significantly enhanced multiple parton interaction probability (\textsc{MPI:alphaSvalue = 0.150} ) in comparison to the measurement at 13~\TeV{} of the ATLAS Collaboration \cite{Aaboud:2016itf} \label{fig:Tune1}\vspace{0.5cm}}
\end{minipage}
\hspace{0.3cm}
\begin{minipage}{7.1cm}
  \centering
  \includegraphics[width=1.0\linewidth]{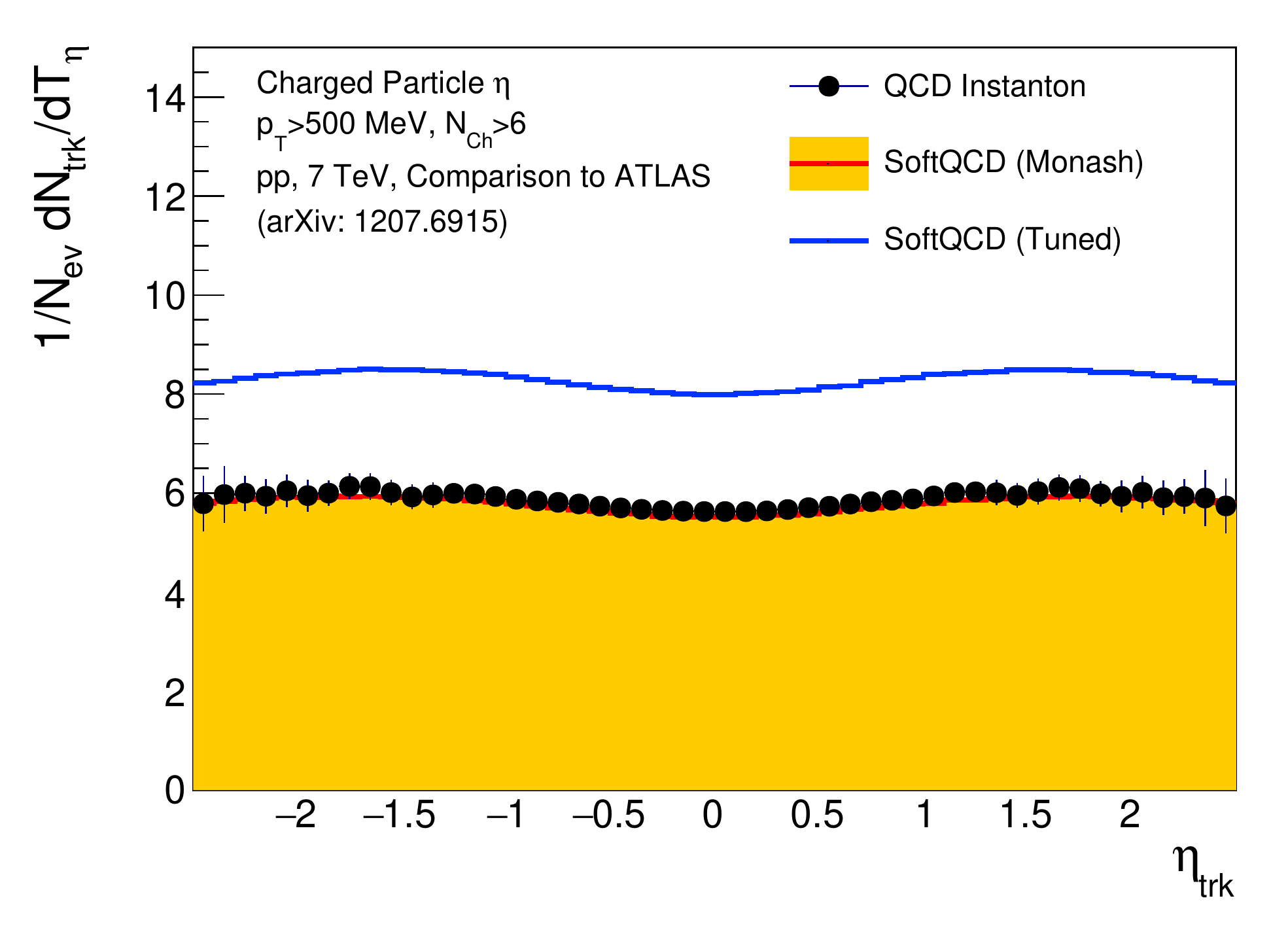}
\caption{Predicted distribution of the charged particle spectrum vs. $\eta$ of the \textsc{Monash} \softQCD tune of \textsc{Pythia8} as well as a modified version with significantly enhanced multiple parton interaction probability (\textsc{MPI:alphaSvalue = 0.150} ) in comparison to the measurement at 7~\TeV{} of the ATLAS Collaboration \cite{Aad:2012fza} \label{fig:Tune2}}
\end{minipage}%
\end{figure}

In this context it is interesting to note that an existing ATLAS measurement of charged particle event shapes~\cite{Aad:2012fza} in Minimum Bias events did show light discrepancy between the data and different generator predictions, indicating data events were slightly more spherical than expected. As a proof-of-principle demonstration, we have tested if the \softQCD Pythia predictions can be made to yield significantly more spherical events, even beyond what data indicates. Starting with the baseline \textsc{Monash} tune~\cite{Skands:2014pea} of \textsc{Pythia8}, we found that increasing the   \textsc{MultipartonInteractions:alphaSvalue = 0.150} in \textit{softQCD} events does produce more spherical events, as seen in Figure~\ref{fig:Tune1}. 

However, such tunes would also alter many other event shape distributions, such as the number of charged particles vs. $\eta$, which are not supported by data as seen for example in Figure \ref{fig:Tune2}). While the latter is based on $\sqrt{s}=7$~\TeV{} data, the same conclusions hold for $\sqrt{s}=13$~\TeV{} data as well. Even more importantly, a tune of multiple parton interactions, would not impact certain distributions, such as the number of tracks with displaced vertices $\NDisplaced$, which could be used to isolate Instanton events. As  the cross section dependence on $\mInst$ of QCD Instanton and \softQCD processes is very different. it is also non trivial to tune QCD Instanton sensitive distributions for  different $\mInst$ regions.  This can be taken as a further motivation for studying Instanton production at low mass in different mass ranges, i.e. $20<\mInst<40$~\GeV{} and $40<\mInst<80$~\GeV.

\subsection{LHC Projections}
Having defined suitable signal regions, we evaluate here the expected 95\% confidence level (CL) upper limits on the instanton cross-section. 
For each range of  Instanton mass, the respective \textit{tight} signal region selection is applied.  The  systematic uncertainty on the background is estimated as described in the previous sections,  and ranges from about 20\% at low invariant masses to about 50\% at high invariant masses.
The limits are then performed as counting experiment with the \verb|pyhf| package~\cite{pyhf}. 
The results are shown in Fig.~\ref{fig:Limit1} for different assumptions on the integrated luminosity of 1~\ipb, 100~\ipb, and 10~\ifb.
We can see how even with only 1~\ipb, the predicted Instanton cross-sections can be excluded for masses up to about 150~\GeV, excluding at low masses cross-section ten times smaller than those predicted by \cite{Khoze:2019jta}. Increasing the collected luminosity to 100~\ipb{} would extend the limit at large invariant masses to 300~\GeV, with a negligible improvement of the limit at low masses. A further integrated luminosity increase by an additional factor of 100 would push the limit to the 400~GeV mass point.
\begin{figure}[th]
\centering
\begin{minipage}{7.1cm}
\centering
\includegraphics[width=1.0\linewidth]{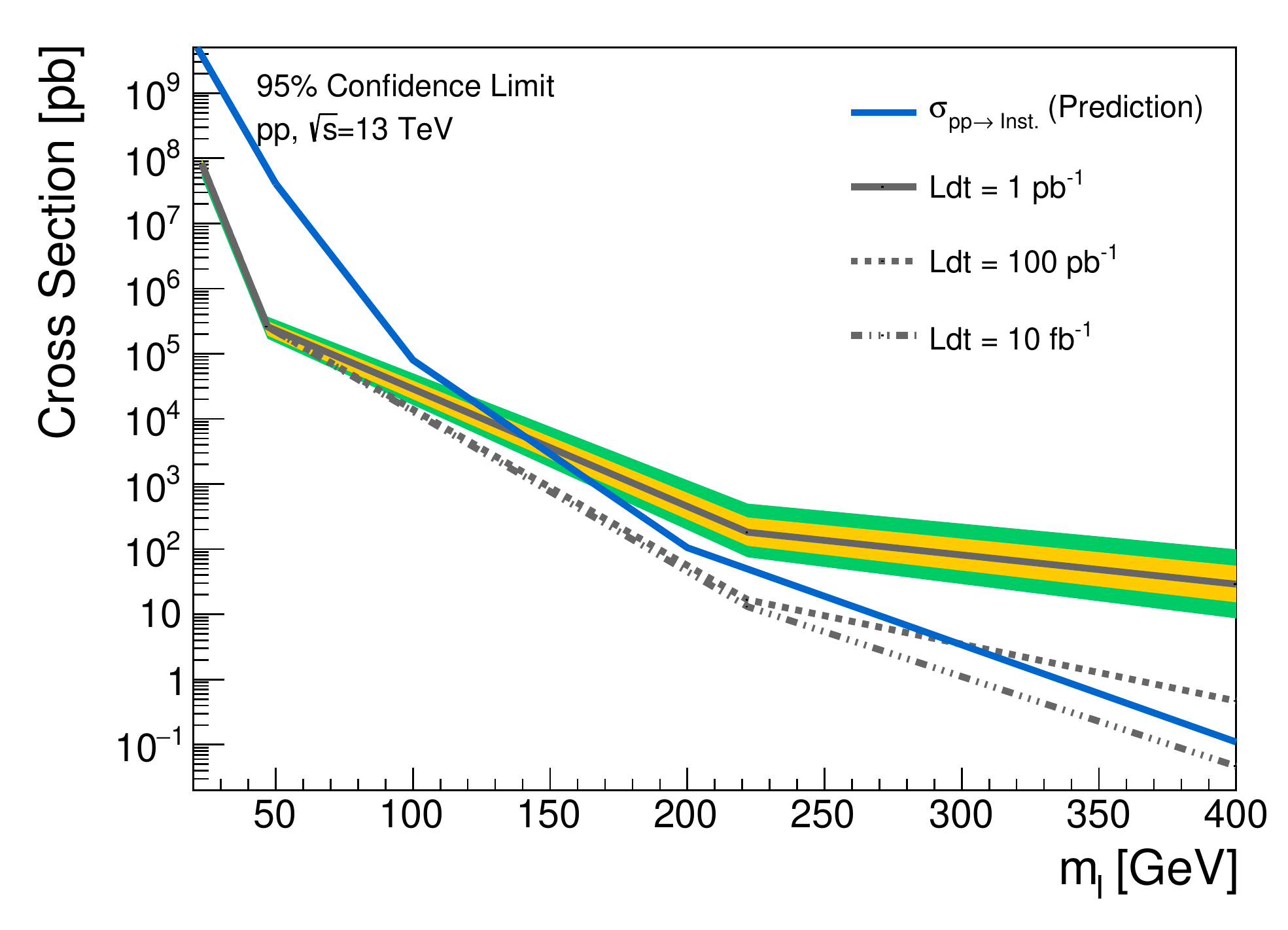}

\caption{Expected exclusion limits at 95\% CL on the cross section for Instanton induced processes for three different assumed integrated luminosities . Conservative systematic uncertainties on the background modelling have been made. \vspace{2.4cm}\label{fig:Limit1}}
\end{minipage}
\hspace{0.3cm}
\begin{minipage}{7.1cm}
  \centering
  \includegraphics[width=1.0\linewidth]{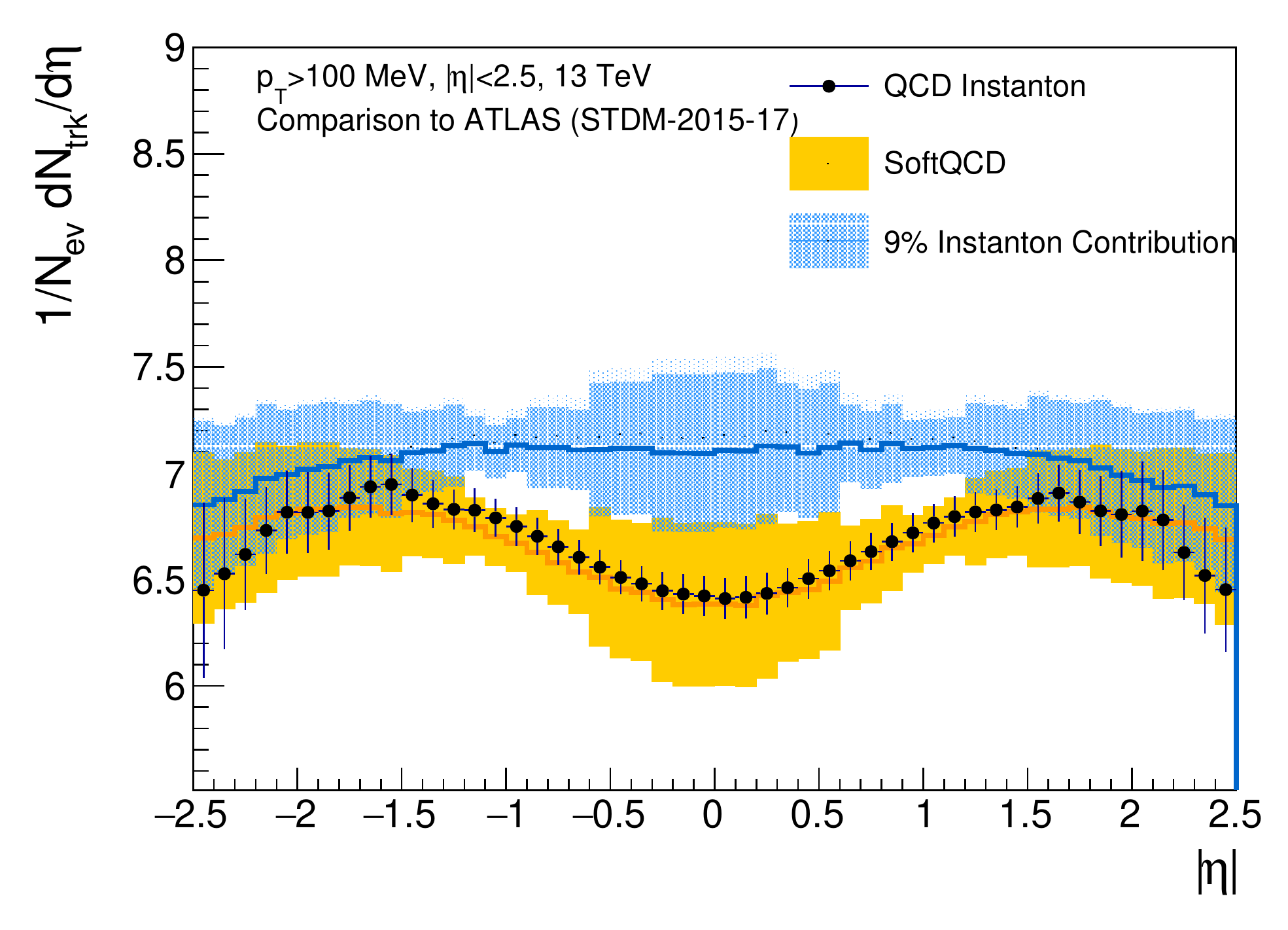}
\caption{Primary charged-particle multiplicities as a function of for events with at least two primary charged particles with \pT>100~\MeV{} and |$\eta$|<2.5, each with a lifetime $\tau$> 300~ps. The black dots represent the measurement of ATLAS \cite{Aaboud:2016itf}, while the \softQCD prediction is produced with \textsc{Pythia8}. Also shown is a prediction which includes a contribution of 9\% on Instanton induced processes. The model uncertainties are indicated by the bands \label{fig:Limit2}}
\end{minipage}%
\end{figure}

\subsection{First limit on Instanton Processes from proton-proton collision data}

As already indicated in Figure \ref{fig:LowMass1Overview}, the distribution of charged particles vs. their pseudorapidity is a sensitive observable for QCD Instanton processes in the low mass regime, which is in fact a standard distribution to be measured in the context of \softQCD studies \cite{Aaboud:2016itf, ALICE:2011ac, Aad:2014jgf, Chatrchyan:2012tb}. This distribution has been therefore also measured previously by the ATLAS Collaboration \cite{Aaboud:2016itf} for events with at least two primary charged particles with \pT>100~\MeV{} and $|\eta|<$2.5, each with a lifetime $\tau$>300~ps. We studied to which extent these measurements can be used to constrain QCD Instanton production. The  \textsc{Rivet}-routine for the ATLAS analysis was applied on the simulated \softQCD sample based on \textsc{Pythia8} as well as the Instanton signal sample with $\smin>25\,$~\GeV. A modelling uncertainty on the \softQCD prediction was estimated by considering the envelope of the \textsc{Pythia}, \textsc{Herwig7} and \textsc{Sherpa} samples. Since no second generator for Instanton processes for this low mass regime is currently available, the differences in the $\Etatrk$ distribution at $\mInst\approx 500$~\GeV{} between the \textsc{Sherpa} and \textsc{Herwig7} predictions have been taken as approximation. The resulting signal and background uncertainties have been treated once fully bin-to-bin correlated uncertainties as well as once fully bin-to-bin uncorrelated. 

The Instanton distribution has been added with a scaling factor $\alpha$ to the predicted \softQCD distribution, normalized accordingly and then fitted via a $\chi^2$ minimization approach to the data distribution, shown in Figure \ref{fig:Limit2}. The fit yields a maximal value of $\alpha=0.09$ and $\alpha=0.03$ at 95\% CL, assuming bin-to-bin correlated as well as bin-to-bin uncorrelated uncertainties on the predictions, respectively. The expected shape for an Instanton contribution of 9\% to the standard \softQCD processes is also indicated in Figure \ref{fig:Limit2}. The fiducial cross section defined by the selection  \pT>100~\MeV{} and $|\eta|<$2.5, each with a lifetime $\tau$> 300~ps can be estimated to be $\sigma = 71$\,mb, when taking the integrated luminosity of $\int L dt = 151\,\mu$b$^{-1}$, the number of selected events $N=9.3\cdot10^6$ and assuming a detector efficiency of $\epsilon=0.87$. Hence an upper limit on Instanton induced processes with $\smin>25$~\GeV{} can be placed between 2.1 and 6.4\,mb, depending on the correlation scenario assumed. In principle, further measured distributions could be used to derive more stringent limits, however, we think a dedicated analysis effort by the LHC collaborations would be the right next step, to shed light on QCD Instanton processes.

\section{\label{Sec:Conclusion}Conclusion}

In this paper we presented  detailed studies towards possible analysis strategies to observe Instanton induced processes in proton-proton collisions at the Large Hadron Collider. Several observables have been identified, which allow to effectively separate signal and background events. In order to study Instanton processes at higher energies, special triggers might have to be implemented for the upcoming LHC runs. However, the situation is different for low energies which have significantly larger cross sections and should be in principle already recorded. It is concluded that the most promising phase-space region for an early observation is therefore at low energies, for Instanton masses below 100~\GeV, where the cross-section is very high.  Since the dominant background in this energy regime is from \softQCD processes, several methods to constrain and validate \softQCD models in dedicated control regions have been discussed. We find that with just 1~\ipb{} of integrated luminosity, the LHC can already probe this low mass Instanton regime. With 10~\ifb{} it would be possible to probe Instanton mass of up to 0.5~\TeV. In addition, available measurements of Minimum Bias data have been used to derive a first upper limit on the cross section of Instanton processes, yielding an upper bound of $6.4$\,mb for Instanton masses above 25~\GeV.

The methods described in this paper will hopefully boost dedicated search efforts at the LHC over the full Instanton mass range by several experiments, leading to a robust result based on different strategies. 



\section*{Acknowledgements}

We would like to thank Frank Krauss for all his help during the generation of Instanton signal samples with the \textsc{Sherpa} generator,  Simon Plaetzer and Andreas Papaefstathiou for providing us with a preliminary implementation of Instanton production in \textsc{Herwig7}, and Daniel Milne for his help in validating our event shape calculations. 


\bibliographystyle{unsrt}
\bibliography{paper}


\end{document}
\endinput